\documentclass[a4paper,12pt]{article}
\usepackage{amsmath,amssymb,amsthm,verbatim}
\usepackage{latexsym,graphicx}

\newcommand{\NF}{N_{\rm f}}

\newcommand{\beq}{\begin{eqnarray}}
\newcommand{\eeq}{\end{eqnarray}}
\newcommand{\non}{\nonumber\\}
\newcommand{\D}{\mathcal{D}}
\newcommand{\p}{\partial}
\newcommand{\Tr}{{\rm Tr}}
\newcommand{\tr}{{\rm tr}}
\newcommand{\diag}{{\rm diag}}

\newcommand{\vb}[1]{\vbox to #1 mm{}}
\newcommand{\vs}[1]{\vspace{#1 mm}}
\newcommand{\hs}[1]{\hspace{#1 mm}}
\newcommand{\bs}{\boldsymbol}
\newcommand{\ba}{\left( \begin{array}} 
\newcommand{\ea}{\end{array} \right)} 

\usepackage[usenames]{color}

\makeatletter
\@addtoreset{equation}{section}

\renewcommand{\thefootnote}{\fnsymbol{footnote}}
\newcommand{\thetablename}{Table}
\def\fnum@table{\thetablename\ \thetable}
\makeatother

\begin{document}

\thispagestyle{empty}
\begin{flushright}
YGHP-11-45; IFUP-TH/2010-11;  KUNS-2362
\end{flushright}
\vspace{10mm}
\begin{center}
{\Large \bf  Vortices and Monopoles \\
   in Mass-deformed $SO$ and $USp$ Gauge Theories
}
\\[15mm]
Minoru Eto${}^{1}$,
Toshiaki Fujimori${}^{2,3}$,
Sven Bjarke Gudnason${}^{4}$,  \\
Yunguo Jiang${}^{5,6}$,
Kenichi Konishi${}^{3,2}$,
Muneto Nitta${}^{7}$,
Keisuke Ohashi${}^{8}$

\vspace{0.5cm}
{\it\small 
$^1$ Department of Physics, Yamagata University, Yamagata 990-8560, Japan}\\
{\it\small
${}^2$ INFN, Sezione di Pisa,
Largo B.~Pontecorvo, 3, 56127 Pisa, Italy}\\
{\it\small
${}^3$ Department of Physics,``E. Fermi`'', University of Pisa,
Largo B.~Pontecorvo, 3, 56127 Pisa, Italy}\\
{\it\small
${}^4$
~Racah Institute of Physics, The Hebrew University,
Jerusalem 91904, Israel}\\
{\it\small
${}^5$ ~Institute of High Energy Physics, Chinese Academy of Sciences,
100049 Beijing, China}\\
{\it\small
${}^6$ ~Theoretical Physics Center for Science Facilities, Chinese Academy of Sciences,
100049 Beijing, China}\\
{\it\small
${}^7$ Department of Physics, and Research and Education
Center for Natural Sciences,}\\
{\it\small
 Keio University, Hiyoshi 4-1-1, Yokohama, Kanagawa 223-8521, Japan}\\
${}^8$ {\it\small
Department of Physics, Kyoto University, Kyoto
606-8502, Japan}

\end{center}

\vskip 6 mm

\bigskip
\bigskip

{\noindent \bf Abstract}

Effects of mass deformations on 
$1/2$ Bogomol'nyi-Prasad-Sommerfield (BPS)
non-Abelian vortices are studied in 
$4d$ ${\cal N}=2$ supersymmetric $U(1) \times SO(2n)$ and 
$U(1) \times USp(2n)$  
gauge theories, with $N_{f}=2n$ quark multiplets. 
The $2d$ ${\cal N}=(2,2)$ effective
worldsheet sigma models on the Hermitian symmetric spaces
$SO(2n)/U(n)$ and  $USp(2n)/U(n)$ found recently which describe the
low-energy excitations of the orientational moduli of the vortices, 
are generalized to the respective massive sigma models. 
The continuous vortex moduli
spaces are replaced by a finite number ($2^{n-1}$ or $2^{n}$) of
vortex solutions.  
The $1/2$ BPS kinks connecting different vortex vacua 
are magnetic monopoles in the $4d$ theory, trapped inside the vortex
core, with total configurations being 1/4 BPS composite states.
These configurations are systematically studied within the
semi-classical regime.

\bigskip

\newpage
\pagenumbering{arabic}
\setcounter{page}{1}
\setcounter{footnote}{0}
\renewcommand{\thefootnote}{\arabic{footnote}}
\tableofcontents
\section{Introduction}

The discovery of non-Abelian vortex solutions
\cite{Hanany:2003hp,Auzzi:2003fs} and the subsequent development \cite{Tong:2003pz}-\cite{Eto:2006pg} have led to a substantial improvement of  
our understanding of non-Abelian solitons in general, appearing in spontaneously broken gauge theories.   
 Among the techniques used to explore these objects are: the string
 theory construction due to Hanany-Tong \cite{Hanany:2003hp} which in
 turn gave rise to the K\"ahler quotient construction, the powerful
 moduli matrix method 
\cite{Eto:2005yh}--\cite{Eto:2006pg} 
(the determination of the moduli space \cite{Eto:2005yh}, 
the moduli space metric \cite{Fujimori:2010fk}, low-energy dynamics 
\cite{Eto:2006db,Eto:2011pj}, generalization to arbitrary gauge groups 
\cite{Eto:2008yi}--\cite{Eto:2009bg}, etc),
apart from the standard field-theoretic analyses
\cite{Auzzi:2003fs}-\cite{GJK}, have all proven invaluable.  

The effective theories of the fluctuations of non-Abelian vortex orientational
modes  have been derived in the field-theoretic approach~\cite{Auzzi:2003fs}, \cite{Shifman:2004dr}-\cite{Gorsky:2004ad} as well as from string theory~\cite{Hanany:2003hp}.  In theories with gauge group $U(N)$ and $N_{f}= N$ flavors
(hypermultiplets), the worldsheet effective action turns out to be a
two-dimensional $\mathbb{C}P^{N-1}$ sigma model. 

A particularly interesting result concerns the precise matching of the BPS spectra of the
four dimensional gauge theories (having $\mathcal{N}=2$ supersymmetry)
under study and the two-dimensional sigma models (having $\mathcal{N}=(2,2)$
supersymmetry) \cite{Dorey}. In a sense the Abelian monopoles in the
Coulomb phase become trapped inside a vortex in the Higgs phase~\cite{Tong:2003pz}, \cite{Hanany:2004ea}-\cite{Gorsky:2004ad}. 
The monopoles are realized as classical kinks along the 
vortex string when the unequal squark mass terms are introduced. This
in turn induces a potential on the 
sigma model, which has now $N$ vacua instead of a continuous
$\mathbb{C}P^{N-1}$ degeneracy.  

In the theory without the bare quark masses, $N$ vacua appear as a result of quantum effects.  The kinks interpolating different (vortex) vacua correspond to the light monopoles appearing as the singularities of the Seiberg-Witten curves.  
More precisely, these $N$ vortex vacua and the light monopoles
connecting them correspond to the physics of the  
  quantum $r=0$ vacua,  arising from the classical $r=N$ vacua of the theory with $N_{f}=N$ massless flavors \cite{CKM}-\cite{DKO}.  

In a recent development  \cite{GJK}, the 
worldsheet low-energy effective actions of non-Abelian vortices 
in gauge theories with $ U(1)\times SO(2n)$ or $ 
 U(1) \times USp(2n)$ gauge symmetry and
$\NF=2n$ flavors (hypermultiplets) 
\cite{Eto:2008yi}-\cite{Eto:2009bg} 
have been found explicitly.  
The effective theories 
found are non-linear sigma models on the Hermitian symmetric spaces (HSS):
  $SO(2n)/U(n)$ and $USp(2n)/U(n)$, respectively \cite{Eto:2008yi}.  
The construction
of  the effective action has been extended to some higher-winding vortices in
$U(1) \times SO(2n)$ as well as in $U(N)$ theories, and in all cases the
results  found agreed with those obtained in a more formal approach based on
symmetry  and group-theoretic considerations \cite{Group}. The results found
reduce  to the known $\mathbb C P^{N-1}$ effective action~\cite{Tong:2003pz,Hanany:2004ea,Shifman:2004dr,Gorsky:2004ad}  in the case of the $U(N)$
theory.

In this paper we will extend these non-linear sigma models to the case
of the mass-deformed theories. As in the $\mathbb{C}P^{N-1}$ models, 
a potential is induced on the sigma model.  Such a potential can be obtained
either in the bulk theory, or alternatively, in the worldsheet
effective theory by using the Scherk-Schwarz (SS) dimensional 
reduction.\footnote{Our analysis follows the method~\cite{Eto:2004rz}
  developed to obtain the mass-deformed theory in four space-time
  dimensions from a five space-time 
dimensional massless theory by the Scherk-Schwarz dimensional 
reduction \cite{Scherk:1978ta,Scherk:1979zr}.   
}
We then construct 1/4 BPS configuration of 
magnetic monopoles confined in a vortex-string 
as kinks in the vortex world-sheet effective theory with mass deformation.
By generalizing the operator method to create domain walls 
in the case of $SU(N)$ gauge theory \cite{Isozumi:2004va} 
(in which the operators are identified with the root vectors \cite{Sakai:2005sp}), 
we find the rule of ordering of kinks (monopoles on the vortex) in 
$SO$ and $USp$ gauge theories.

The paper is organized as follows.  
In Section \ref{sec:bulk}  we review the non-Abelian vortex solutions and their orientational zero modes 
in the bulk gauge theory and derive the vortex worldsheet effective action.  
These serve as the starting point for the discussion of the mass-deformed systems and 
the derivation of the massive sigma models in Section~\ref{Sec:Mdformed}.  
Section~\ref{sec:kinks} is dedicated to the analysis of 
the Abelian monopoles which appear along the vortex string as kinks.\footnote{  
When this paper was being written, a paper by Arai and Shin \cite{Arai:2011gg} appeared which 
discusses the mass-deformed sigma models in four (or three)  dimensions, 
without any reference to the non-Abelian vortices, 
but based on a similar model with $SO(2n)$ and $USp(2n)$ global symmetries.  
Their analysis of the domain walls naturally have some overlap with our analysis of the kink monopoles and, 
where the comparison is possible, our results seem to agree. } 
 A comment on  the solutions involving domain walls in the bulk is
given in Appendix \ref{appendix:wall}. 
The convention used for the Lie algebra is reviewed in Appendix~\ref{appendix:convention}, 
and some useful results are collected in  Appendix \ref{appendix:ordering}.  
The case of the effective worldsheet action for certain doubly wound 
vortices in $SO(2n)$ theory, which is given by a quadric surface
$Q^{2n-2}$ sigma model, is discussed briefly in
Appendix~\ref{appendix:quadric}.

\section{The bulk theory}\label{sec:bulk}

\subsection{The Model}
The theory we are considering is the standard $\mathcal{N}=2$
supersymmetric QCD with gauge group $U(1)\times G$ with
$G=SU(N),SO(2n)$ and $USp(2n)$ (while $N$ denotes in all cases the
dimension of the fundamental representation, i.e.~$N=2n$ for
$SO,USp$).  
The F-terms are given by variations of the following
superpotential
\beq
\mathcal{W} ~=~ \Tr\left[\tilde{Q}\Phi Q 
 - \tilde{Q} Q M \right] \ ,
\eeq
where $Q,\tilde{Q}$ are fundamental and anti-fundamental chiral superfields
comprising the $\NF$ hypermultiplets ($Q$ is an $N$-by-$\NF$ matrix, and
$\tilde{Q}$ is an $\NF$-by-$N$ matrix), 
while $\Phi$ is a chiral superfield belonging to a vector multiplet, 
in the adjoint representation. 
In this paper we will set the $\NF$-by-$\NF$ mass matrix to be real: 
$M^\dagger = M$. 
Using a flavor rotation we can diagonalize the mass matrix as 
$M=\diag(m_1,\cdots,m_{\NF})$. 
Furthermore, the trace part of $M$ can be 
absorbed by a shift of the adjoint scalar, 
so that we can assume that $\Tr \, M = 0$ without loss of generality.

We shall restrict ourselves 
to the case of $\NF=N$ flavors, 
for definiteness. 
This is the minimum number of flavors which allows for
a color-flavor locked vacuum. 
In the massless case, then, 
the theory has an $SU(N)$ flavor symmetry, in all cases, including
$U(1) \times SO(2n)$ or $U(1)\times USp(2n)$, with $N_{f}=N=2n$ flavors. 

Finally,  we can  set $\tilde{Q}=0$
and $\Phi^\dagger = \Phi$ for the consideration of the non-Abelian vortex solutions and small excitations around them.

Due to renormalization group (RG) flow,  the
gauge coupling constants for $U(1)$ and for $G$ would become distinct
at a lower energy scale even if we start with an equal, common
coupling constant.  However, as our results here do not substantially
depend on the $U(1)$ gauge coupling and also because our analysis
remains semi-classical throughout, we shall set $e=g$ for
simplicity.

The bosonic part of the Lagrangian density is then 
\begin{align}
\mathcal{L} ~=~ \Tr\left[
-\frac{1}{2g^2}F_{\mu \nu} F^{\mu\nu}
-\frac{\theta}{16\pi^2} F_{\mu\nu}\tilde{F}^{\mu\nu}
+\frac{1}{g^2}\left(\D_\mu \Phi\right)^2
+\left|\D_\mu Q\right|^2 \right]
-V_D - V_F \ ,
\end{align}
where we denote the scalar components of the superfields with the same
symbols as the superfields themselves. The scalar potentials read
\begin{align}
V_D &~=~ \frac{g^2}{2}\left| \Tr\left(Q Q^\dag t^\alpha\right) 
  -\xi^\alpha\right|^2 \ , \\
V_F &~=~ \Tr\left|\Phi Q - Q M\right|^2 \ , 
\end{align}
where $\alpha=0,1,\ldots,\dim(G)$ and $0$ stands for the
$U(1)$ part of the gauge group. $\xi^0=\xi$ is the only non-zero 
Fayet-Iliopoulos (FI)
parameter as $G$ is a simple group. 
Finally all generators are normalized as $\Tr(t^\alpha
t^\beta) = \delta^{\alpha\beta}/2$. 

Since the mass matrix $M$ is diagonal and traceless, it belongs to an element
of the Cartan subalgebra of $G$. 
There exists a Higgs vacuum in which the vacuum expectation values
(VEVs) of the scalar fields take the form
\beq
\langle \Phi \rangle ~=~ M\ , \hs{10} \langle Q \rangle ~=~ v \mathbf 1_N\ , \hs{10}
\left(v \equiv \sqrt[4]{\frac{2}{N}}\sqrt{\xi} \right)\ . \label{eq:vev}
\eeq
In this paper we focus our attention on this system.

For the massless case $M=0$, the VEV of the adjoint scalar $\Phi$
does not break the $U(1) \times G$ gauge symmetry, while the squark
VEV breaks it completely, bringing the system in the Higgs
phase. However, the global color-flavor diagonal $G_{\rm C+F}$
symmetry remains intact. 
 
In the theory with a generic bare mass-matrix $M$, the gauge group is 
broken to $U(1)^r$ where $r={\rm rank} \, G$. 
As already mentioned, $M$ can be rotated into a diagonal form 
without loss of generality. 
For the $G=SU(N)$ theory, the generic mass matrix $M$ takes the form
$M=\diag(m_1,\cdots,m_{N})$. 
While in the cases of $G=SO(2n)$ and $G=USp(2n)$, 
we will be working in the usual basis $U^{\rm T}J U = J$, with
$J=\sigma^1\otimes\mathbf{1}_n$ and $J=i\sigma^2\otimes\mathbf{1}_n$,
respectively. In this basis the mass matrix reads
$M = \diag(m_1,\cdots,m_n,-m_1,\cdots,-m_n)$. 
The properties of the solitons will be seen to depend crucially on how
the components of the mass matrix $M$ is chosen. 

There are several kinds of topological solitons in our models. Among them
we are interested in vortices and monopoles.
The $U(1)$ symmetry breaking at the energy $\left<Q\right> \sim v$ gives rises 
to vortices, while the breaking of $G \to U(1)^r$ at the energy $\left<\Phi\right> \sim m$ gives
rises to monopoles. 
The static 1/4 BPS equations 
for composite states of the vortex and monopole
\cite{Isozumi:2004vg} can easily be found to be 
\begin{align}
0 &~=~ \D_{\bar z} Q \ , \label{eq:BPS1} \\
0 &~=~ \D_3 Q + \Phi Q - Q M \ , \label{eq:BPS3} \\
0 &~=~ \D_{\bar z} \Phi + i F_{\bar z3}
= [\D_{\bar z}, \Phi+\D_3]
 \ , \label{eq:BPS4}\\
0 &~=~ \D_3 \Phi - F_{12} + g^2\left[\Tr\left(Q Q^\dag t^\alpha\right) t^\alpha 
  - \xi t^0 \right] \ , \label{eq:BPS2} 
\end{align}
where $\D_{\bar z} = \frac{1}{2} (\D_1+i\D_2)$ and 
we have chosen the vortex string to lie on the $x_3$-axis. 
It is well known that the last two equations  describe 
t'Hooft-Polyakov monopoles in the Coulomb phase, $Q=\xi=0$ 
while the first and the third equations describe vortex solutions if $A_3=\Phi=M=0$. The second 
equation, therefore, describes  the kinks regarded as monopoles 
 along the vortex string which
appear in the Higgs phase $\xi\not=0$ of the mass-deformed theory.
When these first order differential equations are obeyed, the Bogomol'nyi bound
will be saturated 
\begin{align}
\mathcal{E} ~=~ \Tr\left[ \frac{1}{g^2} \epsilon_{ijk} \p_i(\Phi F_{jk}) -2 \xi F_{12} t^0 + 2 \xi \p_3 \Phi t^0 
- i\epsilon_{nm} \p_n \left[(\D_m Q )Q^\dag \right]
- \p_3 \left[(\Phi Q - Q M) Q^\dag\right] \right] \ , 
\label{eq:bulk_BPS_energy}
\end{align}
where $i,j,k=1,2,3$; while $n,m=1,2$ denote the directions in
the transverse plane. 
The first two terms in the energy density describe the monopole and vortex,
 respectively; the others are boundary terms.

The 1/4 BPS equations (\ref{eq:BPS1})--(\ref{eq:BPS2}) 
can be formally solved in terms of the moduli
matrix \cite{Eto:2005yh,Isozumi:2004vg,Eto:2006pg}.
The first equation describes distributions of $Q$ in the $x_1$-$x_2$ plane 
and the second one determines the $x_3$-dependence of 
$Q$ for a given set of $A_i$ and $\Phi$. 
Note that the third equation is just an integrability condition
which tells us only that the first two equations can be solved
simultaneously. 
The integration constants for $Q$ can be summarized in an 
$N$-by-$N_{f}$ holomorphic matrix in the complex coordinate 
$z=x_1+ix_2$, which is called the
moduli matrix. With this moduli matrix,
 the last equation can only be solved numerically and tells us about
the magnetic flux distribution in the plane transverse to 
the vortex string. 
In this paper, however, we do not treat the 1/4 BPS equations and
their solutions directly.  
Note that  
the integrability condition (\ref{eq:BPS4})
allows us to regard 
the second equation as a description of the $x_3$-dependence
 for the moduli of the vortices.
We will consider 1/2 BPS 
kink solutions in an effective theory on 1/2 BPS vortices
corresponding to the 1/4 BPS composite states.
We should remark that the 1/4 BPS equations 
(\ref{eq:BPS1})--(\ref{eq:BPS2}) admit, in general, 
domain walls and 1/4 BPS composite states 
of vortices stretched between domain walls 
\cite{Isozumi:2004vg,Eto:2006pg,Eto:2008mf},
when the VEVs of the scalar fields
are different at $x_3 \rightarrow - \infty$ and $x_3 \rightarrow \infty$.
Furthermore, the BPS equations admit 1/2 semi-local \cite{Eto:2008qw} and fractional vortices \cite{Eto:2009bz}.
In the rest of this paper, we do not discuss such configurations
and consider only configurations of local vortices and monopoles 
which are invariant under rotations in the $x_1$-$x_2$ plane.
The precise condition for the absence of domain walls, 
semi-local and fractional vortices can be found in Appendix \ref{appendix:wall}.

\subsection{Orientational zero modes of the non-Abelian vortices\label{zeromodes}}
Let us first consider the 1/2 BPS solutions of local-vortex strings 
without monopoles in the massless
case, $M=0$, for which $\Phi=0$, $A_3=0$ and $\p_3=0$. 
Here only the first and the last equations in the $1/4$ BPS
equations  remain non-trivial.
We impose the following boundary
condition for the scalar field $Q$ and the gauge field $A_{\theta}$ 
such that it has non-trivial winding in the gauge orbit of the scalar
VEV
\beq
Q ~\rightarrow~ v \, \exp ( i \theta \lambda )\ , \hs{10} A_{\theta}
~\rightarrow~ - \frac{1}{r}\, \lambda \hs{10} (r \rightarrow \infty)\ ,
\eeq
where $ r e^{i \theta}=x_1+ix_2 $.  $\lambda$ is an $N$-by-$N$ constant
matrix, which  can be diagonalized as
\beq
\lambda ~~\rightarrow~~ U \lambda \, U^\dagger ~=~ \nu_0 \mathbf 1_N +
\tilde{\boldsymbol \nu} \cdot \mathbf H\ , \hs{10} U \in G\ ,
\eeq
where $\mathbf H = (H_1, H_2, \cdots, H_r)$ is 
a basis of the Cartan subalgebra of $G$ 
(see Appendix \ref{appendix:convention} 
for the conventions for the Lie algebra). 
Since the scalar field $Q$ is a single-valued function, 
all the eigenvalues of $\lambda$ should be integers. 
This condition is satisfied only if 
$\nu_0$ is quantized in the following way \cite{Eto:2008yi}
\beq
{\renewcommand{\arraystretch}{1.5}
\nu_0 ~=~ k \times \left\{ 
\begin{array}{cl} 
\displaystyle \frac{1}{N} & \mbox{for $G=SU(N)$} \\
\displaystyle \frac{1}{2} & \mbox{for $G=SO(2n)$} \\
\displaystyle \frac{1}{2} & \mbox{for $G=USp(2n)$} 
\end{array} \right., \hs{10} k \in \mathbb Z\ .}
\eeq
The integer $k$ corresponds to the vortex winding number 
classifying the topological sectors of the vortex configurations.
In addition to this quantization condition for $\nu_0$, 
the single-valuedness condition is satisfied only when 
the coefficient vector $\tilde{\boldsymbol \nu}$ is a coweight, 
namely $\tilde{\boldsymbol \nu}$ should be 
one of the weight vectors of the dual group $\widetilde G$. 
Here $\widetilde G$ is the dual group whose 
root vectors $\tilde{\bs \alpha}_i$ are related to those of $G$ as
\beq
\tilde{\boldsymbol \alpha}_i ~=~ 2 \, \frac{\boldsymbol \alpha_i}{\boldsymbol
  \alpha_i \cdot \boldsymbol \alpha_i}\ .
\eeq
For the single winding configurations $(k=1)$, 
the vector $\tilde{\boldsymbol \nu}$ is given by\footnote{
In the case of $G=SO(2n)$, there are two disjoint topological sectors
classified by $\mathbb Z_2$ topological charge \cite{Eto:2009bg}.
For $k=1$, the vortices with different $\mathbb Z_2$ charges are  
characterized by the weight vectors of the Weyl spinors with opposite 
chirality.} 
\beq
\tilde{\bs \nu} ~=~ \mbox{weight vector of}~ \left\{ 
\begin{array}{lll} 
\mathbf N &\mbox{(fundamental rep. of $SU(N)$)} & \mbox{for $G=SU(N)$} \\
\mathbf{2^{n-1}} & \mbox{(Weyl spinor rep. of $SO(2n)$)}& \mbox{for $G=SO(2n)$} \\
\mathbf{2^{n}} & \mbox{(spinor rep. of $SO(2n+1)$)}& \mbox{for $G=USp(2n)$} 
\end{array} \right..
\label{eq:weight}
\eeq
If we choose the highest weight vector $\tilde{\boldsymbol \nu}_h$ of each representation, 
the diagonal matrix 
$\lambda_{h} = \nu_0 \mathbf 1_N + \tilde{\boldsymbol \nu}_{h} \cdot \mathbf H$ takes the form
\beq
\lambda_h ~=~ \ba{c|c} \mathbf 1_p & \\ \hline & \mathbf 0_q \ea, \hs{10}
(p,q) ~=~ \left\{ 
\begin{array}{cl} 
(1,N-1) & \mbox{for $SU(N)$} \\
(n,n) & \mbox{for $SO(2n)$} \\
(n,n) & \mbox{for $USp(2n)$} 
\end{array} \right. ,
\label{eq:T_matrix}
\eeq
where the subscript of $\lambda_h$ stands for the highest weight vector,
which we denote by $\tilde{\bs \nu}_h$. 
For the other weight vectors, 
the matrix $\lambda_h$ can be obtained by using 
the transformations which exchange the eigenvalues of $\lambda$, 
i.e.~the Weyl group elements of $G$.
For the matrices $\lambda_h$ given above, 
the basic single winding vortex solution takes the following form
\beq
Q &=& v \ba{c|c} z e^{-\frac{1}{2} \psi} \mathbf 1_p & \\ \hline & \ \mathbf
1_q \ \ea ~=~ v \, \exp \left[ \left( \log z - \frac{1}{2} \psi \right)
  \lambda_h \right]\ , \label{eq:sol_0Q} \\ 
A_{\bar z} &=& -\frac{i}{2} \ba{c|c} \p_{\bar z} \psi \, \mathbf 1_p & \\
\hline & \ \mathbf 0_q \ \ea ~=~  - \frac{i}{2} \p_{\bar z} \psi \, \lambda_h\
, 
\label{eq:sol_0A}
\eeq
where $z=x_1+ix_2$ and $\psi$ is a smooth real function satisfying
\beq
4 \p_z \p_{\bar z} \psi ~=~ g^2 v^2 ( 1 - |z|^2 e^{-\psi} )\ ,
\eeq
with $\psi \rightarrow \log |z|^2$ at spatial infinity. 
From the solution \eqref{eq:sol_0Q}-\eqref{eq:sol_0A} and 
the asymptotic behavior of the function $\psi$, 
we can see that the matrix $\lambda_h$ is nothing but 
the magnetic flux  of the vortex 
\beq
\int d x_1 d x_2 \, F_{12} ~=~ -2 \pi \lambda_h\ , \hs{10} 
\Big( F_{12} = - 2 \p_z \p_{\bar z} \psi \, \lambda_h \Big)\ .
\label{eq:flux}
\eeq
In the massless theory $(M=0)$, there exists an
unbroken color-flavor symmetry $G_{C+F}$ in the vacuum. 
Therefore we can obtain a set of vortex solutions 
by rotating the configuration \eqref{eq:sol_0Q}-\eqref{eq:sol_0A} as 
\beq
Q \rightarrow U^\dagger Q \, U\ , \hs{5} A_{\bar z} \rightarrow U^\dagger
A_{\bar z} \, U\ , \hs{10} U \in G_{C+F}\ .
\eeq
This set of solutions is parametrized by the orientational moduli, 
which are the coordinates of a coset space of the form
\beq
\mathcal M_{\rm orientation} ~\cong~ G/H\ ,
\eeq
where $H$ is a subgroup of $G$ which does not change the solution
\eqref{eq:sol_0Q}-\eqref{eq:sol_0A}, that is
\beq
h \in H ~~\Longleftrightarrow~~ h^\dagger \lambda_h h ~=~ \lambda_h\ .
\eeq
The orientational moduli of the vortices can be interpreted as
the Nambu-Goldstone zero modes arising from the breaking of the
color-flavor symmetry $G_{C+F}$ due to the vortex configuration. 
For the single winding vortex 
in the $G=SU(N)$, $SO(2n)$ and $USp(2n)$ theories, 
the orientational moduli spaces are given by \cite{Eto:2008yi}
\beq
\mathcal M_{\rm orientation} ~\cong~ \frac{SU(N)}{SU(N-1) \times U(1)}\ , \hs{5} 
\frac{SO(2n)}{SU(n) \times U(1)}\ , \hs{5} 
\frac{USp(2n)}{SU(n) \times U(1)}\ .
\eeq

The physical meaning of these orientational moduli can be seen 
from the magnetic flux \eqref{eq:flux}. 
Under the color-flavor global symmetry $G_{C+F}$, 
the magnetic flux transforms as
\beq
\lambda_h ~\rightarrow~ \lambda ~=~ U^\dagger \lambda_h U\ , \hs{10} U \in G\ .
\eeq
This implies that the orientational moduli parametrize 
the internal direction of the vortex magnetic flux 
in the Lie algebra of the gauge group $G$.

Since the subgroup $H$ is unbroken, 
not all parameters in the matrix $U$ correspond to 
physical zero modes of the vortex. 
We can parametrize the coset space as follows. 
Let $E_{-\bs \alpha}$ be the lowering operators in the Lie algebra of $G$ 
for which the generator $\lambda_h$ has negative eigenvalues, namely
\beq
[\lambda_h, E_{-\bs \alpha}] ~=~ - (\tilde{\boldsymbol \nu}_h \cdot \boldsymbol
\alpha) \, E_{-\bs \alpha}\ , \hs{10} \tilde{\boldsymbol \nu}_h \cdot
\boldsymbol \alpha > 0\ .  
\eeq
Note that in our convention 
$[\mathbf H, E_{-\bs \alpha}] = - \boldsymbol \alpha E_{-\bs \alpha}$. 
Then the generic matrix $U \in G$ containing 
only physical parameters (i.e.~called the reducing matrix) 
can be constructed as follows
\beq
U ~=~ g_u g_l\ , \hs{10} 
g_l ~=~ \exp \left( \sum_{\tilde{\boldsymbol \nu}_h \cdot \boldsymbol \alpha_i >
    0} b^i E_{-\bs \alpha_i} \right)\ .
\eeq
The parameters $b^i$, which are associated with the root vectors 
$\bs \alpha_i$, 
can be interpreted as the complex coordinates 
which parametrize the orientational moduli space. 
These parameters cover the coordinate patch containing the point 
on the moduli space which corresponds to 
the BPS configuration \eqref{eq:sol_0Q}-\eqref{eq:sol_0A}. 
The matrix $g_u$ is an element of the group $P$, 
(a parabolic subgroup of $G^{\mathbb C}$) 
generated by the Cartan generators $\mathbf H$, 
all the raising operators $E_{\bs \alpha}$ and 
the lowering operators $E_{-\bs \alpha}$ with
$\tilde{\boldsymbol \nu}_h \cdot \boldsymbol \alpha = 0$. 
Formally, the matrix $g_u$ can be written as
\beq
g_u ~=~ \exp \left[ \bs \theta \cdot \mathbf H + \sum_{\bs \alpha_i \in \Delta^+} c^i E_{\bs \alpha_i} + 
\sum_{\tilde{\bs \nu}_h \cdot \bs \alpha_i = 0} d^i E_{-\bs \alpha_i}
\right]~\in~P\ ,
\eeq
where $\Delta^+$ denotes the set of all positive roots. 
For a given set of parameters $b^i$, 
the matrix $g_u$ can be determined 
up to $H$ transformations $g_u \rightarrow h g_u$ 
from the unitarity condition $U U^\dagger = \mathbf 1_N$, that is
\beq
g_u^\dagger g_u ~=~ (g_l g_l^\dagger)^{-1}\ .
\eeq
Note that the choice of $h \in H$ is not important 
since it does not change the BPS vortex configuration.
For the generator $\lambda_h$ in Eq.\,\eqref{eq:T_matrix}, the matrix 
$\sum_{\tilde{\bs \nu} \cdot \bs \alpha_i > 0} b^i E_{-\bs \alpha_i}$ 
takes the lower triangular matrix  form
\beq
\sum_{\tilde{\bs \nu} \cdot \bs \alpha_i > 0} b^i E_{-\bs \alpha_i} ~=~ \ba{c|c}
\mathbf 0_p & \\ \hline B & \mathbf 0_q \ea\ .
\eeq
The $q$-by-$p$ matrix $B$ is an $(N-1)$-vector for $G=SU(N)$
and an anti-symmetric and symmetric $n$-by-$n$ matrix 
for $G=SO(2n)$ and $USp(2n)$, respectively
\beq
B ~~=~~ 
\left( 
{\renewcommand{\arraystretch}{0.8}
{\arraycolsep .5mm
\begin{array}{c} b_1 \\ b_2 \\ \vdots \\ b_{N-1} \end{array}}} \right)\ ,
\hs{5}
\left( 
{\renewcommand{\arraystretch}{0.7}
{\arraycolsep .5mm
\begin{array}{cccc} 
0 & b_{12} & \cdots & b_{1n} \\
-b_{12} & 0 & \ddots & \vdots \\
\vdots & \ddots & \ddots & b_{n-1,n} \\
-b_{1n} & \cdots & -b_{n-1,n} & 0  
\end{array}}} \right)\ , 
\hs{5}
\left( 
{\renewcommand{\arraystretch}{0.8}
{\arraycolsep .7mm
\begin{array}{cccc} 
b_{11} & b_{12} & \cdots & b_{1n} \\
b_{12} & b_{22} & \ddots & \vdots \\
\vdots & \ddots & \ddots & b_{n-1,n} \\
b_{1n} & \cdots & b_{n-1,n} & b_{nn} 
\end{array}}} \right)\ .
\eeq
Then, the matrix $g_l$ is given by
\beq
g_l ~=~ \mathbf 1_N + \sum_{\tilde{\bs \nu} \cdot \bs \alpha_i > 0} b^i
E_{-\bs \alpha_i} ~=~ \left( \begin{array}{c|c} \mathbf 1_p & 0 \\ \hline B &
    \mathbf 1_q \end{array} \right)\ ,
\eeq
where we have used that $(\sum b^i E_{-\bs \alpha_i})^2 = 0$.
The corresponding matrix $g_u$ takes the following form
\beq
g_u ~=~ \left( \begin{array}{c|c} \mathbf 1_p & -B^\dagger \\ \hline 0 & \
    \mathbf 1_q \ \end{array} \right) \left( \begin{array}{c|c}
    X^{\frac{1}{2}} & \\ \hline & Y^{-\frac{1}{2}} \end{array} \right)\ , 
\eeq
where $X^{\frac{1}{2}}$ and $Y^{-\frac{1}{2}}$ are 
invertible Hermitian matrices defined by\footnote{
The square root of the matrices $X$ and $Y$ always exists. 
For $G=SO(2n)$ and $USp(2n)$, it can be written as 
$X^{\frac{1}{2}} = U_X (1+|D|^2)^{\frac{1}{2}} U_X^\dagger$ and 
$Y^{\frac{1}{2}} = U_Y (1+|D|^2)^{\frac{1}{2}} U_Y^\dagger$ 
where $U_X$ and $U_Y$ are unitary matrices and 
$D$ is a diagonal matrix such that $B = U_Y D \, U_X^\dagger$.}
\beq
X ~=~ \mathbf 1_{p} + B^\dagger B\ , \hs{10} Y ~=~ \mathbf 1_{q} + B B^\dagger\ .
\label{eq:XY}
\eeq
Therefore the matrix $U$ containing only the physical parameters $b^i$ 
takes the form
\beq
U ~=~ \left( \begin{array}{c|c} \mathbf 1_p & - B^\dagger \\ \hline 0 & \
    \mathbf 1_q \ \end{array} \right) \left( \begin{array}{c|c}
    X^{\frac{1}{2}} & \\ \hline & Y^{-\frac{1}{2}} \end{array} \right)
\left( \begin{array}{c|c} \mathbf 1_p & 0 \\ \hline B & \mathbf
    1_q \end{array} \right)\ .
\label{eq:matrix_U}
\eeq
Such a matrix has been introduced earlier and termed the reducing matrix \cite{Delduc:1984sz}.

Next, let us discuss the effective action for the orientational moduli 
by using the matrix given in Eq.\,\eqref{eq:matrix_U}. 
At a sufficiently low energy scale, the dynamics of the zero modes 
is described by a non-linear sigma model on the moduli space and 
its effective action can be obtained as follows. 
The fluctuations of the orientational zero modes 
along the vortex worldsheet are represented 
by the matrix $U(t,x_3) \in G$ which depends weakly on 
the worldsheet coordinate $t$ and $x_3$. 
It induces the fluctuation fields 
around the ``slowly moving'' vortex background.
Up to first order in the derivatives $\p_\alpha~(x_\alpha=t,x_3)$, 
the scalar field $Q$ and $A_{\bar z}$ are not modified
\beq
Q &=& U(x_\alpha)^\dagger Q_0 \, U(x_\alpha) + \mathcal O(\p_\alpha^2)\ , \\
A_{\bar z} &=& U(x_\alpha)^\dagger A_{0 \bar z} \, U(x_\alpha) + \mathcal
O(\p_\alpha^2)\ ,
\eeq
where $Q_0$ and $A_{0 \bar z}$ are the static BPS configurations. 
For notational simplicity, we use the following singular gauge fixing 
\beq
Q_0 ~=~ v \exp \left[ \left( \log |z| - \frac{1}{2} \psi \right) \lambda_h
\right]\ , \hs{10}
A_{0 \bar z} ~=~ - \frac{i}{2} \left( \p_{\bar z} \psi - \frac{1}{\bar{z}}
\right) \lambda_h\ .  
\eeq
The fluctuations of the gauge fields along the vortex worldsheet $A_\alpha$ 
can be determined by solving the equations of motion 
\beq
\frac{2}{g^2} \D^\mu F_{\mu \alpha} ~=~ i \left[ ( \D_\alpha Q) Q^\dagger - Q
  (\D_\alpha Q^\dagger)\right]\ .
\eeq
In the slowly moving background, the solution is given by\footnote{
A general formula for the solution can be obtained by using
the moduli matrix approach, 
see \cite{Eto:2006uw,Eto:2006pg} for the $U(N)$ case.
} 
\beq
A_{\alpha} ~=~ - i U^\dagger \p_\alpha U + i U^\dagger \left[ Q_0
(\delta_\alpha^\dagger U) U^\dagger Q_0^{-1} - {\rm h.c.} \right] U 
+ \mathcal{O}(\p_\alpha^3)\ ,
\label{eq:A_alpha}
\eeq 
where we have decomposed the derivative $\p_\alpha$ into
the holomorphic and anti-holomorphic parts as
\beq
\p_\alpha = \delta_{\alpha} + \delta_{\alpha}^\dagger\ , \hs{10} \delta_{\alpha}
= \p_\alpha b^i \frac{\p}{\p b^i}\ , \hs{5} \delta_{\alpha}^\dagger = \p_\alpha
\bar b^{\bar i} \frac{\p}{\p \bar b^{\bar i}}\ .
\eeq
Although $Q_0^{-1}$ is singular at the vortex center,
we can check that the gauge field $A_\alpha$ is non-singular 
by using the explicit form of $U$ given in Eq.\,\eqref{eq:matrix_U}
\beq
A_\alpha ~=~ i \big(1 - |z| e^{-\frac{1}{2} \psi} \big) \, U^\dagger
\ba{c|c} 0 & X^{-\frac{1}{2}} \p_\alpha B^\dagger Y^{- \frac{1}{2}} \\ \hline 
- Y^{-\frac{1}{2}} \p_\alpha B X^{- \frac{1}{2}} & 0 \ea U\ , 
\eeq
where we have used $\delta_\alpha B = \p_\alpha B$ and 
$\delta_\alpha B^\dagger = \p_\alpha B^\dagger$.
Substituting the scalar field $Q$ and gauge field $A_\mu$ 
into the bulk action and integrating over the $(x_1,x_2)$-plane, 
we obtain the effective action of the form
\beq
S_{\rm eff} ~=~ \int dt dx_3 \left( - T_v + g_{i \bar j} \p_\alpha b^i \p^\alpha
  \bar b^{\bar{j}} \right)\ ,
\eeq
where the first term is the tension of the vortex string and 
$g_{i \bar j}$ is the metric of the orientational moduli space, 
which is given in terms of the matrices $g_l$ and $g_u$ by
\beq
g_{i \bar j} ~=~ \frac{4\pi}{g^2} \frac{\p}{\p \bar b^{\bar j}} \Tr \left[ \left( g_l
    \frac{\p}{\p b^i} g_l^{-1} \right) \, g_u^{-1} \lambda_h g_u \right]\ .
\eeq 
If we use the matrix $B$, 
the effective action on the vortex worldsheet 
takes the form \cite{GJK}
\beq
S_{\rm eff} ~=~ \frac{4\pi}{g^2} \int dt dx_3 \; \Tr \Big[ (\mathbf{1}_q + B
B^\dagger)^{-1} \p_\alpha B (\mathbf{1}_p + B^\dagger B)^{-1} \p^\alpha B^\dagger
\Big]\ ,
\label{eq:massless_S}
\eeq
where we have ignored the string tension $T_v$.

\subsection{Mass deformation\label{massdeform}}
So far we have discussed the orientational moduli and 
their effective action in the massless theory. 
Next let us discuss the effect of the bulk mass term 
on the effective action of the vortex string. 

The mass matrix $M$ is taken to be in the Cartan subalgebra of $G$
\beq
M ~=~ \mathbf m \cdot \mathbf H\ ,\quad
\mathbf m = \left(m_1,m_2,\cdots,m_r\right).
\eeq
If $\mathbf m$ is a generic vector, 
all the mass elements are different and 
the color-flavor symmetry is broken to the product of the Cartan
subgroups. Then, the $F$-term condition or Eq. (\ref{eq:BPS3}) requires
\beq
 \left[\mathbf m \cdot \mathbf H, Q\right] = 0.
\eeq
Therefore, the magnetic flux is forced to be oriented in the particular directions specified by
the weight vectors $\tilde{\boldsymbol \nu}$.
Note that these solutions are invariant under the Cartan subgroups.

The mass deformation breaks the color-flavor symmetry, and the
remaining global symmetry 
of the system depends on the mass matrix. 
We will however assume that $\Lambda_{\rm NLSM} \ll m_i \ll g v$,
where 
$\Lambda_{\rm NLSM}$ is the intrinsic scale of the sigma model while
$g v$  
is the mass of the particles in the bulk theory.
This condition allows us to treat the deformation as a shallow
potential on the string worldsheet. In general we shall take all
the elements in the mass matrix to be distinct;  when some of them
coincide, a non-Abelian subgroup of the color-flavor group will
emerge, giving non-Abelian moduli to 
kinks \cite{Shifman:2003uh} and monopoles \cite{Nitta:2010nd}.

The potential on the moduli space induced by the mass term
can be calculated in a way similar to the kinetic term 
of the effective action \eqref{eq:massless_S} discussed 
in the previous section. 
First let us determine the modification of the fields 
perturbatively in terms of the mass parameter. 
Up to first order in the masses, 
the background vortex configuration is not modified. 
The adjoint scalar field $\Phi$ can be determined by 
solving the equation of motion in the vortex background 
\beq
\frac{2}{g^2} \D_\mu \D^\mu \Phi ~=~ - (\Phi Q - Q M) Q^\dagger - Q ( Q^\dagger
\Phi - M Q^\dagger)\ .
\eeq
Since $\Phi$ is of order $m_i$,  $\D_\alpha \D^\alpha \Phi$ is of order $
\mathcal O(m\p_\alpha^2)$, which is small. 
We obtain the following solution to the equation of motion
\beq
\Phi ~=~ M + i U^\dagger ( \delta_m + \delta_m^\dagger ) U - i U^\dagger 
\Big[ Q_0 (\delta_m^\dagger U )U^\dagger Q_0^{-1} - {\rm h.c.} \Big] U + \mathcal
O(m^3,m \p_\alpha^2)\ ,
\label{eq:adjoint_sol}
\eeq
where we have defined the derivative operators\footnote{There is no summation
on $i$ in $k^i = i (\boldsymbol \alpha_i \cdot \mathbf m) \, b^i$, 
and $\bar k^{\bar i} = -i (\boldsymbol \alpha_{\,\bar i} \cdot \mathbf m) \,
\bar b^{\bar i}$.}
\beq
\delta_m = k^i \frac{\p}{\p b^i}\ , \hs{10} 
\delta_m^\dagger = \bar k^{\bar i} \frac{\p}{\p \bar b^{\bar i}}\ , \hs{10} 
k^i = i (\boldsymbol \alpha_i \cdot \mathbf m) \, b^i\ , \hs{5}
\bar k^{\bar i} = -i (\boldsymbol \alpha_{\,\bar i} \cdot \mathbf m) \, \bar
b^{\bar i}\ .
\eeq
By using the explicit form of the matrix $U$, we obtain 
\beq
\Phi ~=~ M - i \big(1 - |z| e^{-\frac{1}{2} \psi} \big) \, U^\dagger
\ba{c|c} 0 & X^{-\frac{1}{2}} (\delta_m^\dagger B^\dagger) Y^{- \frac{1}{2}} \\ \hline 
- Y^{-\frac{1}{2}} (\delta_m B) X^{- \frac{1}{2}} & 0 \ea U\ . 
\eeq
Note that $k^i$ can be interpreted as 
the holomorphic Killing vector on the moduli space.
The corresponding isometry, which we call $U(1)_M$, 
is a subgroup of $G$ and acts on the matrix $U$ 
and the complex coordinates $b^i$ as
\beq
U \rightarrow e^{i M \vartheta} U e^{-i M \vartheta} ~~~\Longleftrightarrow~~~
b^i \rightarrow e^{i (\bs \alpha_i \cdot \mathbf m) \, \vartheta} \, b^i \qquad
({\rm Def.} \ \ U(1)_M)\ .
\eeq
Then substituting the configuration into the bulk action 
and integrating over the $(x_1,x_2)$-plane, 
we obtain the potential of the form 
\cite{AlvarezGaume:1983ab,Gates:1983py}\footnote{A similar result has been
obtained for 1/4 BPS monopoles in ${\cal N}=4$ theories in
Ref.~\cite{Tong:1999mg}.}
\beq
V_{\rm eff} ~=~ g_{i \bar j} k^i \bar k^{\bar j} \ .  
\label{eq:potential}
\eeq
This is the squared norm of the Killing vector 
and has minima at the zeros of $k^i$, viz.~the fixed points of the
$U(1)_M$ isometry.  
This implies that the vortex in the massive theory has
minimum energy if it is invariant under $U(1)_M \subset G_{C+F}$.

The potential of this form can also be obtained 
in the following way.
First let us consider the massless sigma model 
with one additional compact direction $\vartheta$ which has period
$2\pi R$ 
\beq \label{eq:massiveKK}
S_{\rm eff} ~=~ \frac{1}{2\pi R} \int dt dx_3 d \vartheta \; g_{i \bar j} (
\p_\alpha b^i \p^\alpha \bar b^{\bar j} - \p_\vartheta b^i
\p_{\vartheta} \bar b^{\bar j} )\ .
\eeq
Then we impose the following twisted boundary condition with respect
to the $U(1)_M$ symmetry 
\beq
b^i(\vartheta+2\pi R) ~=~ e^{2 \pi i R \, (\bs \alpha_i \cdot \mathbf m)}
b^i(\vartheta)\ .
\eeq
where there is no summation on $i$. If we ignore the infinite tower of
the Kaluza-Klein modes, the $\vartheta$-dependence of the field can be
determined as 
\beq
b^i(t,x_3,\vartheta) ~=~ e^{i (\bs \alpha_i \cdot \mathbf m) \vartheta} b^i
(t,x_3)\ ,
\eeq
Substituting this lowest mode into the massless effective action
(\ref{eq:massiveKK}), we obtain the same potential term as
that of Eq.\,\eqref{eq:potential} 
\beq
S_{\rm eff} ~=~ \int dt dx_3 \; g_{i \bar j} ( \p_\alpha b^i \p^\alpha
\bar b^{\bar j} - k^i \bar k^{\bar j} )\ .
\eeq
In the next section, we will use this method 
to find the explicit potentials 
on the vortex worldsheet for $G=SU(N)$, $SO(2n)$ and $USp(2n)$.

In the $G=SU(N)$ case, the generic mass matrix breaks the gauge
symmetry to $U(1)^{N-1}$ and one observes that the breaking gives rise to
monopoles. Turning on the FI parameter $\xi$,
the monopoles are still there but will be
confined to live on the vortex string. They turn out to be kinks along
the vortex ~\cite{Tong:2003pz,Hanany:2004ea,Gorsky:2004ad}. We will study
the generalization of this phenomenon for the $G=SO(2n)$ and $USp(2n)$
theories in section \ref{sec:kinks}.

\section{Mass-deformed  sigma model  \label{Sec:Mdformed}}

\subsection{$\mathbb{C}P^{N-1}$ as a warm up}

Let us take a simple example to illustrate the method. 
In the case of $G=SU(2)$, 
the orientational moduli space is 
the complex projective space $\mathbb C P^1 \cong SU(2)/U(1)$. 
The K\"ahler potential for the $\mathbb{C}P^1$ model is
\beq
K ~=~ \frac{4\pi}{g^2} \log\left(1+|b|^2\right) \ , 
\eeq
giving rise to the sigma model
\beq
\mathcal{L} ~=~ \frac{4\pi}{g^2} \frac{|\p_\alpha b|^2}
  {\left(1+|b|^2\right)^2} \ ,
  \label{eq:CP1Lagrangian}
\eeq
where $b\in\mathbb{C}$ is the inhomogeneous coordinate on
$\mathbb{C}P^1$.  
Now let us use the method of Ref.~\cite{Eto:2004rz} 
to generate the twisted mass potential on the vortex worldsheet 
induced by the bulk mass term
\beq
M ~=~ 2 \ba{cc} m & 0 \\ 0 & -m \ea.
\eeq
First we have to use the $U(1)_M$ global symmetry of the system at
hand. In the case of the $\mathbb{C}P^1$ model,  
the $U(1)_M$ symmetry acts on the coordinate $b$ 
as a global phase rotation
\beq 
b \to e^{- i 2 m \vartheta} b \ , 
\eeq
which leaves the Lagrangian \eqref{eq:CP1Lagrangian} invariant. 
Taking advantage of an argument similar to that of the last section,
we obtain 
\beq 
b(t,z,\vartheta) ~=~ e^{-i 2 m \vartheta} b(t,z) \ . 
\eeq
Plugging this field back into the Lagrangian \eqref{eq:CP1Lagrangian}
leaves us with the following mass-deformed theory 
\cite{Abraham:1992vb}, \cite{Shifman:2004dr} 
\beq
\mathcal{L} ~=~ \frac{4\pi}{g^2} \frac{|\p_\alpha b|^2 
  - 4 m^2 |b|^2}{\left(1+|b|^2\right)^2} \ . 
\label{eq:CP1_massdeform}
\eeq
The theory is known to have two vacua. The description we
have used above uses the inhomogeneous coordinates on $\mathbb{C}P^1$
and hence we need two patches to describe the theory. On each patch
the vacuum is seen to be given by $b=0$, which corresponds to
$b\to\infty$ on the other patch. 
As we have seen in the previous section, 
the potential is the squared norm of the Killing vector of $U(1)_M$.
Indeed the vacua $b=0$ and $b \rightarrow \infty$ are 
the fixed points of the $U(1)_M$ symmetry. 
Hence we have checked (trivially)
that the number of vacua found is indeed two, in accord with the
literature. 

Generalizing the above discussion to the case of $G=SU(N)$, 
we have the following K\"ahler potential for the sigma model on
$\mathbb{C}P^{N-1}$
\beq 
K ~=~ \frac{4\pi}{g^2} \,\log\left(1 + b^\dag b\right) \ , 
\eeq
where $b$ is an ($N-1$)-component complex column vector. 
The Lagrangian reads \cite{Gauntlett:2000ib,Hanany:2003hp,Gorsky:2004ad}
\beq
\mathcal{L} ~=~ \frac{4\pi}{g^2} \left[
\frac{\p_\alpha b^\dag \p^{\alpha} b}{1+b^\dag b}
  -\frac{\left(b^\dag\p_\alpha b\right)\left(\p ^{\alpha} b^\dag b\right)}
  {\left(1+b^\dag b\right)^2}\right] \ .
\eeq
The $U(1)_M$ global symmetry of the Lagrangian is expressed as 
\beq
b \to \exp \left( -i m_1 \vartheta \right) \exp \left( i M_{N-1} \vartheta \right) b \ , 
\eeq
where we have assumed that the bulk mass is
\beq
M ~=~ \ba{c|c} m_1 & \\ \hline & M_{N-1} \ea. 
\eeq
Keeping only the lowest mode
\beq 
b(t,z,\vartheta) ~=~ e^{i M_0 \vartheta} b_0(t,z) \ ,
\eeq
where $M_0=-m_1 \mathbf 1_{N-1} + M_{N-1}$.
Insertion of this field into the Lagrangian and dropping the suffix of
$b$ gives us the deformed sigma model
\beq
\mathcal{L} ~=~ \frac{4\pi}{g^2} \left[
\frac{\p_\alpha b^{\dag} \p^{\alpha} b - b^{\dag} M_0^2 b}{1+b^{\dag} b}
  -\frac{\left(b^\dag\p_\alpha b\right)\left(\p^{\alpha} b^\dag b\right) 
  - \left(b^\dag M_0 b\right)^2}
  {\left(1+b^\dag b\right)^2}\right] \ .
\label{eq:CPN_massdeform}
\eeq
A vacuum satisfying the fixed point condition $M_0 b=0$ is 
just the origin of this coordinate patch.
Hence, there will be $N$ vacua corresponding to 
$N$ patches covering $\mathbb{C}P^{N-1}$,
\beq
n_{\rm vacua}^{SU(N)} ~=~ N\ .
\eeq

\subsection{The $SO(2n)/U(n)$ and $USp(2n)/U(n)$ sigma models}

Let us now apply the aforementioned technique to 
the $SO(2n)/U(n)$ and $USp(2n)/U(n)$ sigma models on the vortex worldsheet. 
We will treat them on the same footing in the following. 
For $SO(2n)/U(n)$ the field $B^{\rm T} = - B$ is an anti-symmetric
matrix valued field while for $USp(2n)/U(n)$ it is symmetric 
$B^{\rm T}=B$.  
The K\"ahler potential 
\beq
K = \frac{4\pi}{g^2} \Tr \log \left(\mathbf{1}_n + B B^\dag\right) \ , 
\eeq
gives rise to the Lagrangian which we found in the previous section,
i.e.,
\beq
\mathcal{L} ~=~ \frac{4\pi}{g^2} \,\Tr\left\{
\left(\mathbf{1}_n + B^\dag B\right)^{-1}
\p_\alpha B^\dag 
\left(\mathbf{1}_n + B B^\dag\right)^{-1}
\p^\alpha B
\right\} \ . \label{eq:HSS_SOUSpLagrangian}
\eeq
Since we have assumed that $M$ is in the Cartan subalgebra of $G$, 
the mass matrix takes the form
\beq
M = \ba{c|c} M_n & \\ \hline & -M_n \ea, \hs{10} M_n = {\rm
  diag}(m_1,m_2,\cdots,m_n)\ .
\eeq
The $U(1)_M$ action on $B$ can be seen from 
$U \rightarrow e^{-i \vartheta M} U  e^{i \vartheta M}$ to be
\beq
B \to e^{iM_n \vartheta} B e^{i M_n \vartheta} \ .  
\eeq  
As above, we expand the field in modes and keep just the lowest mode
giving rise to
\beq
B(t,z,\vartheta) ~=~ e^{i M_n \vartheta} B_0(t,z) e^{i M_n \vartheta} \ .
\eeq
Upon inserting this field in the Lagrangian
\eqref{eq:HSS_SOUSpLagrangian} and dropping the suffix, we obtain the
following mass-deformed sigma model
\begin{align}
\mathcal{L} ~=~ \frac{4\pi}{g^2} \,\Tr\Big\{
&\left(\mathbf{1}_n + B^\dag B\right)^{-1}
\p_\alpha B^\dag 
\left(\mathbf{1}_n + B B^\dag\right)^{-1}
\p^\alpha B \non
-&\left(\mathbf{1}_n + B^\dag B\right)^{-1}
\left\{M_n,B^\dag\right\}
\left(\mathbf{1}_n + B B^\dag\right)^{-1}
\left\{M_n,B\right\}
\Big\} \ . \label{eq:HSS_SOUSpLagrangian_massdeform}
\end{align}
As the mass matrix is Hermitian, the vacuum equation reads
\beq
\left\{M_n, B \right\} ~=~ 0 \ , 
\eeq
which in general can only be satisfied for $B=0$.\footnote{We assume
  that $m_i \ne \pm m_j$ to break the color-flavor group to Cartan
  generators.}  
However, there exist other vacua in the coordinate patches
which are not covered by $B$.
In the next section we will see that the numbers of vacua 
of the $SO(2n)/U(n)$ and $USp(2n)/U(n)$ sigma models are 
\beq
n_{\rm vacua}^{SO(2n)} = 2^{n-1}\ , \hs{10} 
n_{\rm vacua}^{USp(2n)} = 2^n \ ,
\eeq
respectively.
Although our analysis here is limited to 
the worldsheet action of the single winding vortex ($k=1$), 
we can generalize the discussion to higher winding cases. 
In the case of $k \geq 2$, we have various choices 
for the representation of the coweight $\tilde{\bs \nu}$, 
which determines the orbit of the $G_{C+F}$ symmetry.
As an example, the mass-deformed sigma model 
on the quadric surface $Q^{2n-2}$ 
($G=SO(2n)$, $k=2$, $\tilde{\bs \nu}={\rm vector\; representation}$) 
is discussed in Appendix \ref{appendix:quadric}.

We have now considered a few worldsheet sigma models 
which are all low-energy effective descriptions of non-Abelian vortex
systems.  
For the sigma models we are considering, 
the number of vacua in the classical regime is equal to 
the Euler number of the target space 
$\chi(\mathcal M_{\rm orientation})$ \cite{GJK},
see Table \ref{tab:eulercharacteristics}.
This is consistent with the Witten index \cite{Witten:1981nf} and 
expected to remain the same in the quantum regime. 
In the next section we will consider the kinks
interpolating the different vacua of the vortex worldsheet theory,
viz.~the sigma model.
\begin{table}
\begin{center}
\begin{tabular}{l||c}
moduli space $\mathcal{M}_{\rm orientation}$ &
$\chi(\mathcal{M}_{\rm orientation})$ \\
\hline\hline
$\tfrac{SO(2n)}{U(n)}$ & $2^{n-1}$ \\
$\tfrac{USp(2n)}{U(n)}$ & $2^n$ \\
$\mathbb{C}P^{N-1}=\tfrac{SU(N)}{SU(N-1)\times U(1)}$ & $N$ \\
$Gr_{N,k}=\tfrac{SU(N)}{S(U(k)\times U(N-k))}$ &
{\tiny$\begin{pmatrix}N\\k\end{pmatrix}$} \\
$Q^{2n-2}=\tfrac{SO(2n)}{SO(2)\times SO(2n-2)}$ & $2n$
\end{tabular}
\caption{The Euler characteristics of various orientational moduli
  spaces. }
\label{tab:eulercharacteristics}
\end{center}
\end{table}

\section{Monopoles as kinks on the vortex}\label{sec:kinks}

In this section we discuss kink configurations 
in the effective action on the worldsheet 
of the $U(1)\times SO(2n)$ and $U(1)\times USp(2n)$ vortices.
We will see that the kinks on the vortex worldsheet can be 
interpreted as the $SO(2n)$ and $USp(2n)$ monopoles 
from the viewpoint of the bulk theory. 

\subsection{$SU(2)$ monopole}
For illustration, let us first review 
the 1/4 BPS configuration of the kink monopole 
in the $G=SU(2)$ case \cite{Tong:2003pz}. 

\begin{figure}[!ht]
\begin{center}
\begin{tabular}{ccc}
\includegraphics[width=50mm]{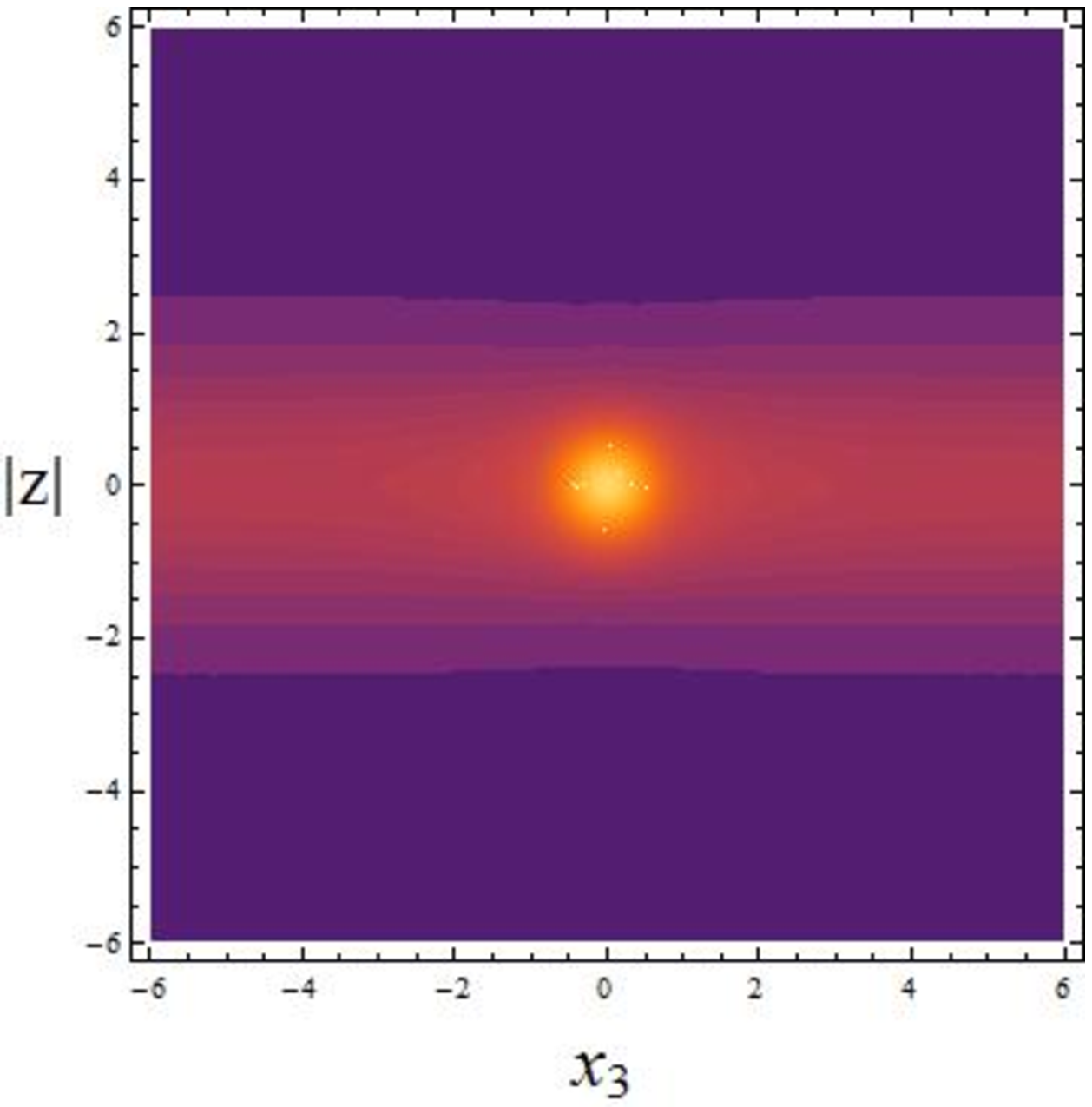} &
\includegraphics[width=50mm]{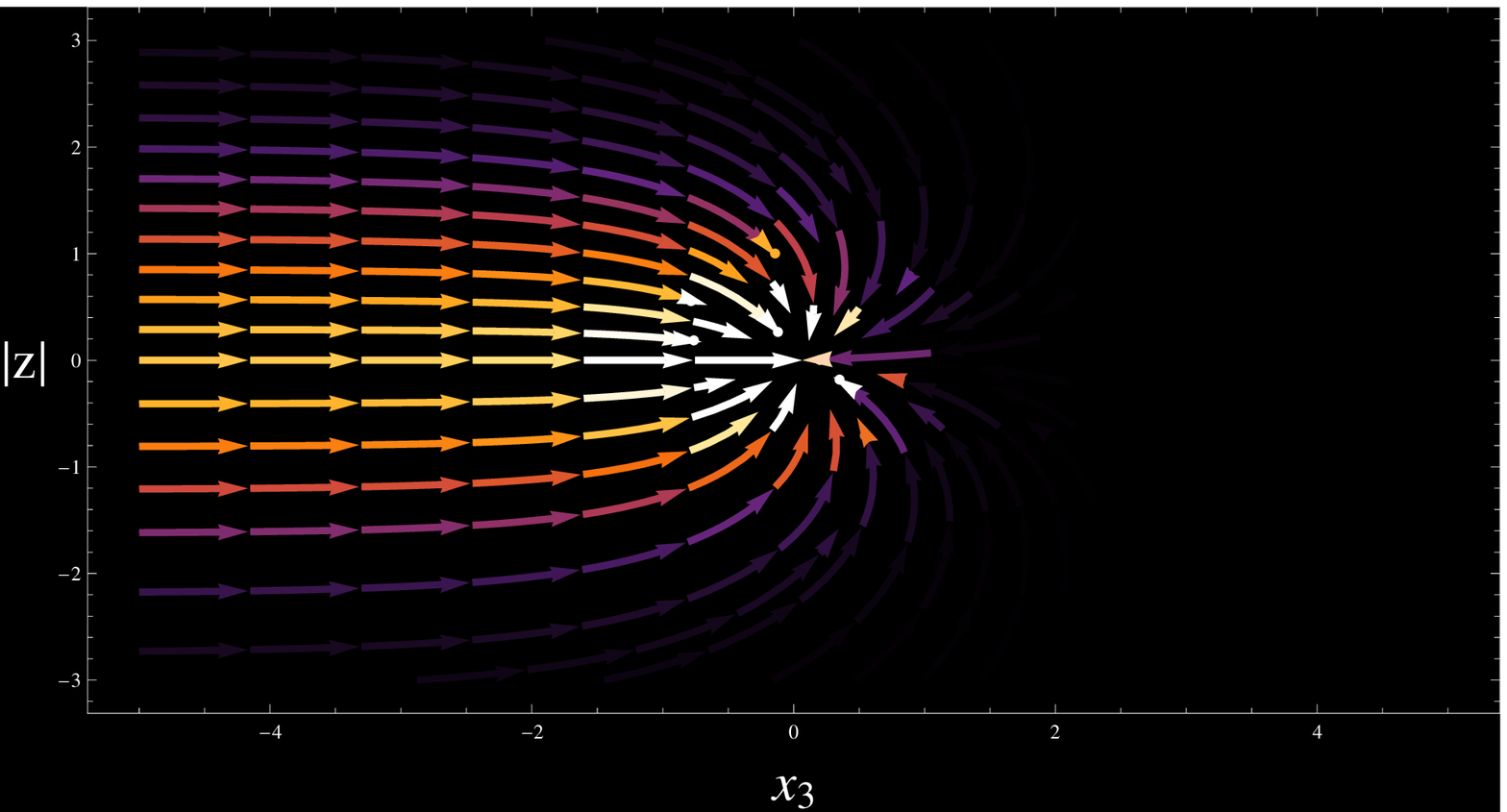} &
\includegraphics[width=50mm]{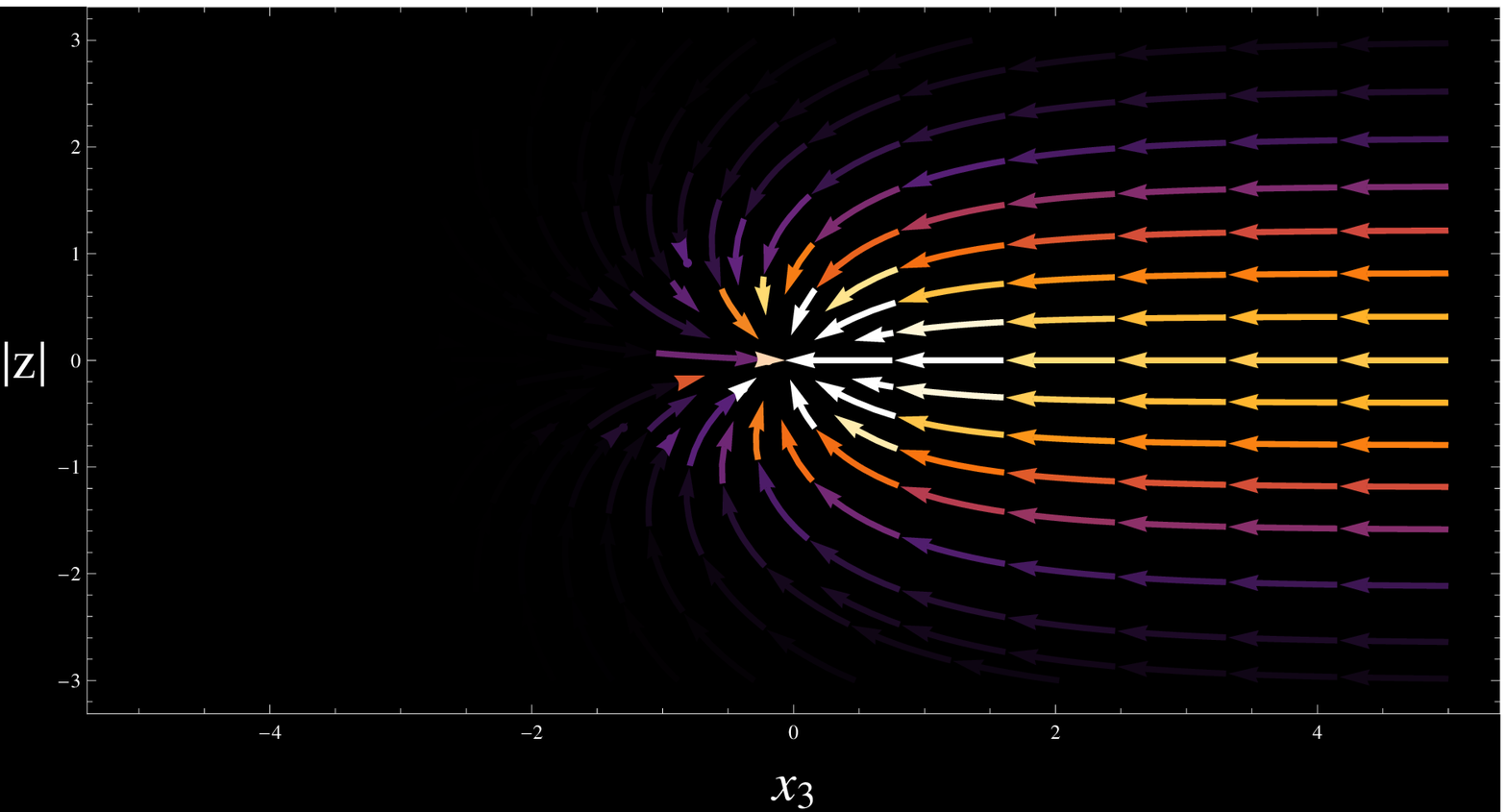} \\
(a) energy density & (b) $\Tr[ B_i (\mathbf 1 + \sigma_3)]$ & (c) $\Tr[ B_i (\mathbf 1 - \sigma_3)]$
\end{tabular}
\end{center}
\caption{(a) The energy density profile of the vortex-monopole
configuration (numerical solution of the 1/4 BPS equations).
The energy is localized along the vortex $(|z| = |x_1 + i x_2| = 0)$ 
and around the monopole $(|z| = x_3 = 0)$. 
(b, c) The magnetic flux projected onto 
$\mathbf 1 + \sigma_3 \propto \lambda_{\rm highest}$ and 
$\mathbf 1 - \sigma_3 = \lambda_{\rm lowest}$. 
The monopole is attached to two vortices with magnetic flux 
$\lambda_{\rm highest}$ and $\lambda_{\rm lowest}$, respectively. The
plots for negative $|z|$ are simply mirror images in order to
illustrate the cross section of the configuration.}
\label{fig:SU(2)_monopole}
\end{figure}

The orientational moduli of the $U(2)$ vortex is 
$\mathbb{C}P^1 \cong SU(2)/U(1)$. 
In the presence of the mass term $M = m \, \sigma_3$,
the potential is induced and only two configurations 
are left to be the minimal energy configurations 
(which we call the vortex vacua).
At  these points, the magnetic flux 
$\lambda=-\frac{1}{2\pi}\int d x_1 d x_2\, F_{12}$ is given by
\beq
\lambda_{\rm highest} = \ba{cc} 1 & \\ & 0 \ea\ , \hs{10}
\lambda_{\rm lowest} =  \ba{cc} 0 & \\ & 1 \ea\ .
\eeq
The 1/4 BPS equations \eqref{eq:BPS1}-\eqref{eq:BPS4} 
do admit configurations of vortices 
which approach the vortex vacua at $x_3 \rightarrow \pm \infty$ as
\beq
\lim_{x_3 \rightarrow - \infty} \lambda = \lambda_{\rm highest}\ , \hs{10}
\lim_{x_3 \rightarrow   \infty} \lambda = \lambda_{\rm lowest}\ .
\eeq
For such a configuration, the energy \eqref{eq:bulk_BPS_energy} 
is given by the difference between the magnetic fluxes
at $x_3 \rightarrow \pm \infty$, i.e.~the magnetic charge inside the
vortex 
\beq
E ~=~ -\frac{4\pi}{g^2} \int dx_3 \, \p_{x_3} \Tr \left[ \lambda \Phi \right] +
\int dx_3 \, T_v ~=~ \frac{8\pi m}{g^2} + \int dx_3 \, T_v\ , 
\label{eq:SU(2)_energy}
\eeq
where the second term represents the vortex tension.
Note that the adjoint scalar $\Phi$ approaches the VEV 
$\langle \Phi \rangle = M$ at spatial infinity
\beq
\lim_{x_3 \rightarrow \pm \infty} \Phi ~=~ m\, \sigma_3\ .
\eeq
The energy of Eq.\,\eqref{eq:SU(2)_energy} is 
given by the mass of the monopole and the vortex tension. 
Hence we can interpret this as the energy of the 1/4 BPS configuration  
of the confined monopole attached to two vortices having
magnetic flux $\lambda_{\rm highest}$ and $\lambda_{\rm lowest}$,
respectively (see Fig.\,\ref{fig:SU(2)_monopole}). 
Fig.\,\ref{fig:SU(2)_monopole} is the first full numerical solution 
of a confined monopole.

We can discuss the same configuration using
the worldsheet effective action of the vortex. 
Using the effective action \eqref{eq:CP1_massdeform} for the $U(2)$
vortex in the massive theory, we can easily find the BPS equation for
the kink by rewriting the energy of the static configuration as
follows 
\beq
E ~=~ \int d x_3 \; \frac{4\pi}{g^2} \frac{|\p_{x_3} b - 2m b|^2}{\left(1+|b|^2\right)^2} + \int d x_3 \;
\frac{4\pi}{g^2} \p_{x_3} \sigma ~\geq~ \frac{4\pi}{g^2}
[\sigma(\infty)-\sigma(-\infty)]\ ,
\label{eq:SU(2)_kink_energy}
\eeq
where $\sigma$ is the standard height function of $\mathbb C P^1$ 
which is given by
\beq
\sigma ~=~ - m \frac{1-|b|^2}{1+|b|^2}\ .
\eeq
The BPS equation and its solution are given by
\beq
\p_{x_3} b - 2m b = 0 ~~~\Longrightarrow~~~ b(x_3)= b_0\,  e^{2m {x_3}}\ ,
\eeq
where $b_0$ is a complex constant which corresponds to
the position and phase moduli of the kink.
Although the solution $b \to \infty$ at $x_3\to\infty$, this is an
artifact of the choice of the coordinate. 
The moduli field $b$ is the inhomogeneous
coordinate on $\mathbb{C}P^1$. In order to see the other vacuum, one needs to
change the patch, say $b' = 1/b$ ($b\neq0$).
This BPS configuration saturates the bound
\eqref{eq:SU(2)_kink_energy} and the kink mass is given by 
\beq
E ~=~ \frac{4\pi}{g^2} [\sigma(\infty)-\sigma(-\infty)] ~=~ \frac{8\pi
  m}{g^2}\ .
\eeq
This kink mass is in precise agreement with the monopole mass. 
In this way, we can identify the kink interpolating between the
distinct vortex vacua with the monopole confined inside the vortex \cite{Tong:2003pz}. 

\subsection{$SO(2n)$ and $USp(2n)$ monopoles} 
\subsubsection{BPS kink equations}
We will use the following convention for the Cartan generators
\beq
 \mathbf H ~=~ \ba{c|c} \mathbf H_n & \\ \hline & - \mathbf H_n \ea\ , \hs{10} 
\mathbf H_n ~=~ 
\ba{ccc} \bs e_1 & & \\ & \ddots & \\ & & \bs e_n \ea\ ,
\label{remember}\eeq 
where $\{\bs e_i\}$ are the standard orthonormal basis.
With this normalization, the highest weight vector of 
the (Weyl) spinor representation of $SO(2n)$ and $SO(2n+1)$ is given by
\beq
\tilde{\bs \nu}_h ~=~ \frac{1}{2}( \bs e_1 + \cdots + \bs e_n )\ .
\eeq
The mass matrix takes the form
\beq
M ~=~ \mathbf m \cdot \mathbf H ~=~ 
\left( \begin{array}{c|c} M_n & \\ \hline
    & - M_n \end{array} \right)    \ .
\label{masses}  \eeq
Without loss of generality, we can always choose the ordering of the
masses by using the $SO(2n)$ and $USp(2n)$ rotations as 
\beq
\mathbf m = (m_1,m_2,\cdots,m_n)\ , \hs{10} 
m_1 \geq m_2 \geq \cdots \geq m_n \geq 0\ .
\eeq 
If some of the masses are equal, 
a non-Abelian subgroup of $G_{C+F}$ is unbroken, 
while the symmetry is maximally broken to $U(1)^n$ 
for a non-degenerate mass matrix.
For simplicity, we restrict ourselves to the case of 
maximal symmetry breaking $m_1 > m_2 > \cdots > m_n > 0$, 
which implies that
\beq
\mathbf m \cdot \boldsymbol \alpha_i > 0\ , 
\eeq
for all positive root vectors $\bs \alpha_i$. 
In other words, $\mathbf m$ is a vector 
in the interior of the positive Weyl chamber.

The time-independent energy density of the effective sigma model 
can be decomposed into a positive semi-definite term 
and a total derivative term
\beq
\mathcal E &=& \frac{4\pi}{g^2} \Tr \bigg[ \, X^{-1} \big( \p_{x_3} B^\dagger
- \{M_n, B^\dagger \} \big) Y^{-1} \big( \p_{x_3} B - \{M_n, B \} \big) \,
\bigg]  + \frac{4\pi}{g^2} \p_{x_3} \sigma \ , 
\label{eq:energy}
\eeq
where $X$ and $Y$ are the matrices given in Eq.\,\eqref{eq:XY}. 
The function $\sigma$, which is called the moment map of the $U(1)_M$
action, is given by
\beq
\sigma ~=~ \Tr \left[ M_n - X^{-1} M_n - Y^{-1} M_n \right]\ .
\label{eq:sigma}
\eeq
Note that $\sigma$ satisfies
\beq
\p_{x_3} \sigma ~=~ \Tr \left[ X^{-1} \p_{x_3} B^\dagger Y^{-1} \{ M_n, B \} +
  X^{-1} \{M_n,B^\dagger \} Y^{-1} \p_{x_3} B \right]\ .
\label{kore}  \eeq
The first term in the energy density \eqref{eq:energy} 
is positive semi-definite and vanishes if the matrix $B$ satisfies
\beq
\p_{x_3} B ~=~ \{ M_n , B \}\ .
\label{eq:kink_BPSeq}
\eeq
This is the BPS equation for the kinks on the vortex worldsheet. 
If the matrix $B$ satisfies the BPS equation, 
the energy of the configuration, 
which can be interpreted as the total mass of the BPS kinks, 
is given by the boundary values of the function $\sigma$ 
\beq
M_{\rm kink} ~=~ \frac{4\pi}{g^2} \int_{-\infty}^{\infty} dx_3 \; \p_{x_3} \sigma
~=~ \frac{4\pi}{g^2} \big[ \sigma(\infty) - \sigma(-\infty) \big]\ .
\label{eq:BPS_mass}
\eeq
The physical meaning of the function $\sigma$ can be seen by 
using Eqs.\,\eqref{eq:flux}, \eqref{eq:adjoint_sol} and \eqref{eq:matrix_U} 
\beq
\frac{1}{2\pi} \int dx_1 d x_2 \, \Tr \left[ F_{12} \Phi \right] ~=~ - \Tr
\left[ U^\dagger \lambda_h U M \right] ~=~ \sigma\ .
\eeq
This implies that the function $\sigma$ is the magnetic flux of
the vortex projected onto the internal direction specified by 
the adjoint scalar $\Phi$. 
The total mass \eqref{eq:BPS_mass} is proportional to 
the difference of the magnetic flux, i.e.~the magnetic charge inside
the vortex string 
\beq
M_{\rm kink} ~=~ \frac{4\pi}{g^2} \mathbf m \cdot \mathbf g\ ,
\eeq
where we have defined the magnetic charge vector $\mathbf g$ as
\beq
\mathbf g &\equiv& \int dx_3 \; \Tr \left[ X^{-1} \p_{x_3} B^\dagger Y^{-1} \{
  \mathbf H_n, B \} + X^{-1} \{\mathbf H_n ,B^\dagger \} Y^{-1} \p_{x_3} B
\right] \notag  \\
&=& \int dx_3 \; \p_{x_3} \Tr \left[ \mathbf H_n - X^{-1} \mathbf H_n - Y^{-1}
  \mathbf H_n \right]\ .
\label{magcha} \eeq
Therefore the kinks on the vortex worldsheet are 
magnetically charged objects, i.e.~they are the magnetic monopoles. 

The general solution to the BPS equation \eqref{eq:kink_BPSeq} 
can be easily obtained as
\beq
B ~=~ e^{M_n {x_3}} \, B_0 \, e^{M_n {x_3}}\ ,
\label{eq:kink_sol}
\eeq
where the matrix elements of $B_0$ are the integration constants, 
namely the moduli parameters of the BPS configurations.
Although the solution \eqref{eq:kink_sol} 
diverges at spatial infinity $x_3 \rightarrow \infty$, 
this is an artifact of the choice of the coordinates 
as in the example of the $SU(2)$ monopole.

In order to see the configuration at $x_3 \rightarrow \infty$, 
we need to change the patch specified by $B$ to other patches.
To this end,
let us use the fact that 
the target space $SO(2n)/U(n)$ and $USp(2n)/U(n)$ can be 
embedded into the complex Grassmannian $Gr(2n,n)$.
As is well known, the complex Grassmannian can be described 
by an $n$-by-$2n$ matrix $\Lambda$ with the following equivalence relation
\beq
\Lambda \sim V \Lambda\ , \hs{10} V \in GL(n,\mathbb C)\ . \label{eq:Vequivalence}
\eeq
The coset spaces $SO(2n)/U(n)$ and $USp(2n)/U(n)$ are subspaces
in the complex Grassmannian $Gr(2n,n)$ defined by the following 
constraint\footnote{
In the case of $SO(2n)/U(n)$, 
the set of the solutions to the constraint \eqref{eq:constraint}
consists of two disjoint copies of $SO(2n)/U(n)$, 
one of which can be identified with the moduli space of vortices of definite chirality.} on the matrix $\Lambda$
\cite{Higashijima:1999ki} 
\beq
\Lambda \, J \Lambda^{\rm T} = 0\ , \hs{10} 
J = \left( \begin{array}{c|c} \mathbf 0 & \mathbf 1_n \\ \hline \epsilon
    \mathbf 1_n & \mathbf 0 \end{array} \right)\ ,
\label{eq:constraint}
\eeq
where $\epsilon=1$ for $SO(2n)$ and $\epsilon=-1$ for $USp(2n)$. 
Since the group $G$ transitively acts on the coset space, 
any point on the coset space can be obtained 
from a specific matrix $\Lambda$ by the $G$ transformations 
\beq
\Lambda \rightarrow \Lambda U\ , \hs{10} U \in G\ .
\eeq
Let us take the base point at $B=0$ as the corresponding matrix
\beq
\Lambda_h \equiv \left( \begin{array}{c|c} \mathbf 0_n & \ \mathbf
    1_n \end{array} \right)\ ,
\label{eq:Lambda_h}
\eeq
where the suffix of $\Lambda_h$ stands for the vortex configuration 
\eqref{eq:sol_0Q}-\eqref{eq:sol_0A} 
which is associated with the highest weight vector $\tilde{\bs \nu}_h$.
Then, any point on the coset space is given by
\beq
\Lambda ~=~ \Lambda_h U\ , \hs{10} U \in G\ , 
\eeq
Note that the action of the subgroup $H \cong U(n)$ is trivial on $\Lambda_h$, 
so that the whole set of matrices $\Lambda$ is identified with 
the coset space $G/H$. 
If we multiply by the group element $U \in G$ given in
Eq.\,\eqref{eq:matrix_U}, the matrix $\Lambda$ takes the form
\beq
\Lambda ~=~ \Lambda_h g_u g_l ~\sim~ \Lambda_h g_l ~=~
\left( \begin{array}{c|c} B \ & \ \mathbf 1_n \end{array} \right)\ .
\eeq
This shows that the element of the parabolic subgroup $g_u \in P$ 
trivially acts on $\Lambda_h$, 
while the action of the complexified group element $g_l \in G^{\mathbb C}$, 
generated by the lowering operators 
$E_{-\bs \alpha}$ with $\tilde{\bs \nu}_h \cdot \bs \alpha > 0$, 
is non-trivial. 

In the following, we discuss the vacua and kink configurations 
in the $SO(2n)/U(n)$ and $USp(2n)/U(n)$ sigma models 
in terms of the $n$-by-$2n$ matrix $\Lambda$. 
Since these sigma models can be regarded as the low energy effective
theories of a $U(n)$ gauge theory with $2n$ flavors obeying 
the F-term constraint \eqref{eq:constraint}, 
the discussion below can be viewed as a generalization of 
the  construction of the domain walls in the $U(n)$ gauge theory 
(or the Grassmann sigma model)
discussed in Ref.\,\cite{Isozumi:2004jc,Isozumi:2004va,Eto:2004vy,Arai:2003tc}.

\subsubsection{Vacua on the vortex worldsheet}
Now let us discuss the vacuum configurations 
in terms of the matrix $\Lambda$. 
The vacuum condition $\{M_n, B\} = 0$ implies that 
the vacua are fixed points of the $U(1)_M$ transformation
\beq
B ~\rightarrow~ e^{i M_n \vartheta} B \, e^{i M_n \vartheta} ~=~ B\ .
\eeq
In terms of the matrix $\Lambda$, this condition of 
the fixed points can be rewritten as
\beq
\Lambda ~\rightarrow~ \Lambda \, e^{i M \vartheta} ~\sim~ \Lambda\ .
\eeq
We can easily show that $\Lambda_h$ is one of the fixed points 
of $U(1)_M$
\beq
\Lambda_h e^{i M \vartheta} ~=~ e^{-i M_n \vartheta} \Lambda_h ~\sim~
\Lambda_h\ .
\eeq
The other matrices corresponding to fixed points 
can be found by multiplying matrix $\Lambda_h$ by Weyl group
transformations of $G$ as follows
\beq
\Lambda~=~ \Lambda_h \, w \ , \hs{10} 
w ~=~ w_{\bs \alpha_i} w_{\bs \alpha_j} \cdots w_{\bs \alpha_k} ~\in~ G\ ,
\label{eq:fixed_points}
\eeq
where $w_{\bs \alpha_i}$ are the generators of the Weyl group corresponding to 
reflections with respect to the simple roots $\bs\alpha_i$
\beq
w_{\bs \alpha_i} \equiv \exp \left[ -\frac{\pi}{2} ( E_{\bs \alpha_i} -
  E_{-\bs \alpha_i}) \right] \in G\ .
\eeq
We can check that the $\Lambda$'s given by
Eq.\,\eqref{eq:fixed_points} are fixed points by using the fact that
the mass matrix transforms as\footnote{
Note that $w_{\bs \alpha_i}$ acts on any element of the Cartan subalgebra as 
the Weyl reflection of the coefficient vector 
$w_{\bs \alpha_i} ( \mathbf v \cdot \mathbf H ) w_{\bs \alpha_i}^\dagger =
\left[ \mathbf v - (\mathbf v \cdot \tilde{\bs \alpha}_i) \bs \alpha_i \right]
\cdot \mathbf H$. 
}
\beq
M ~\rightarrow~ w_{\bs \alpha_i} M w_{\bs \alpha_i}^\dagger ~=~ \Big[ \mathbf
m - (\mathbf m \cdot \tilde{\bs \alpha}_i) \bs \alpha_i \Big] \cdot \mathbf H\ , 
\hs{10}
\tilde{\bs \alpha}_i = 2 \, \frac{\bs \alpha_i}{\bs \alpha_i \cdot \bs \alpha_i}\
.
\label{eq:Weyl}
\eeq
For example, $\Lambda_h w_{\bs \alpha_i}$ transforms under $U(1)_M$ as
\beq
\Lambda_h w_{\bs \alpha_i} \, e^{i M \vartheta} ~=~ \Lambda_h e^{i w_{\bs
    \alpha_i}(M) \vartheta} w_{\bs \alpha_i} ~=~ e^{-i w_{\bs \alpha_i}(M_n)
  \vartheta} \Lambda_h w_{\bs \alpha_i} ~\sim~ \Lambda_h w_{\bs \alpha_i}\ ,
\eeq
where $w_{\bs \alpha_i}(M)$ and $w_{\bs \alpha_i}(M_n)$ are $2n$-by-$2n$ and $n$-by-$n$ matrices given by
\beq
w_{\bs \alpha_i}(M) ~=~ \ba{c|c} w_{\bs \alpha_i}(M_n) & \\ \hline & - w_{\bs
  \alpha_i}(M_n) \ea ~=~ \Big[ \mathbf m - (\mathbf m \cdot \tilde{\bs
  \alpha}_i) \bs \alpha_i \Big] \cdot \mathbf H\ .
\eeq

Let us now calculate the magnetic flux $\sigma$ 
defined in Eq.\,\eqref{eq:sigma} at each vacuum point. 
In terms of $\Lambda$, the flux $\sigma$ can be rewritten as
\beq
\sigma ~=~ \Tr \left[ (\Lambda \Lambda^\dagger)^{-1} \Lambda M \Lambda^\dagger
\right]\ . 
\eeq
For the highest weight vacuum $\Lambda_h$, 
the flux $\sigma$ is given by 
\beq
\sigma ~=~ - \Tr \, M_n ~=~ - 2 \, \mathbf m \cdot \tilde{\bs \nu}_h \ ,
\eeq
Similarly, the flux $\sigma$ in the vacua $\Lambda_h w_{\bs \alpha_i}$ 
can be obtained by using Eq.\,\eqref{eq:Weyl} as
\beq
\sigma &=& \Tr \left[ (\Lambda_h \Lambda_h^\dagger)^{-1} \Lambda_h w_{\bs
    \alpha_i}(M) \Lambda_h^\dagger \right] ~=~ - 2 \, \mathbf m \cdot w_{\bs
  \alpha_i}(\tilde{\bs \nu}_h)\ ,
\eeq 
where $w_{\bs \alpha_i}(\tilde{\bs \nu}_h)$ is the coweight vector 
obtained from $\tilde{\bs \nu}_h$ 
by the Weyl reflection with respect to the simple root $\bs \alpha_i$
\beq
w_{\bs \alpha_i}(\tilde{\bs \nu}_h) ~\equiv~ \tilde{\bs \nu}_h - (\tilde{\bs
  \nu}_h \cdot \bs \alpha_i) \tilde{\bs \alpha}_i\ .
\eeq
As this example shows, the values of the magnetic flux 
at the vacuum points are specified 
by the coweights related to $\tilde{\bs \nu}_h$ via Weyl reflections. 
In the case of a single vortex in $G=SO(2n) , ~USp(2n)$ theory, 
the orbit of $\tilde{\bs \nu}_h$ under Weyl reflections 
coincides with the whole set of weight vectors of 
the (Weyl) spinor representation of $\tilde G = SO(2n)$, $SO(2n+1)$.
Therefore the number of the vacua is 
the dimension of the representation\footnote{
We can also see that the number of vacua is 
the Euler characteristics of the target manifold \cite{GJK}
\beq
\chi \left( \frac{SO(2n)}{U(n)} \right) = 2^{n-1}\ , \hs{10}
\chi \left( \frac{USp(2n)}{U(n)} \right) = 2^n\ . \notag
\eeq} 
\beq
n_{\rm vacua} = \left\{ 
\begin{array}{lll} 
2^{n-1} & \mbox{(Weyl spinor rep. of $\tilde G =SO(2n)$)} 
& \mbox{for $G=SO(2n)$} \\ 
2^n & \mbox{(spinor rep. of $\tilde G =SO(2n+1)$)}
& \mbox{for $G=USp(2n)$} 
\end{array} \right..
\eeq 

\paragraph{Example 1: vacua of the $USp(4)/U(2)$ sigma model \\}
Let us take the $USp(4)/U(2)$ case as an example.\footnote{
Note that $SO(6)/U(3) \cong USp(4)/U(2) \cong \mathbb C P^3$.}
The Weyl group is generated by 
the following elements corresponding to 
the simple roots $\bs \alpha_1 = \bs e_1 - \bs e_2$ 
and $\bs \alpha_2 = 2 \bs e_2$
\beq
w_{\bs \alpha_1} ~=~ 
\ba{cc|cc} 
0 & -1 & 0 &  0 \\
1 &  0 & 0 &  0 \\ \hline
0 &  0 & 0 & -1 \\
0 &  0 & 1 &  0 
\ea\ , 
\hs{10} 
w_{\bs \alpha_2} ~=~ 
\ba{cc|cc} 
1 &  0 & 0 &  0 \\
0 &  0 & 0 & -1 \\ \hline
0 &  0 & 1 &  0 \\
0 &  1 & 0 &  0 
\ea\ .
\eeq
By using these group elements and $GL(n,\mathbb C)$ transformations, 
we obtain the following vacuum matrices 
\beq
\Lambda_{++} &=& \ba{cc|cc} 0 & 0 & 1 & 0 \\ 0 & 0 & 0 & 1 \ea\ , \hs{10}
\Lambda_{+-} ~=~ \ba{cc|cc} 0 & 0 & 1 & 0 \\ 0 & 1 & 0 & 0 \ea\ , \label{eq:USp_vac_1} \\
\Lambda_{-+} &=& \ba{cc|cc} 1 & 0 & 0 & 0 \\ 0 & 0 & 0 & 1 \ea\ , \hs{10}
\Lambda_{--} ~=~ \ba{cc|cc} 1 & 0 & 0 & 0 \\ 0 & 1 & 0 & 0 \ea\ ,  \label{eq:USp_vac_2}
\eeq
where we have used the Weyl group elements in the following way
\beq
\Lambda_{++}   =  \Lambda_h ~~~\overset{\bs \alpha_2}{\rightarrow}~~~
\Lambda_{+-} \sim \Lambda_{++} w_{\bs \alpha_2} ~~~\overset{\bs \alpha_1}{\rightarrow}~~~
\Lambda_{-+} \sim \Lambda_{+-} w_{\bs \alpha_1} ~~~\overset{\bs \alpha_2}{\rightarrow}~~~
\Lambda_{--} \sim \Lambda_{-+} w_{\bs \alpha_2}\ .
\eeq
The symbol $\sim$ denotes that we have used the equivalence relation
\eqref{eq:Vequivalence}. For each vacuum point on the target space,  
the coweight $\tilde{\bs \nu}$ and the magnetic flux, respectively,
are given by 
\beq
\tilde{\bs \nu}_{\pm \pm} ~=~ \frac{\pm \bs e_1 \pm \bs e_2}{2}\ , \hs{10}
\sigma_{\pm \pm} ~=~ - ( \pm m_1 \pm m_2 )\ .
\eeq
Note that the signatures $\pm$ correspond 
to the spins of the $SO(5)$ spinor representation.
In general, the vacuum matrices for $G=USp(2n)$ has the following form
\beq
\Lambda_{\pm \cdots \pm} ~=~
\ba{ccc|ccc} 
a_{1-} & & & a_{1+} & & \\
& \ddots & & & \ddots & \\
& & a_{n-} & & & a_{n+} 
\ea\ , 
\eeq
where $a_{i+} = 1$ ($a_{i+} = 0$) and $a_{i-} = 0$ ($a_{i-} = 1$) 
if the $i$-th signature of $\Lambda_{\pm \cdots \pm}$ is $+(-)$. 
\paragraph{Example 2: vacua of the $SO(6)/U(3)$ sigma model \\}
In the case of $G=SO(6)$, 
the Weyl group is generated by 
the following elements corresponding to 
the simple roots $\bs \alpha_1 = \bs e_1 - \bs e_2$, 
$\bs \alpha_2 = \bs e_2 - \bs e_3$ and $\bs \alpha_3 = \bs e_2 + \bs e_3$
\beq
{\renewcommand{\arraystretch}{0.7}
{\arraycolsep = 1mm
w_{\bs \alpha_1} \! = \!
\ba{ccc|ccc} 
0  & -1 &  0 &   &    &   \\
1 &   0 & 0  &   &    &   \\
 0 & 0   & 1 &   &    &   \\ \hline
  &    &   &  0 & -1 & 0  \\
  &    &   & 1 &   0 &  0 \\
  &    &   &  0 &    0& 1 \ea}}\ , \hs{1}
{\renewcommand{\arraystretch}{0.7}
{\arraycolsep = 1mm
w_{\bs \alpha_2} \! = \!
\ba{ccc|ccc} 
1 &  0 &   0 &   &   &    \\
0  &   0& -1 &   &   &    \\
 0 & 1 &  0  &   &   &    \\ \hline
  &   &    & 1 &  0 & 0   \\
  &   &    &  0 & 0  & -1 \\
  &   &    & 0  & 1 &  0  \ea}}\ , \hs{1}
{\renewcommand{\arraystretch}{0.7}
{\arraycolsep = 1mm
w_{\bs \alpha_3} \! = \!
\ba{ccc|ccc} 
1 &   &    &   &   &    \\
  &  0 &    &   &   & -1 \\
  &   &   0 &   & 1 &    \\ \hline
  &   &    & 1 &   &    \\
  &   & -1 &   & 0  &    \\
  & 1 &    &   &   &   0 \ea}}\ . \notag
\eeq
Combining these group elements with $GL(n,\mathbb C)$ transformations 
(i.e.~the equivalence relation), 
we obtain the following vacuum matrices 
\beq
\Lambda_{+++} &=& 
{\renewcommand{\arraystretch}{0.7}
{\arraycolsep = 1mm
\ba{ccc|ccc} 
0 &   &   & 1 &   &   \\ 
  & 0 &   &   & 1 &   \\ 
  &   & 0 &   &   & 1 \ea}}\ , \hs{10}
\Lambda_{+--} ~=~ 
{\renewcommand{\arraystretch}{0.7}
{\arraycolsep = 1mm
\ba{ccc|ccc} 
0 &   &   & 1 &   &   \\ 
  & 1 &   &   & 0 &   \\ 
  &   & 1 &   &   & 0 \ea}}\ , \label{eq:SO_vac_1} \\
\Lambda_{-+-} &=& 
{\renewcommand{\arraystretch}{0.7}
{\arraycolsep = 1mm
\ba{ccc|ccc} 
1 &   &   & 0 &   &   \\ 
  & 0 &   &   & 1 &   \\ 
  &   & 1 &   &   & 0 \ea}}\ , \hs{10}
\Lambda_{--+} ~=~ 
{\renewcommand{\arraystretch}{0.7}
{\arraycolsep = 1mm
\ba{ccc|ccc} 
1 &   &   & 0 &   &   \\ 
  & 1 &   &   & 0 &   \\ 
  &   & 0 &   &   & 1 \ea}}\ . \label{eq:SO_vac_2}
\eeq
Here we have used the Weyl group elements in the following way
\beq
\Lambda_{+++}   =  \Lambda_h ~~\overset{\bs \alpha_3}{\rightarrow}~~
\Lambda_{+--} \sim \Lambda_{+++} w_{\bs \alpha_3} ~~\overset{\bs \alpha_1}{\rightarrow}~~
\Lambda_{-+-} \sim \Lambda_{+--} w_{\bs \alpha_1} ~~\overset{\bs \alpha_2}{\rightarrow}~~
\Lambda_{--+} \sim \Lambda_{-+-} w_{\bs \alpha_2}\ .
\eeq
In these vacua, 
the coweight $\tilde{\bs \nu}$ and the magnetic flux are given by
\beq
\tilde{\bs \nu}_{\pm \pm \pm} ~=~ \frac{\pm \bs e_1 \pm \bs e_2 \pm \bs
  e_3}{2}\ , \hs{10}
\sigma_{\pm \pm} ~=~ - ( \pm m_1 \pm m_2 \pm m_3)\ ,
\eeq
where the signatures $\pm$ are the spins of 
the $SO(6)$ Weyl spinor representation.
In general, the vacuum matrices for $G=SO(2n)$ have the following form
\beq
\Lambda_{\pm \cdots \pm} =
\ba{ccc|ccc} 
a_{1-} & & & a_{1+} & & \\
& \ddots & & & \ddots & \\
& & a_{n-} & & & a_{n+} 
\ea\ , 
\eeq
where the number of ``$-$''s in $\Lambda_{\pm \cdots \pm}$ is even
while $a_{i+} = 1$ ($a_{i+} = 0$) and $a_{i-} = 0$ ($a_{i-} = 1$) 
if the $i$-th signature is $+(-)$. 

\subsubsection{Single monopole configurations}
We have seen that the total mass of the BPS kinks is 
determined by the difference of the magnetic flux $\sigma$ 
at $x_3 \rightarrow \pm \infty$,
which is specified by the coweights, 
i.e.~the weight vectors of the dual group $\tilde G$. 
It follows that the total mass of the kinks interpolating 
between two vacua labeled by $\tilde{\boldsymbol \nu}_{\pm}$ is given by
\beq
M_{\rm kink} ~=~ \frac{4\pi}{g^2} \mathbf m \cdot \mathbf g ~=~ 
- \frac{8\pi}{g^2} \mathbf m \cdot (\tilde{\boldsymbol \nu}_{+} -
\tilde{\boldsymbol \nu}_{-}) \ .
\label{eq:total_mass}
\eeq
This shows that the magnetic charge vector 
$\mathbf g$ is proportional to the difference of the vectors
$\tilde{\boldsymbol \nu}_{-} - \tilde{\boldsymbol \nu}_{+}$, 
which is an element of the coroot lattice,~i.e. the lattice generated 
by the root vector of the dual group $\tilde G$.
\begin{figure}[ht]
\begin{center}
\includegraphics[width=10cm]{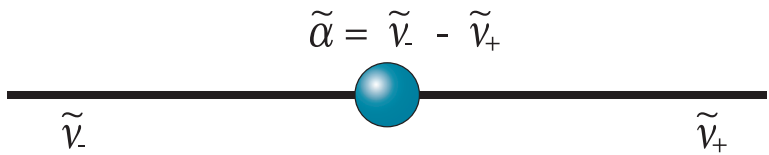}
\caption{A schematic picture of the composite of the vortices labeled by the
coweight vectors $\tilde{\boldsymbol \nu}_+$ and $\tilde{\boldsymbol \nu}_-$
and the monopole with the coroot $\tilde{\boldsymbol \alpha} = \tilde{\boldsymbol \nu}_-
- \tilde{\boldsymbol \nu}_+$.}
\label{fig:vm}
\end{center}
\end{figure}
Since any element of the coroot lattice is a linear combination of 
the simple roots of $\tilde G$ with integer coefficients, 
the total mass can be decomposed into the masses of ``the elementary kinks'', 
which are given by the simple roots of the dual group 
\beq
M_i = \frac{8\pi}{g^2} \, \mathbf m \cdot \tilde{\boldsymbol \alpha}_i\ , \hs{10}
(i=1,2,\ldots, \tilde{r} = {\rm rank} \, \tilde G)\ .
\eeq
These BPS masses of the elementary kinks coincide with 
those of the elementary monopoles appearing in the $4d$ gauge theory
with maximal gauge symmetry breaking  
\beq
SO(2n) \rightarrow U(1)^n\ , \hs{10} USp(2n) \rightarrow U(1)^n\ .
\eeq
A schematic picture is shown in Fig.~\ref{fig:vm}.

As we can see from the BPS solution \eqref{eq:kink_sol}, 
the kink configuration is a one parameter flow 
on the target space parametrized by $x_3$. 
Similarly, this BPS flow can be rewritten in terms of $\Lambda$ as
\beq
\Lambda(x_3) ~=~ \Lambda_0 \, e^{M x_3}\ ,
\eeq
where $\Lambda_0$ is a constant $n$-by-$2n$ matrix which specifies 
the kink configuration.\footnote{
This $n$-by-$2n$ matrix $\Lambda_0$ is called 
the moduli matrix for domain walls 
\cite{Isozumi:2004jc,Isozumi:2004va}.} 
As well as $\Lambda$, the matrix $\Lambda_0$ should obey 
the constraint and equivalence relation
\beq
\Lambda_0 J \Lambda_0^{\rm T} = 0\ , \hs{10} \Lambda_0 \sim V \Lambda_0\ , \hs{5} V \in
GL(n,\mathbb C)\ .
\eeq
For example, the matrix $\Lambda_0$ for the BPS solution
\eqref{eq:kink_sol} is given by 
\beq
\Lambda_0 ~=~ \left( \begin{array}{c|c} B_0 \ & \ \mathbf 1_n \end{array}
\right)\ .
\eeq
By using the matrix $\Lambda_0$, the magnetic flux $\sigma$
for the kink configurations can be expressed as
\beq
\sigma ~=~ \Tr \left[ (\Lambda \Lambda^\dagger)^{-1} \Lambda M \Lambda^\dagger
\right] ~=~ \frac{1}{2} \p_{x_3} \log \det \left( \Lambda_0 e^{2M x_3}
  \Lambda_0^\dagger \right)\ .
\eeq

In the following, we will discuss the kink configurations 
for a given matrix $\Lambda_0$. 
First let us consider the case with $\Lambda_0 = \Lambda_h$.
Since $e^{M x_3}$ acts trivially on $\Lambda_h$, 
the configuration is independent of $x_3$
\beq
\Lambda(x_3) ~=~ \Lambda_h \, e^{M x_3} ~\sim~ \Lambda_h\ .
\eeq
Therefore $\Lambda_0 = \Lambda_h$ corresponds to the highest weight vortex vacuum. 
In general, the fixed points of $U(1)_M$ 
(generated by the Killing vectors $k^i$) 
correspond to those of the BPS flow (generated by $i k^i$) 
\beq
\Lambda_0 \, e^{i M \vartheta} ~\sim~ \Lambda_0 ~~~\Longleftrightarrow~~~ 
\Lambda_0 \, e^{M x_3} ~\sim~ \Lambda_0\ .
\eeq
This means that if $\Lambda_0$ is one of the vacuum matrices,
there is no kink in the configuration. 
The first example of the matrix $\Lambda_0$ which represents
single kink configurations can be obtained 
by multiplying the highest weight vortex vacuum by the lowering
operators $E_{-{\bs \alpha}}$ as  
\beq
\Lambda_0 ~=~ \Lambda_h \, \exp \left[ b \, E_{-\bs \alpha} \right]\ , 
\label{eq:L_0}
\eeq
where $b$ is a complex parameter. 
There is a caveat though, i.e.~the action of the lowering operator
$E_{-{\bs \alpha}}$ is non-trivial if and only if the inner product of  
$\tilde{\bs \nu}_h$ and $\bs \alpha$ is positive 
\beq
\tilde{\bs \nu}_h \cdot \bs \alpha > 0\ .
\label{eq:condition_1}
\eeq
In that case, the matrix $\Lambda(x_3)$ is given by
\beq
\Lambda(x_3) ~=~ \Lambda_h \, \exp \left[ b \, E_{-\bs \alpha} \right] e^{Mx_3}
~\sim~ \Lambda_h \exp \left[ b \, e^{(\boldsymbol \alpha \cdot \mathbf m) \,
    x_3} E_{-\bs \alpha} \right]\ ,
\label{eq:La}
\eeq
where we have used that
\beq
e^{- Mx_3} E_{-\bs \alpha} e^{Mx_3} ~=~ e^{(\boldsymbol \alpha \cdot \mathbf m)
  \, x_3} E_{-\bs \alpha}\ .
\eeq
Eq.\,\eqref{eq:La} shows that the vortex configuration at 
$x_3 \rightarrow -\infty$ is in the highest weight vacuum $\Lambda_h$.  
Note that we have fixed the ordering of the masses so that
$\mathbf m \cdot \bs \alpha > 0$ for all the positive roots $\bs \alpha$.
To see the vortex configuration at $x_3 \rightarrow \infty$, 
let us use the following decomposition of the group element
\beq
\exp \left[ a \, E_{-\bs \alpha} \right] ~=~ \exp \left[a^{-1} E_{\bs \alpha}
\right] \exp \left[ - \log a \ \tilde{\bs \alpha} \cdot \mathbf H \right]
w_{\bs \alpha} \exp \left[ a^{-1} E_{\bs \alpha} \right]\ .
\label{eq:Bruhat}
\eeq
Since the raising operator $E_{\bs \alpha}$ and 
the generator of the Cartan subalgebra $\mathbf H$ 
act on $\Lambda_h$ trivially, 
the matrix $\Lambda(x_3)$ can be rewritten as
\beq
\Lambda(x_3) ~\sim~ (\Lambda_h w_{\bs \alpha}) \exp \left[ b^{-1}
  e^{-(\boldsymbol \alpha \cdot \mathbf m) \, x_3} E_{\bs \alpha} \right]\ .
\eeq
Therefore the matrix $\Lambda(x_3)$ represents 
the kink between the vacua specified by $\tilde{\bs \nu}_h$ and 
$w_{\bs \alpha} (\tilde{\boldsymbol \nu}_h)$
\beq
\lim_{x_3 \rightarrow -\infty} \Lambda(x_3) \sim \Lambda_h\ , \hs{10} 
\lim_{x_3 \rightarrow \infty} \Lambda(x_3) \sim \Lambda_h w_{\bs \alpha}\ .
\eeq
Note that if $\tilde{\bs \nu}_h \cdot \bs \alpha \leq 0$, 
there does not exist any kink in the configuration since $\Lambda_h
w_{\bs \alpha} \sim \Lambda_h$.
As we have seen in Eq.\,\eqref{eq:total_mass}, 
the mass of the kink is given by the difference of the coweights $\tilde{\bs \nu}_{\pm}$
specifying the vacua at $x_3 \rightarrow \pm \infty$. 
For $x_3 \rightarrow -\infty$, 
the vacuum is specified by the highest weight vector $\tilde{\bs \nu}_+ =
\tilde{\bs \nu}_h$, 
while $\tilde{\bs \nu}_-$ is given by Weyl reflection of $\tilde{\bs \nu}_h$ 
with respect to $\bs \alpha$. 
Since $\tilde{\bs \nu}_h \cdot \bs \alpha \leq 1$ for any root $\bs \alpha$,
the Weyl reflection $w_{\bs \alpha} (\tilde{\boldsymbol \nu}_h)$ 
with respect to $\bs \alpha$ satisfying $\tilde{\bs \nu}_h \cdot \bs \alpha > 0$ 
is given by
\beq
w_{\bs \alpha} (\tilde{\boldsymbol \nu}_h) ~=~ \tilde{\bs \nu}_h - (\tilde{\bs
  \nu}_h \cdot \bs \alpha) \tilde{\bs \alpha} ~=~ \tilde{\bs \nu}_h -
\tilde{\bs \alpha}\ .
\eeq
Therefore, $\Lambda(x_3)$ represents 
the set of configurations of kinks interpolating between
the vacua $\tilde{\bs \nu}_h$ and $\tilde{\bs \nu}_h - \tilde{\bs \alpha}$,
whose mass is given by
\beq
M_{\rm kink} ~=~ \frac{8\pi}{g^2} \ \tilde{\boldsymbol \alpha} \cdot \mathbf m\ .
\label{eq:m_alpha} 
\eeq
The physical meaning of the parameter $b$ 
can be read off the profiles of the magnetic flux $\sigma$
and the energy density $\mathcal E = \frac{4\pi}{g^2} \p_{x_3} \sigma$
\beq
\sigma &=& \tilde{\bs \alpha} \cdot \mathbf m \Big[ 1 + \tanh \big[ \bs \alpha
\cdot \mathbf m (x_3-r_0) \big] \Big] - 2 \ \tilde{\bs \nu}_h \cdot \mathbf m
\ , \\
\mathcal E &=& \frac{4\pi}{g^2} (\tilde{\bs \alpha} \cdot \mathbf m)(\bs
\alpha \cdot \mathbf m) \, {\rm sech}^2 \big[ \bs \alpha \cdot \mathbf m
(x_3-r_0) \big] \ ,
\eeq 
where we have redefined the parameter $b$ as (See Eq.(\ref{eq:L_0}))
\beq
b ~=~ \exp \Big[ - (\boldsymbol \alpha \cdot \mathbf m) \, r_0 + i \eta \Big]\ , 
\hs{10} r_0,~ \eta \in \mathbb R\ .
\label{eq:kink_moduli}
\eeq 
Since the kink profile of $\sigma$ 
and the energy density are functions of $x_3 - r_0$ 
and independent of $\eta$, 
the parameters $r_0$ and $\eta$ can be interpreted as 
the position and internal phase moduli of the kink. 
The phase can be interpreted as the Nambu-Goldstone zero mode 
of the $U(1)_M$ symmetry broken by the kink.
Note that the phase $\eta$ transforms under the $U(1)^n$ symmetry as
\beq
\exp \left[ i \bs \theta \cdot \mathbf H \right] :~ 
\eta ~\rightarrow~ \eta + \bs \theta \cdot \bs \alpha\ .
\eeq

As already mentioned, the kinks specified by the simple roots 
correspond to the elementary monopoles in the $4d$ theory. 
There exists only one simple root\footnote{
$\bs \alpha_n = \bs e_{n-1} + \bs e_{n}$
and $\bs \alpha_n = 2 \bs e_n$ for $G=SO(2n)$ and $G=USp(2n)$, respectively.}
which satisfies the non-triviality condition 
$\tilde{\bs \nu}_h \cdot \bs \alpha >0$. 
Therefore only $\bs \alpha = \bs \alpha_n$ corresponds to 
the single elementary monopole for $\Lambda_0$ of the form \eqref{eq:L_0}.
As we will see in the next section, 
the other cases are the coincident monopoles, 
which can be decomposed into several elementary monopoles.

Next, let us consider the action of the lowering operator on another vacuum 
\beq
\Lambda_0 ~=~ (\Lambda_h w) \, \exp \left[ b \, E_{-\bs \alpha} \right] ~=~
\Lambda_h \, \exp \left[ b \, E_{-w(\bs \alpha)} \right] w\ ,
\eeq
where $w(\bs \alpha)$ is the root vector obtained by the Weyl reflection.
The action of $E_{-\bs \alpha}$ on $\Lambda_h w$ is non-trivial if and only if
\beq
\tilde{\bs \nu}_h \cdot w(\bs \alpha) ~=~ 
w(\tilde{\bs \nu}_h) \cdot \bs \alpha ~>~ 0\ .
\label{eq:condition_2}
\eeq
As in the previous case, we can show that
\beq
\Lambda(x_3) ~\sim~ \Lambda_h w \, \exp \left[ b \, e^{(\bs \alpha \cdot \mathbf
    m) x_3} E_{-\bs \alpha} \right] ~\sim~ \Lambda_h w \, w_{\bs \alpha} \, \exp
\left[ b^{-1} e^{-(\bs \alpha \cdot \mathbf m) x_3} E_{-\bs \alpha} \right]\ .
\eeq
Therefore the matrix $\Lambda(x_3)$ corresponds to
the kink between the vacua specified 
by the vectors $w(\tilde{\bs \nu}_h)$ and 
$w(\tilde{\boldsymbol \nu}_h) - \tilde{\bs \alpha}$
\beq
\lim_{x_3 \rightarrow -\infty} \Lambda(x_3) \sim \Lambda_h w\ , \hs{10} 
\lim_{x_3 \rightarrow \infty} \Lambda(x_3) \sim \Lambda_h w \, w_{\bs \alpha}\ .
\eeq
The difference between this and the previous cases is that 
the vacua $\Lambda_h$ and $\Lambda_h w$ can admit 
different types of elementary monopoles 
(see Eqs.\,\eqref{eq:condition_1} and \eqref{eq:condition_2}). 
In general, the $i$-th elementary monopole 
can exist if and only if $i$-th Dynkin label 
$\tilde l_i$ of the coweight $\tilde{\bs \nu}$ at $x_3 \rightarrow - \infty$ 
is positive 
\beq
\tilde l_i ~\equiv~ \tilde{\bs \nu} \cdot \bs \alpha_i ~\in~ \mathbb Z_+\ . 
\eeq
Indeed, we can easily show that if the $i$-th Dynkin label is not positive, 
the action of the lowering operator $\exp \left[ b E_{\bs \alpha_i} \right]$ is trivial
\beq
\tilde l_i \leq 0 ~~~\Longleftrightarrow~~~ (\Lambda_h w) \exp \left[
  b \, E_{\bs
    \alpha_i} \right] \sim \Lambda_h w\ .
\eeq

\paragraph{Example 1: elementary kinks in the $USp(4)/U(2)$ sigma model \\}
Let us first consider the elementary kinks in the $USp(4)/U(2)$ sigma model. 
The Dynkin labels of the vacua 
$\tilde l_i = {\bs \alpha}_i \cdot \tilde{\bs \nu}$ 
are given by
\beq
\tilde{\bs \nu}_{++} = \frac{ \bs e_1 + \bs e_2}{2} ~~&\leftrightarrow&
[\tilde l_1,\tilde l_2] = [0,1]\ , \hs{10}
\tilde{\bs \nu}_{+-} = \frac{ \bs e_1 - \bs e_2}{2} ~~\leftrightarrow~~
[\tilde l_1,\tilde l_2] = [1,-1]\ ,\\ 
\tilde{\bs \nu}_{-+} = \frac{-\bs e_1 + \bs e_2}{2}  &\leftrightarrow&
[\tilde l_1,\tilde l_2] = [-1,1]\ ,\hs{7}
\tilde{\bs \nu}_{--} = \frac{-\bs e_1 - \bs e_2}{2}  ~\leftrightarrow~
[\tilde l_1,\tilde l_2] = [0,-1]\ .
\eeq
Therefore there is no elementary BPS kink between
$(--)$ and another vacuum.\footnote{
The kinks between
$(--)$ and the other vacua are anti-BPS monopoles.}
The lowering operators for the simple roots 
$\bs \alpha_1 = \bs e_1 - \bs e_2$ and $\bs \alpha_2 = 2 \bs e_2$ 
are given by
\beq
{\renewcommand{\arraystretch}{0.8}
\exp \left[ b \, E_{-\bs \alpha_1} \right] = 
\ba{cc|cc} \,
1 & 0 & 0 &  0 \\ \,
b & 1 & 0 &  0 \\ \hline \,
0 & 0 & 1 & -b \\ \,
0 & 0 & 0 &  1 \ea, \hs{10}
\exp \left[ b \, E_{-\bs \alpha_2} \right] = 
\ba{cc|cc} \,
1 & 0 & 0 &  0 \\ \,
0 & 1 & 0 &  0 \\ \hline \,
0 & 0 & 1 &  0 \\ \,
0 & b & 0 &  1 \ea.
}
\eeq
By using the vacuum matrices given in 
Eqs.\,\eqref{eq:USp_vac_1} and \eqref{eq:USp_vac_2}, 
we find the following matrices $\Lambda_0$ 
for single elementary monopole configurations
\beq
\Lambda_{(++,+-)} &=& \Lambda_{++} \exp \left[ b \, E_{-\bs \alpha_2} \right] ~=~
\ba{cc|cc} 0 & 0 & 1 & 0 \\ 0 & b & 0 & 1 \ea\ ,  \\
\Lambda_{(+-,-+)} &=& \Lambda_{+-} \exp \left[ b \, E_{-\bs \alpha_1} \right] ~=~
\ba{cc|cc} 0 & 0 & 1 & -b \\ b & 1 & 0 & 0 \ea\ , \\
\Lambda_{(-+,--)} &=& \Lambda_{-+} \exp \left[ b \, E_{-\bs \alpha_2} \right] ~=~
\ba{cc|cc} 1 & 0 & 0 & 0 \\ 0 & b & 0 & 1 \ea\ . 
\eeq
For these matrices, $\Lambda(x_3) = \Lambda_0 e^{M x_3}$ correspond to
the kinks interpolating between the vacua given 
in Eqs.\,\eqref{eq:USp_vac_1} and \eqref{eq:USp_vac_2}.
For example, $\Lambda_{(++,+-)}$ represents the elementary kink 
between the $(++)$ and the $(+-)$ vacua
\beq
\Lambda(x_3) ~=~ \Lambda_{(++,+-)} e^{Mx_3} ~\sim~ 
\left\{ \begin{array}{lcl} 
\ba{cc|cc} 0 & 0 & 1 & 0 \\ 0 & a & 0 & 1 \ea & \underset{x_3 \rightarrow -\infty}{\longrightarrow} & \Lambda_{++} \\ \vs{-5} \\
\ba{cc|cc} 0 & 0 & 1 & 0 \\ 0 & 1 & 0 & \frac{1}{a} \ea & \underset{x_3
  \rightarrow \infty}{\longrightarrow} & \Lambda_{+-} \end{array} \right.\ ,
\hs{10} a \equiv b \, e^{2 m_2 x_3}\ ,
\eeq
where $M={\rm diag}(m_1,m_2,-m_1,-m_2)$.

\paragraph{Example 2: elementary kinks in the $SO(6)/U(3)$ sigma model \\}
In the case of the $SO(6)/U(3)$ sigma model, 
the Dynkin labels of the vacua 
$\tilde l_i = {\bs \alpha}_i \cdot \tilde{\bs \nu}$ are
\beq
\tilde{\bs \nu}_{+++} = \frac{ \bs e_1 + \bs e_2 + \bs
  e_3}{2}~~~\,&\leftrightarrow& ~[\tilde l_1,\tilde l_2,\tilde l_3] = [0,0,1]\
, \\
\tilde{\bs \nu}_{+--} = \frac{ \bs e_1 - \bs e_2 - \bs
  e_3}{2}~~~\,&\leftrightarrow& ~[\tilde l_1,\tilde l_2,\tilde l_3] =
[1,0,-1]\ ,\\ 
\tilde{\bs \nu}_{-+-} = \frac{-\bs e_1 + \bs e_2 - \bs e_3}{2}
~&\leftrightarrow&  ~[\tilde l_1,\tilde l_2,\tilde l_3] = [-1,1,0]\ , \\
\tilde{\bs \nu}_{--+} = \frac{-\bs e_1 - \bs e_2 - \bs e_3}{2}
~&\leftrightarrow&  ~[\tilde l_1,\tilde l_2,\tilde l_3] = [0,0,-1]\ .
\eeq
The lowering operators for the simple roots 
$\bs \alpha_1 = \bs e_1 - \bs e_2$, 
$\bs \alpha_2 = \bs e_2 - \bs e_3$ and 
$\bs \alpha_3 = \bs e_2 + \bs e_3$ 
are given by
\beq
{\renewcommand{\arraystretch}{0.7}
{\arraycolsep = 1mm
e^{b E_{-\bs \alpha_1}} \! = \!
\ba{ccc|ccc} 
1 &    &   &   &    &   \\
b &  1 &   &   &    &   \\
  &    & 1 &   &    &   \\ \hline
  &    &   & 1 & -b &   \\
  &    &   &   &  1 &   \\
  &    &   &   &    & 1 \ea}}\ , \hs{1}
{\renewcommand{\arraystretch}{0.7}
{\arraycolsep = 1mm
e^{b E_{-\bs \alpha_2}} \! = \!
\ba{ccc|ccc} 
1 &   &    &   &   &    \\
  & 1 &    &   &   &    \\
  & b &  1 &   &   &    \\ \hline
  &   &    & 1 &   &    \\
  &   &    &   & 1 & -b \\
  &   &    &   &   &  1 \ea}}\ , \hs{1}
{\renewcommand{\arraystretch}{0.7}
{\arraycolsep = 1mm
e^{b E_{-\bs \alpha_3}} \! = \!
\ba{ccc|ccc} 
1 &   &    &   &   &    \\
  & 1 &    &   &   &    \\
  &   &  1 &   &   &    \\ \hline
  &   &    & 1 &   &    \\
  &   & -b &   & 1 &    \\
  & b &    &   &   &  1 \ea}} . \notag
\eeq
By using these group elements and 
the vacuum matrices given in 
Eqs.\,\eqref{eq:SO_vac_1} and \eqref{eq:SO_vac_2}, 
we find the following $\Lambda_0$ corresponding to 
the elementary monopoles
{\renewcommand{\arraystretch}{0.7}
{\arraycolsep = 1.2mm
\beq
\Lambda_{(+++,+--)} &=& \Lambda_{+++} \exp \left[ b \, E_{-\bs \alpha_3} \right] ~=~
\ba{ccc|ccc} 
0  &   &    & 1 &   &   \\ 
  &  0 & -b &   & 1 &   \\ 
  & b &   0 &   &   & 1 \ea, \\
\Lambda_{(+--,-+-)} &=& \Lambda_{+--} \exp \left[ b \, E_{-\bs \alpha_1} \right] ~=~
\ba{ccc|ccc} 
0  &   &   & 1 & -b &   \\ 
b & 1 &   &   & 0   &   \\ 
  &   & 1 &   &    &0 \ \ea, \\
\Lambda_{(-+-,--+)} &=& \Lambda_{-+-} \exp \left[ b \, E_{-\bs \alpha_2} \right] ~=~
\ba{ccc|ccc} 
1 &   &   &0   &   &    \\ 
  &  0 &   &   & 1 & -b \\ 
  & b & 1 & \ &   &  0  \ea \ .
\eeq
}}
\subsubsection{Multi-monopole configurations}
As we have seen in the previous section, 
the lowering operators for the simple roots 
are ``creation operators'' of the elementary monopoles. 
Now let us consider the matrix $\Lambda_0$ 
with two lowering operators for different simple roots 
\beq
\Lambda_0 ~=~ (\Lambda_h w )
\exp \left[ b_i E_{\bs \alpha_i} \right] 
\exp \left[ b_j E_{\bs \alpha_j} \right]\ , \hs{10} i \not = j\ .
\eeq
In this case, we have the following three expressions 
for the matrix $\Lambda(x_3)$
\beq
\Lambda(x_3) &\sim& \left\{ 
{\arraycolsep 0.5mm
\begin{array}{cll} 
(\Lambda_hw) & \exp \left[ b_i e^{(\bs \alpha_i \cdot \mathbf m) x_3} E_{-\bs
    \alpha_i}  \right] & \exp \left[ b_j e^{(\bs \alpha_j \cdot \mathbf m)
    x_3} E_{-\bs \alpha_j} \right]  \\
(\Lambda_h w \, w_{\bs \alpha_i}) & \exp \left[ b_i^{-1} e^{-(\bs \alpha_i
    \cdot \mathbf m) x_3} E_{\bs \alpha_i} \right] & \exp \left[ b_j e^{(\bs
    \alpha_j \cdot \mathbf m) x_3} E_{-\bs \alpha_j} \right]  \\
(\Lambda_h w \, w_{\bs \alpha_i} w_{\bs \alpha_j}) & \exp \left[ b_j^{-1}
  e^{-(\bs \alpha_j \cdot \mathbf m) x_3} E_{\bs \alpha_j} \right] & \exp
\left[ b_i^{-1} e^{-(\bs \alpha_i \cdot \mathbf m) x_3} E_{\bs \alpha_i}
\right] \end{array}} \right.\ , 
\eeq
where we have used the fact that $[E_{\bs \alpha_i},E_{-\bs \alpha_j}]=0$ 
for any simple roots $\bs \alpha_i$ and $\bs \alpha_j$ $(i \not = j)$.
If we set the moduli parameters as\footnote{There is no summation on $i$ and
  $j$ in Eq.\,(\ref{eq:logb}).}
\beq \label{eq:logb}
\log b_i = - (\bs \alpha_i \cdot \mathbf m) r_i + i \eta_i\ , \hs{10}
\log b_j = - (\bs \alpha_j \cdot \mathbf m) r_j + i \eta_j\ ,
\eeq
we find that the matrix $\Lambda(x_3)$ flows though the following three vacua
\beq
\Lambda(x_3) \sim \left\{ 
\begin{array}{cl} 
\Lambda_h w & x_3 \ll r_i \\ 
\Lambda_h w \, w_{\bs \alpha_i} & r_i \ll x_3 \ll r_j \\ 
\Lambda_h w \, w_{\bs \alpha_i} w_{\bs \alpha_j} & r_j \ll x_3 \end{array}
\right. ,
\label{eq:two_m}
\eeq 
where we have assumed $r_i \ll r_j$.
Eq.\,\eqref{eq:two_m} shows that 
$\Lambda(x_3)$ corresponds to 
the configuration of two elementary kinks 
$\tilde{\bs \alpha}_i$ and $\tilde{\bs \alpha}_j$
located at $x_3 = r_i$ and $x_3 = r_j$, respectively
\beq
\begin{array}{ccccc}
\underset{x_3 \ll r_i}{(\Lambda_h w)\mbox{-vortex}} & 
\rightarrow &
\underset{x_3 = r_i}{\tilde{\bs \alpha}_i\mbox{-monopole}} &
\rightarrow &
\underset{r_i \ll x_3 \ll r_j}{(\Lambda_hw \, w_{\bs \alpha_i})\mbox{-vortex}} 
\end{array} \hs{8} \notag \\
\begin{array}{cccc}
\rightarrow &
\underset{x_3 = r_j}{\tilde{\bs \alpha}_j\mbox{-monopole}} &
\rightarrow & 
\underset{r_j \ll x_3}{(\Lambda_h w \, w_{\bs \alpha_i} w_{\bs \alpha_j})\mbox{-vortex}}
\end{array}\ . \notag 
\eeq
If the lowering operators $E_{-\bs \alpha_i}$ and $E_{-\bs \alpha_j}$ 
commute with each other, 
we have another expression for the matrix $\Lambda(x_3)$
\beq
\Lambda(x_3) &\sim& (\Lambda_h w \, w_{\bs \alpha_j}) \exp \left[ b_j^{-1}
  e^{-(\bs \alpha_j \cdot \mathbf m) x_3} E_{\bs \alpha_j} \right] \exp \left[
  b_i e^{(\bs \alpha_i \cdot \mathbf m) x_3} E_{-\bs \alpha_i} \right]  \\
&\approx& \Lambda_h w \, w_{\bs \alpha_j}~~~(r_i \gg x_3 \gg r_j)\ .
\eeq
Therefore the ordering of the monopoles can be exchanged 
if the corresponding operators satisfy 
$[E_{-\bs \alpha_i},E_{-\bs \alpha_j}]=0$
\beq
\begin{array}{ccccc}
\underset{x_3 \ll r_j}{(\Lambda_h w)\mbox{-vortex}} & 
\rightarrow &
\underset{x_3 = r_j}{\tilde{\bs \alpha}_j\mbox{-monopole}} &
\rightarrow &
\underset{r_j \ll x_3 \ll r_i}{(\Lambda_hw \, w_{\bs \alpha_j})\mbox{-vortex}} 
\end{array} \hs{8} \notag \\
\begin{array}{cccc}
\rightarrow &
\underset{x_3 = r_i}{\tilde{\bs \alpha}_i\mbox{-monopole}} &
\rightarrow & 
\underset{r_i \ll x_3}{(\Lambda_h w \, w_{\bs \alpha_i} w_{\bs \alpha_j})\mbox{-vortex}}
\end{array}\ . \notag 
\eeq
On the other hand, if the lowering operators do not commute
$[E_{-\bs \alpha_i},E_{-\bs \alpha_j}] \not =0$,
the monopoles have a fixed ordering, 
which was found for the Grassmannian sigma model \cite{Isozumi:2004va}. 
To see this, let us consider the following example 
in the $USp(4)/U(2)$ sigma model
\beq
\Lambda(x_3) ~=~ \Lambda_{++} 
\exp \left[ b_2 E_{-\bs \alpha_2} \right] 
\exp \left[ b_1 E_{-\bs \alpha_1} \right] e^{M x_3} \phantom{\bigg[} ~\sim~ 
\ba{cc|cc} 
a_1^2 a_2 & a_1 a_2 & 1 & 0 \\
a_1 a_2 & a_2 & 0 & 1 \ea\ ,
\eeq
where $a_1$ and $a_2$ are given by
\beq
a_1 &=& b_1 e^{(m_1-m_2) x_3} ~=~ e^{(m_1-m_2) (x_3-r_1) + i \eta_1}\ , \\
a_2 &=& b_2 e^{2 m_2 x_3} ~~~~~\,~=~ e^{2 m_2 (x_3-r_2) + i \eta_2}\ , 
\eeq
In this configuration, there are two monopoles 
at $x_3 = r_2$ and $x_3 = r_1$ if $r_2 < r_1$. 
Let us take the limits $r_2 \rightarrow \infty$ and $r_1 \rightarrow - \infty$ 
while keeping $a_1^2 a_2$ fixed 
\beq
\Lambda(x_3) &\sim& 
\ba{cc|cc} 
a_{1+2} & 0 & 1 & 0 \\
0 & 0 & 0 & 1 \ea 
~\sim~ \Lambda_{++} \exp \left[ a_{1+2} E_{-(\bs \alpha_1+\bs \alpha_2)}
\right]\ ,
\label{eq:Lambda_a0}
\eeq
where $a_{1+2}$ takes the form 
$a_{1+2} = \exp \left[ (m_1 + m_2) (x_3 - r_{1+2} ) + i \eta_{1+2} \right]$. 
Eq.\,\eqref{eq:Lambda_a0} corresponds to a single monopole 
with root vector $\tilde{\bs \alpha}_1 + \tilde{\bs \alpha}_2$. 
Therefore the monopoles with $\tilde{\bs \alpha}_{1,2}$ cannot be exchanged 
and become a single monopole with 
$\tilde{\bs \alpha}_1 + \tilde{\bs \alpha}_2$ in the coincident limit.

\begin{figure}[!ht]
\begin{center}
\begin{tabular}{ccc}
\includegraphics[width=50mm]{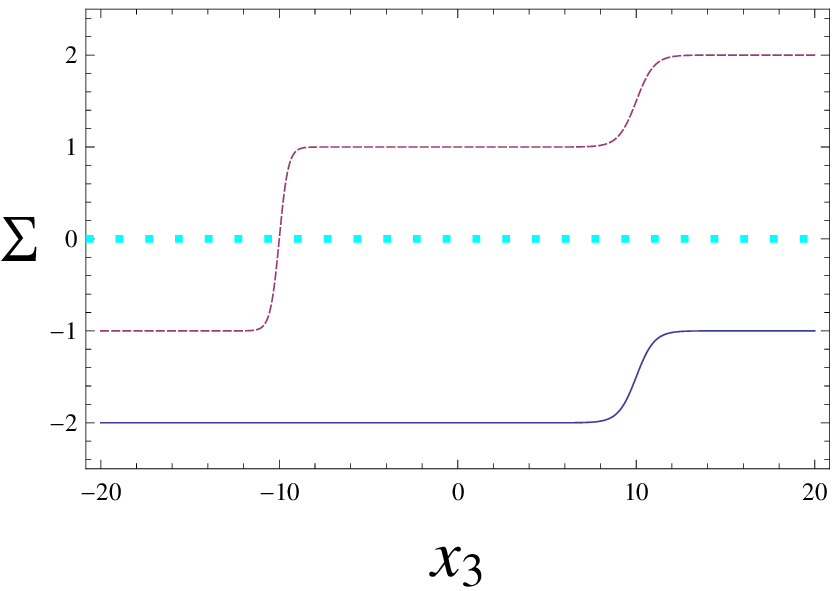} &
\includegraphics[width=50mm]{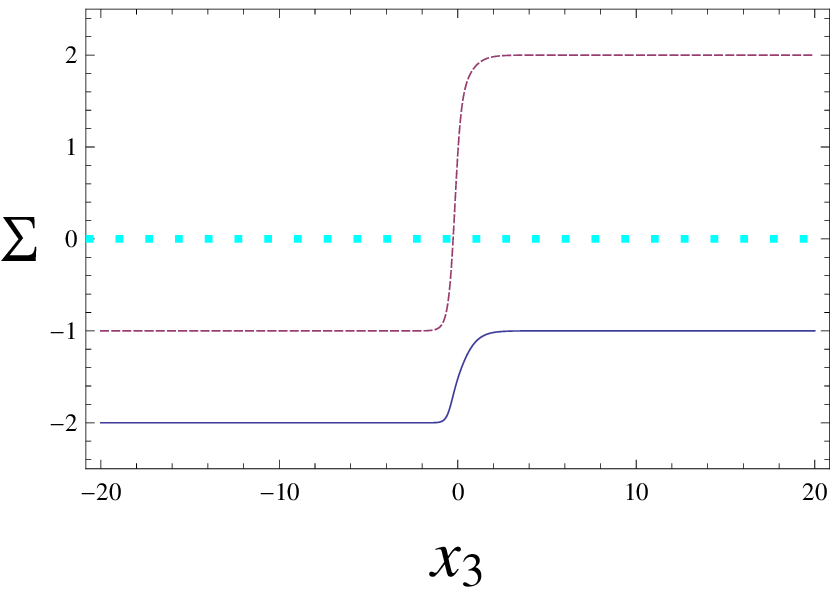} &
\includegraphics[width=50mm]{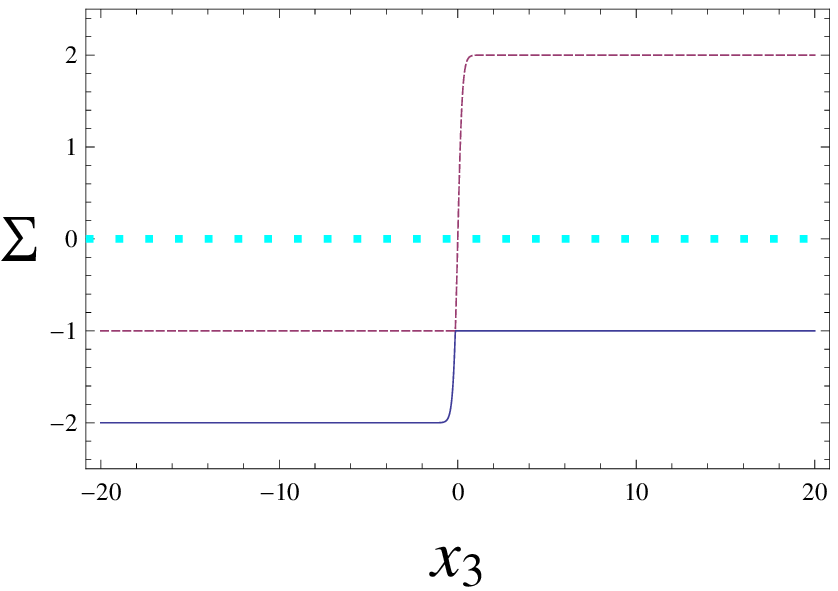} \\
$r_1 =10,~ r_2 = -10$ &
$r_1 =0 ,~ r_2 = 0$  &
$r_1 =-10,~r_2 = 10$
\end{tabular}
\end{center}
\caption{The eigenvalues of $\Sigma$ as functions of $x_3$ 
($G=USp(4),~m_1=2,m_2=1$).}
\label{fig:Sigma}
\end{figure}

To understand the ordering of the monopoles, 
it is convenient to introduce the following matrix\footnote{
The definition of $\Sigma$ here is different from that 
is used in the Grassmannian sigma model 
\cite{Isozumi:2004jc,Isozumi:2004va}
by a matrix conjugation $\Sigma \rightarrow S^{-1} \Sigma S$. }
\beq
\Sigma(x_3) ~\equiv~ (\Lambda \Lambda^\dagger)^{-1} \Lambda M \Lambda^\dagger\
.
\eeq
Since $\Tr \, \Sigma = \sigma$ is the magnetic flux which has a kink profile, 
the eigenvalues of $\Sigma$ also have kink profiles, 
which contain more information on the ordering of the kinks, 
as in the case of Grassmannian sigma model \cite{Eto:2004vy}.
Fig.\,\ref{fig:Sigma} shows the coincident limit of 
the $\bs \alpha_1$- and $\bs \alpha_2$-monopoles in the $USp(4)$ case. 

\begin{figure}[!ht]
\begin{center}
\includegraphics[width=170mm]{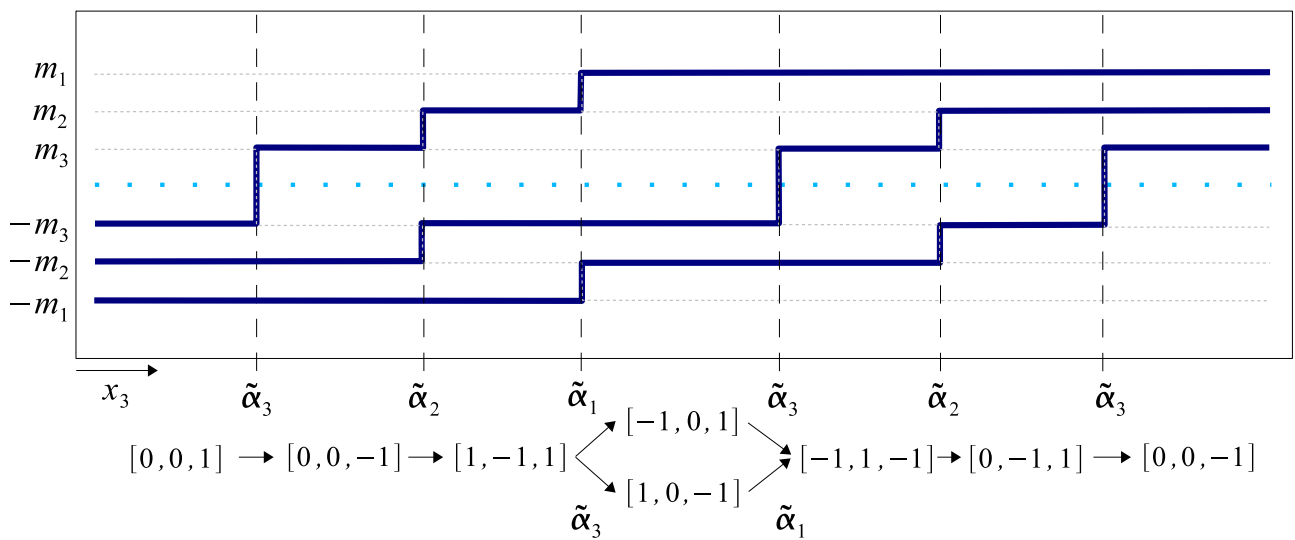}
\caption{
The $USp(6)$ maximal kink configuration in the thin wall limit. 
The eigenvalues of $\Sigma$ (solid lines) and their mirror image 
(dashed lines) with respect to the line $\Sigma=0$ (dotted line) 
are all different in each vortex vacuum 
(the regions are separated by vertical dashed lines). 
The sequence of Dynkin labels of the $SO(7)$ spinor is assigned 
so that the subsequent labels are obtained by subtracting 
the row of the Cartan matrix $\tilde C$ corresponding to 
the coroot of the monopole between the vacua. 
The branch of the sequence corresponds to commutative monopoles.}
\label{fig:USp(6)}
\end{center}
\end{figure}

\begin{figure}[!ht] 
\begin{center}
\includegraphics[width=170mm]{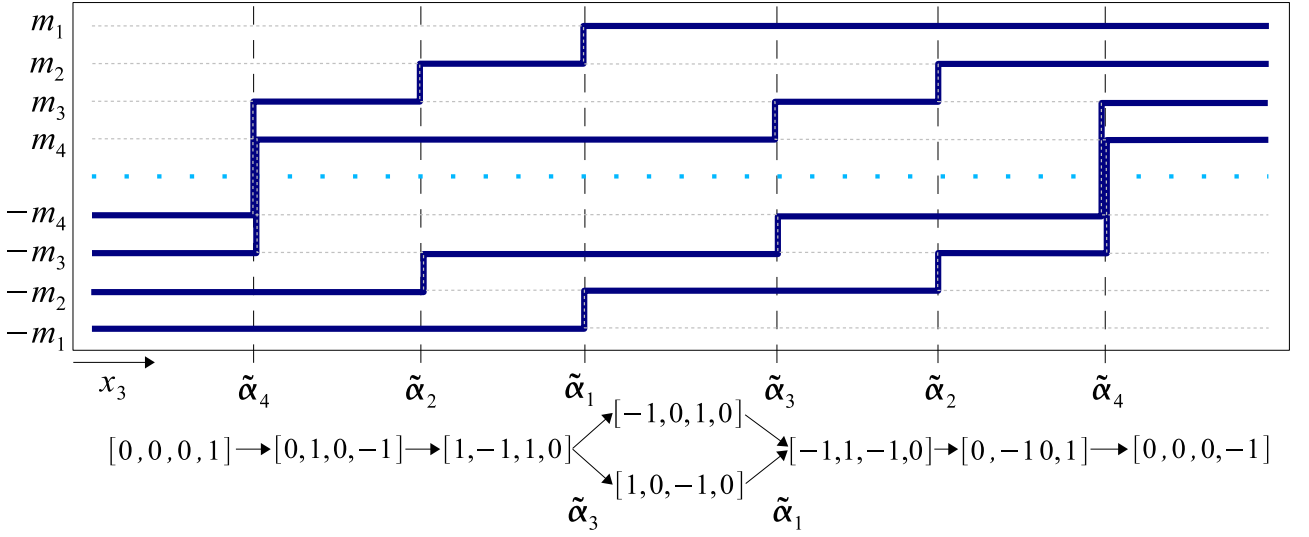}
\caption{
The $SO(8)$ maximal kink configuration in the thin wall limit and
the sequence of Dynkin labels of the $SO(8)$ Weyl spinor. 
Each eigenvalue (solid lines) can cross the dotted line $\Sigma=0$ 
only if it is paired with another eigenvalue.
Because of the vacuum condition, 
two kinks must go up simultaneously across the dotted line.}
\label{fig:SO(8)}
\end{center}
\end{figure}

In general, the matrix $\Sigma$ has the following properties: 
\begin{enumerate}
\item
The eigenvalues of $\Sigma$ are increasing functions of $x_3$.
\item
At the vacuum points, 
all of the eigenvalues are different and 
given by the mass parameters $\pm m_i$. 
\item
If $m_i$ $(-m_i)$ is one of the eigenvalues, 
$-m_i$ $(m_i)$ is not contained in the set of eigenvalues.
\item
For the vacuum $\Lambda_{s_1 \cdots s_n}~(s_i=\pm)$, 
the eigenvalues are $-s_1 m_1\ ,\cdots\ , -s_n m_n$.
\end{enumerate}
With these rules in mind, 
we can draw the generic kink profiles of eigenvalues diagrammatically
in the thin wall limit\footnote{ 
The thin wall limit can be interpreted as the large mass limit in which
the kink profile becomes a step function $\lim_{m \rightarrow \infty} \left[
  1 + \tanh (m x) \right] = 2 \theta(x)$ where $\theta(x)$ stands for the step
function.
In the $SU(N)$ case 
a different limit was taken in Ref.~\cite{Eto:2007aw} to obtain 
kinks with constant slopes, 
in order to study the statistical mechanics of 
non-Abelian vortices.
}
(see $USp(6)$ and $SO(8)$ examples in Figs.\,\ref{fig:USp(6)} and \ref{fig:SO(8)}).
In the $SU(N)$ case, 
the profiles of eigenvalues of $\Sigma$ were interpreted as 
kinky D-brane configurations \cite{Eto:2004vy}.

Finally, let us discuss the most generic configurations of kinks.
The matrix $\Lambda_0$ for the generic configuration can be obtained 
from $\Lambda_h$ by multiplying as many lowering operators as possible
\beq
\Lambda_0 ~=~ \Lambda_h 
\exp [ b_{i_1} E_{-\bs \alpha_{i_1}} ] \cdots
\exp [ b_{i_p} E_{-\bs \alpha_{i_p}} ]\ ,
\eeq
where $\bs \alpha_{i_j}$ are simple roots. 
Since the multiplication of the lowering operators 
becomes trivial at a finite number $p$, 
there exist ``maximal kink configurations'' 
containing the maximal number of kinks.
As shown in Appendix \ref{appendix:ordering}, 
we can always rewrite the matrix $\Lambda_0$ as
\beq
\Lambda_0 &\sim& (\Lambda_h w_{\bs \alpha_{i_1}}) 
\exp [ b_{i_2}' E_{-\bs \alpha_{i_2}} ] \cdots
\exp [ b_{i_p}' E_{-\bs \alpha_{i_p}} ]
\exp [ {b_{i_1}'}^{-1} E_{\bs \alpha_{i_1}} ] \\
&\sim& (\Lambda_h w_{\bs \alpha_{i_1}} w_{\bs \alpha_{i_2}}) 
\exp [ b_{i_3}'' E_{-\bs \alpha_{i_3}} ] \cdots
\exp [ b_{i_p}'' E_{-\bs \alpha_{i_p}} ]
\exp [ {b_{i_2}''}^{-1} E_{\bs \alpha_{i_2}} ]
\exp [ {b_{i_1}''}^{-1} E_{\bs \alpha_{i_1}} ] \notag \\
&\sim& \cdots\ . \notag 
\eeq
Repeating this procedure, 
we can read off the vortex-monopole configuration 
from $x_3 \rightarrow - \infty$ to $x_3 \rightarrow \infty$
\beq
\Lambda_h\mbox{-vortex} ~\rightarrow~ 
\tilde{\bs \alpha}_{i_1}\mbox{-monopole} ~\rightarrow~ 
(\Lambda_h w_{\bs \alpha_i})\mbox{-vortex} ~\rightarrow \cdots \rightarrow~ 
(\Lambda_h w_{\bs \alpha_{i_1}} \cdots w_{\bs \alpha_{i_p}})
\mbox{-vortex}\ . \notag
\eeq
The lowering operator $\exp [ b_i E_{-\bs \alpha_i} ]$ 
creates the $\tilde{\bs \alpha}_i$-monopole 
if the vortex has the coweight $\tilde{\bs \nu}$ 
with a positive Dynkin label $\tilde l_i = \tilde{\bs \nu} \cdot \bs \alpha_i$. 
Then it is connected to the vortex 
with the coweight $\tilde{\bs \nu} - \tilde{\bs \alpha}_i$ 
whose Dynkin labels are $\tilde l_j' = \tilde l_j - \tilde C_{ij}$. 
Here $\tilde C_{ij}$ is an element of 
the Cartan matrix 
\beq
\tilde C_{ij} \equiv \tilde{\bs \alpha}_i \cdot \bs \alpha_j\ .
\eeq
Therefore, the most generic 
$SO(2n)$ and $USp(2n)$ vortex-monopole configurations can be 
constructed in the same way as the (Weyl) spinor representation of 
$SO(2n)$ and $SO(2n+1)$ (see the $USp(6)$ and $SO(8)$ examples given in
Figs.\,\ref{fig:USp(6)} and \ref{fig:SO(8)}).

\paragraph{Example 1: maximal kink configuration in the $USp(6)/U(3)$ sigma model \\}
Let us consider the maximal kink configuration 
in the $USp(6)/U(3)$ sigma model.
The most generic form of the matrix $\Lambda_0$ is given by
\beq
\Lambda_0 &=& 
\Lambda_h \, 
\exp [b_{3,1} E_{-\bs \alpha_3}] \, 
\exp [b_{2,1} E_{-\bs \alpha_2}] \, 
\exp[b_{1,1} E_{-\bs \alpha_1}] \notag \\
&{}& \times ~\,
\exp [b_{3,2} E_{-\bs \alpha_3}] \, 
\exp [b_{2,2} E_{-\bs \alpha_2}] \notag \\
&{}& \times ~\,
\exp [b_{3,3} E_{-\bs \alpha_3}]\ .
\eeq
The number of moduli parameters $b_{i,j}$ is 
in accordance with the number of the matrix elements of 
the symmetric matrix $B_0$. 
The corresponding kink profile and 
the sequence of Dynkin labels of the $SO(7)$ spinor representation
are given in Fig.\,\ref{fig:USp(6)}.

In general, the matrix $\Lambda_0$ for the maximal kink configurations 
in the $USp(2n)/U(n)$ sigma model is given by
\beq
\Lambda_0 &=& \Lambda_h 
\exp[b_{n,1} E_{-\bs \alpha_n}] \cdots 
\exp[b_{2,1} E_{-\bs \alpha_2}] 
\exp[b_{1,1} E_{-\bs \alpha_1}] \notag \\ &{}& \times \
\exp[b_{n,2} E_{-\bs \alpha_n}] \cdots 
\exp[b_{2,2} E_{-\bs \alpha_2}] \notag \\ &{}& \vdots \notag \\
&{}& \times \
\exp[b_{n,n} E_{-\bs \alpha_n}]\ . 
\eeq
There are $i~(i=1,\ldots,n)$ $\tilde{\bs \alpha}_i$-monopoles 
in the maximal configuration
and the complex parameters $b_{i,j}~(j=1,\ldots,i)$ are 
the position and phase moduli of
$j$-th $\tilde{\bs \alpha}_i$-monopole.

\paragraph{Example 2: maximal kink configuration in the $SO(8)/U(4)$ sigma model \\}
The matrix $\Lambda_0$ for the maximal kink configuration 
in the $SO(8)/U(4)$ sigma model is given by
\beq
\Lambda_0 &=& \Lambda_h \, \exp [ b_{4,1} E_{-\bs \alpha_4} ] \, \exp [ b_{2,1} E_{-\bs \alpha_2} ] \, \exp [ b_{1,1} E_{-\bs \alpha_1} ] \notag \\
&{}& \times ~\, \exp [ b_{3,2} E_{-\bs \alpha_3} ] \, \exp [ b_{2,2} E_{-\bs \alpha_2} ] \notag \\
&{}& \times ~\, \exp [ b_{4,3} E_{-\bs \alpha_4} ]\ .
\eeq
As in the case of $USp(6)/U(3)$, 
the number of moduli parameters $b_{i,j}$ is 
in accordance with that of the matrix elements of 
the anti-symmetric matrix $B_0$. 
The kink profiles and 
the sequence of Dynkin labels of the $SO(8)$ 
Weyl spinor representation
are given in Fig.\,\ref{fig:SO(8)}.

The generic form of the matrix $\Lambda_0$ 
for the maximal kink configurations 
in the $SO(2n)/U(n)$ sigma model is given by
\beq
\Lambda_0 &=& \Lambda_h 
\exp[b_{n,1} E_{-\bs \alpha_n}] 
\exp[b_{n-2,1} E_{-\bs \alpha_{n-2}}] \cdots 
\exp[b_{2,1} E_{-\bs \alpha_2}] 
\exp[b_{1,1} E_{-\bs \alpha_1}] \notag \\ &{}& \times \
\exp[b_{n-1,2} E_{-\bs \alpha_{n-1}}] \cdots 
\exp[b_{2,2} E_{-\bs \alpha_2}] \notag \\ &{}& \times \
\exp[b_{n,3} E_{-\bs \alpha_n}] 
\exp[b_{n-2,3} E_{-\bs \alpha_{n-2}}] \cdots 
\exp[b_{3,3} E_{-\bs \alpha_3}] \notag \\ &{}& \times \
\exp[b_{n-1,4} E_{-\bs \alpha_{n-1}}] \cdots 
\exp[b_{4,4} E_{-\bs \alpha_4}] \notag \\ &{}& \vdots \notag \\
&{}& \times \
L\ , 
\eeq
where the last operator $L$ is 
\beq
L ~=~ \left\{ 
\begin{array}{cl} 
\exp[b_{n,n} E_{-\bs \alpha_{n}}] & \mbox{for $SO(4n'+4)$} \\
\exp[b_{n-1,n}   E_{-\bs \alpha_{n-1}}]   & \mbox{for $SO(4n'+2)$} 
\end{array} \right. \ .
\eeq
For $SO(4n'+2)~(SO(4n'+4))$,
the numbers of $\tilde{\bs \alpha}_i$-monopoles 
in the maximal configuration are 
\beq
\#\mbox{$\tilde{\bs \alpha}_i$-monopoles} ~=~ \left\{ 
\begin{array}{ll} 
i & \mbox{for $i=1,\cdots,n-2$} \\
n' & \mbox{for $i=n-1$} \\
n'~(n'+1) & \mbox{for $i=n$}
\end{array} \right. .
\eeq 

\begin{figure}[!ht]
\begin{center}
\includegraphics[width=100mm]{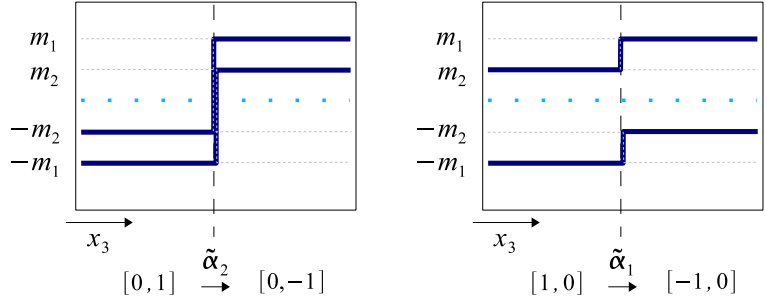}
\caption{The $SO(4)$ kink configurations on the vortices with opposite $\mathbb Z_2$ topological charge. }
\label{fig:SO(4)}
\end{center}
\end{figure}

We can also discuss the kinks on the vortex 
with opposite $\mathbb Z_2$ topological charge (chirality) by 
starting from the following highest weight vacuum 
\beq
\Lambda_h' ~=~ \ba{cc|cc} \mathbf 0_{n-1} & & \mathbf 1_{n-1} & \\ & 1 & & 0 \ea
.
\eeq
In the case of $SO(4)$, the vortices with different 
$\mathbb Z_2$ topological charge admit different types of 
elementary monopoles (see Fig.\,\ref{fig:SO(4)})
\beq
\Lambda_{(++,--)} &=& \Lambda_h \exp [ b \, E_{-\bs \alpha_1} ] ~=~ \ba{cc|cc} 0 & -b & 1 & 0 \\ b & 0 & 0 & 1 \ea, \\
\Lambda_{(+-,-+)} &=& \Lambda_h' \exp [ b \, E_{-\bs \alpha_2} ] ~=~ \ba{cc|cc} 0 & 0 & 1 & -b \\ b & 1 & 0 & 0 \ea,
\eeq
where $\bs \alpha_1 = \bs e_1 - \bs e_2$ and 
$\bs \alpha_2 = \bs e_1 + \bs e_2$.

\section{Witten's effect on the vortex worldsheet}

In this section we consider the dyonic configuration. 
The bulk $\theta$-term induces 
a $\theta$-term on the vortex worldsheet, 
which can be obtained by substituting 
the solution \eqref{eq:A_alpha} into the bulk $\theta$-term
\beq
\mathcal L_\theta ~=~ \frac{\theta}{32\pi^2} \epsilon^{\mu \nu \rho \sigma}
\int d^4 x \, \Tr ( F_{\mu \nu} F_{\rho \sigma} ) ~=~ \frac{\theta}{2\pi}
\int d^2 x \, \frac{i g^2}{4\pi } g_{i \bar j} \epsilon^{\alpha \beta}
\p_\alpha b^i \p_\beta \bar b^{\bar j}\ , 
\eeq
where $\epsilon^{t x_3} = - \epsilon^{x_3 t} = 1$. 
This induced $\theta$-term on the vortex worldsheet is 
proportional to the pullback of the K\"ahler form onto the vortex worldsheet.
For the $SO(2n)/U(n)$ and $USp(2n)/U(n)$ cases, 
the explicit form of the induced $\theta$-term is given by
\beq
\mathcal L_{\theta} ~=~ - i\frac{\theta}{2\pi}  \epsilon^{\alpha \beta} \Tr
\left[ X^{-1} \p_\alpha B^\dagger Y^{-1} \p_\beta B \right]\ .
\label{eq:theta}
\eeq
For the $U(N)$ theory  this reduces to the $\theta$-term in the
worldsheet $\mathbb{C}P^{N-1}$ sigma model,  discussed by Gorsky
et.~al.~\cite{Gorsky:2004ad}. 
Thus the total effective Lagrangian becomes
\beq
\mathcal L &=& \Tr \left[ \left( \frac{4\pi}{g^2} \eta^{\alpha \beta} - i
    \frac{\theta}{2\pi} \epsilon^{\alpha \beta} \right) X^{-1} \p_\alpha
  B^\dagger Y^{-1}  \p_\beta B - \frac{4\pi}{g^2} X^{-1} \{M_n, B^\dagger \} Y^{-1} \{M_n, B \} \right] \notag \\
&=& \Tr \left[ {\rm Im} \left( \tau \, X^{-1} \p_- B^\dagger Y^{-1} \p_+ B
  \right) - \frac{4\pi}{g^2} X^{-1} \{M_n, B^\dagger \} Y^{-1} \{M_n, B \}
\right]\ ,
\eeq
where $\tau$ is the complex coupling constant
\beq
\tau \equiv \frac{\theta}{2\pi} + i \frac{4\pi}{g^2}\ , \hs{10} \p_\pm \equiv
\p_t \pm \p_{x_3}\ .
\eeq
Although the $\theta$-term \eqref{eq:theta} 
does not change the equation of motion, 
it shifts the conserved Noether charges of the $U(1)^n$ global
symmetry ($\mathbf H_n$ is defined in Eq.\,(\ref{remember})) 
\beq
\mathbf q &=& \,\, i \frac{4\pi}{g^2} \int dx_3 \Tr \Big[ X^{-1} \p_t
B^\dagger  Y^{-1} \{{\mathbf H}_n,B\} - X^{-1} \{\mathbf H_n,B^\dagger\} Y^{-1} \p_t B \Big] \notag \\
&{}& - \frac{\theta}{2\pi} \int dx_3 \Tr \Big[ X^{-1} \p_{x_3} B^\dagger
Y^{-1} \{\mathbf H_n,B\} + X^{-1} \{\mathbf H_n,B^\dagger\} Y^{-1} \p_{x_3} B \Big]\ .
\label{eq:charge_density} 
\eeq
Recalling Eqs.\,(\ref{masses}) and (\ref{magcha}), the shift of the $U(1)^{n}$ charges, can be written as
\beq
\Delta  \mathbf q   ~=~ - \frac{\theta}{2\pi}  \mathbf g\ .
\eeq
By using Eq.\,(\ref{kore})  the shift of the  $U(1)_M$ charge  
can be expressed as 
\beq
\Delta (\mathbf m \cdot \mathbf q) ~=~ - \frac{\theta}{2\pi} \int d x_3 \, \p_{x_3} \sigma ~\,.
\eeq
These represent the Witten effect \cite{Witten:1979ey} for the
monopoles on the vortex worldsheet.  

Now let us consider the dyonic configuration. 
For given values of the conserved Noether charge $\mathbf q$ and 
the topological charge $\mathbf g$, 
the energy can be rewritten as
\beq
E &=& \phantom{+} \frac{4\pi}{g^2} \int dx_3 \Tr \left[ X^{-1} \Big( \p_t
  B^\dagger  + i \sin \mu \{ M_n , B^\dagger \} \Big) Y^{-1} \Big( \p_t B - i \sin \mu \{ M_n , B \} \Big) \right] \notag \\
&{}& + \frac{4\pi}{g^2} \int dx_3 \Tr \left[ X^{-1} \Big( \p_{x_3} B^\dagger -
  \cos \mu \{ M_n , B^\dagger \} \Big) Y^{-1} \Big( \p_{x_3} B - \cos \mu \{ M_n , B \} \Big) \right] \notag \\
&{}& + \frac{4\pi}{g^2} \mathbf m \cdot \mathbf g \cos \mu + \mathbf m \cdot
\left( \mathbf q + \frac{\theta}{2\pi} \mathbf g \right) \sin \mu\ .
\eeq
The quantity in the third line is extremized  by 
\beq
\tan \mu ~=~ \frac{g^2}{4\pi} \left( \frac{\mathbf m \cdot \mathbf q}{\mathbf m
    \cdot \mathbf g} + \frac{\theta}{2 \pi} \right)\ .
\eeq
Therefore, the energy is bounded from below by the central charge as
\beq
E &\geq& \sqrt{ \left[ \frac{4\pi}{g^2} \mathbf m \cdot \mathbf g \right]^2 +
  \left[ \mathbf m \cdot \left( \mathbf q + \frac{\theta}{2\pi} \mathbf g
    \right) \right]^2 } ~~=~~ \big| \mathbf m \cdot ( \mathbf q + \tau \mathbf
g ) \big|\ ~=~ |Z|.
\label{eq:bound_dyon}
\eeq
The inequality \eqref{eq:bound_dyon} is saturated 
if the following BPS equations are satisfied
\beq
\p_t B - i \sin \mu \{ M_n , B \} ~=~ 0\ , \hs{10} \p_{x_3} B - \cos \mu \{
M_n , B \} ~=~ 0\ .
\eeq
The general solution is given by
\beq
B(t,x_3) ~=~ 
\exp \left[ M_n ( i t \sin \mu + x_3 \cos \mu ) \right] B_0 \, 
\exp \left[ M_n ( i t \sin \mu + x_3 \cos \mu ) \right]\ .
\eeq
We can also express this solution in terms of the matrix $\Lambda$
\beq
\Lambda(t,x_3) ~=~ 
\ba{c|c} B_0 & \mathbf 1_n \ea 
\exp \left[ M_n ( i t \sin \mu + x_3 \cos \mu ) \right] 
~=~ \Lambda_0 \, \exp \left[ M_n ( i t \sin \mu + x_3 \cos \mu)
  \right]\ . \notag
\eeq
As an example, let us take the matrix $\Lambda_0$ 
for the single monopole configuration
\beq
\Lambda(t,x_3) &=& \Lambda_h \exp \left[ b E_{-\bs \alpha} \right] \exp \left[ M (i t \sin \mu  + x_3 \cos \mu) \right] \notag \\
&\sim& \Lambda_h \exp \big[ b \exp \left[ \mathbf m \cdot \bs \alpha (i t
    \sin \mu + x_3 \cos \mu) \right] E_{-\bs \alpha} \big]\ . 
\eeq
Since $\arg b$ is the phase moduli of the monopole, 
this dyonic solution can be interpreted 
as the $\tilde{\bs \alpha}$-monopole with a rotating phase
\beq
\eta(t) ~=~ \arg b + \sin \mu \, (\mathbf m \cdot \bs \alpha) \, t\ .
\eeq
We have seen in the previous section that the monopole phase 
transforms under the $U(1)^n$ symmetry as 
$\eta \rightarrow \eta + \bs \theta \cdot \bs \alpha$. 
This transformation property implies that 
the conjugate momentum of the phase $p_\eta$ and 
the Noether charge $\mathbf q$ are related by $\mathbf q = p_\eta \, \bs
\alpha$. 
Upon semi-classical quantization of the monopole moduli, 
the conjugate momentum of the phase, which has period $2\pi$, 
is quantized as $p_\eta \in \mathbb Z$. 
Therefore, the Noether charge $\mathbf q$ 
should be an integer multiple of the root $\bs \alpha$
\beq
\mathbf q ~=~ n_q \, \bs \alpha\ , \hs{10} n_q \in \mathbb Z\ . 
\eeq
The BPS spectrum of the monopoles and dyons with 
the magnetic charge $\mathbf g = 2 \tilde{\bs \alpha}$ is therefore given by
\beq
M_{\bs \alpha,n_q} ~=~ \left| \mathbf m \cdot \bs \alpha \left( n_q +
    \frac{4}{\bs \alpha \cdot \bs \alpha} \tau \right) \right|\ .
\eeq
This spectrum is invariant under a $2\pi$ rotation of 
the $\theta$-angle ($\tau \rightarrow \tau + 1$) since 
the length of the root vectors are normalized by
\beq
\frac{4}{\bs \alpha \cdot \bs \alpha} ~=~ \left\{ 
\begin{array}{cl} 
1 & ~~\mbox{for the long roots of $USp(2n)$}\;; \\
2 & ~~\mbox{for the other roots}\;. 
\end{array} \right .  
\eeq

\begin{figure}[!ht]
\begin{center}
\begin{tabular}{ccc}
\includegraphics[width=50mm]{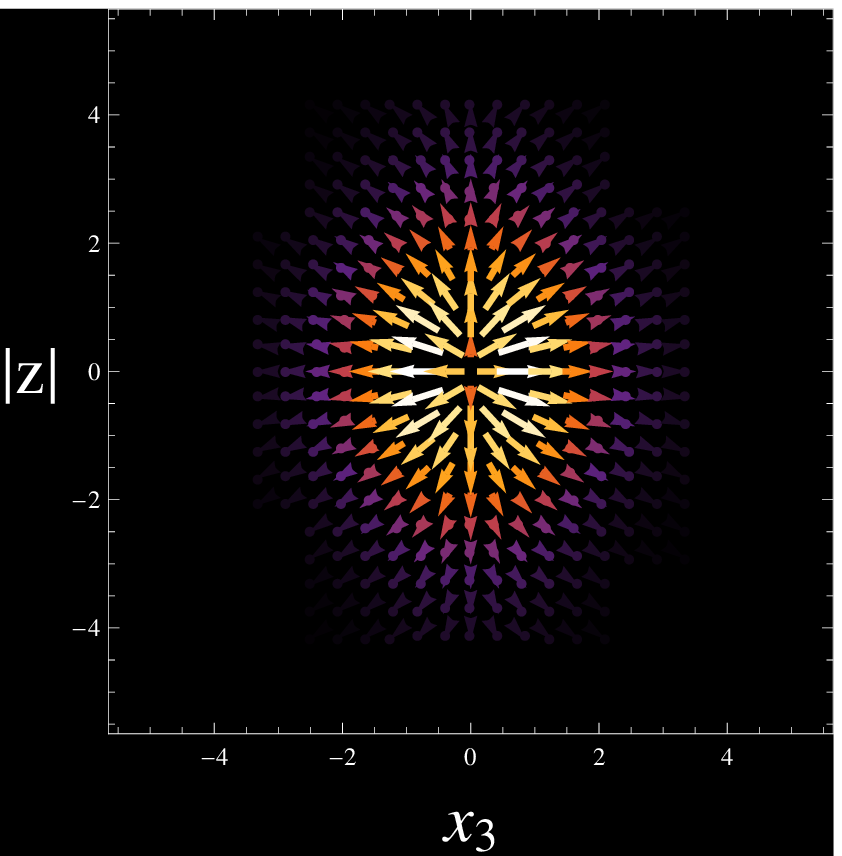} &
\includegraphics[width=50mm]{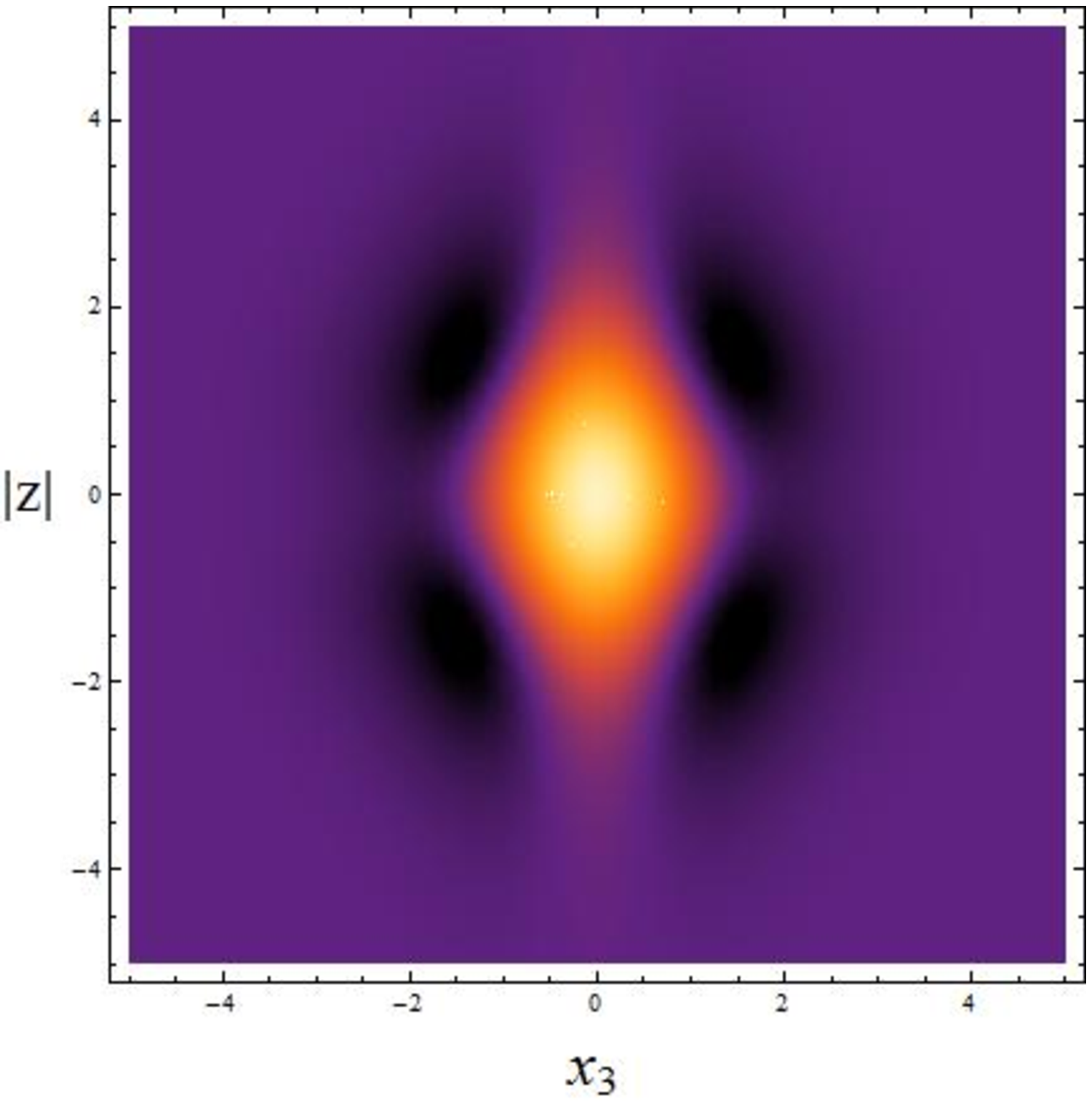} &
\includegraphics[width=50mm]{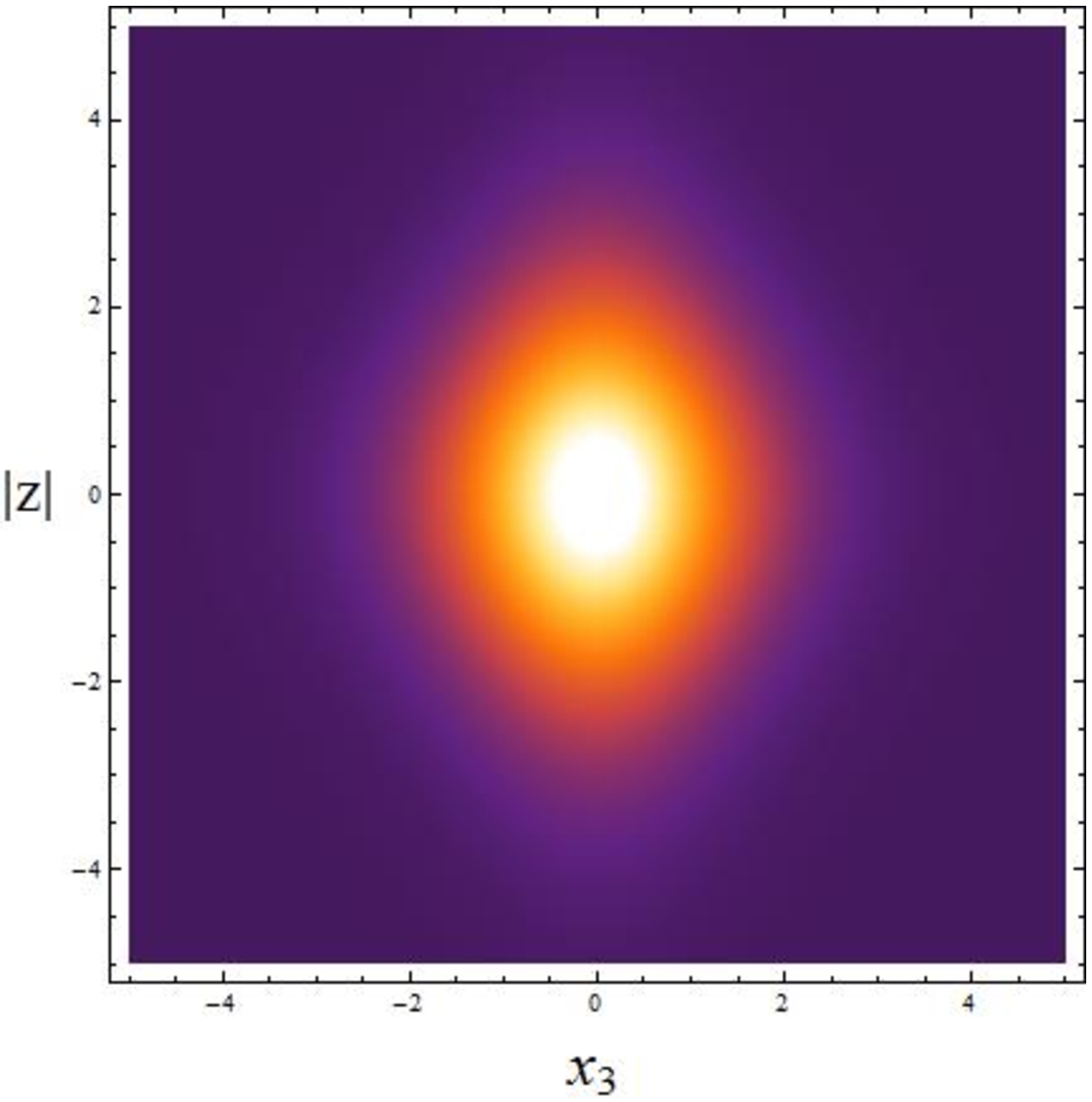} \\
(a) $\Tr[E_i \Phi]$ & (b) $\p_i \Tr[E_i \Phi]$ & (c) $i \Tr[Q M \D_0 Q^\dagger - \D_0 Q M Q^\dagger]$
\end{tabular}
\end{center}
\caption{(a) The electric flux of the $SU(2)$ vortex-dyon configuration.
The electric flux decays with the Yukawa type behavior $e^{-g v x}$. 
(b) The electric charge density.  
(c) The global $U(1)$ charge.
The plots for negative $|z|$ are simply mirror images in order to
illustrate the cross section of the configuration.}
\label{fig:SU(2)_dyon}
\end{figure}

Note that the Noether charge $\mathbf q$ cannot be interpreted as 
the electric charge of a dyon, whether or not it gets Witten's
correction due to the vacuum angle $\theta$.  
This is because the electric flux is screened by the scalar field $Q$
and decays with the Yukawa type behavior $e^{-g v x}$, 
so that the total electric charge should be zero. 
On the other hand, the scalar field is charged under the unbroken 
{\it global}~  $U(1)^n$  symmetry and 
accordingly the vector $\mathbf q$ can be interpreted as 
the associated conserved charges.
Indeed, we can see that the global $U(1)^n$ charges contribute 
to the BPS mass by rewriting the energy in the following way:
\beq
E &=& \int d^3 x \, \Tr \bigg[ \frac{1}{g^2} \Big| B_i - \cos \mu \, \D_i \Phi
- \delta_{i3} g^2 ( \Tr [Q^\dagger t^\alpha Q] t^\alpha - \xi t^0 )  \Big|^2 +
\frac{1}{g^2} \Big| E_i - \sin \mu \, \D_i  \Phi \Big|^2 \notag  \\
&& \hs{18} + 4\, |\D_{\bar z} Q|^2 + |\D_3 Q + \cos \mu (\Phi Q - Q M)|^2 + |\D_0
Q + i \sin \mu (\Phi Q - Q M)|^2 \phantom{\bigg]}  \notag \\
&& \hs{18}  + \frac{1}{g^2} | \D_0 \Phi |^2 - \sin \mu \, \Phi \left( \frac{2}{g^2}  
\D_i E_i + \frac{i2}{g^2} [\Phi, \D_0 \Phi] 
+ i [ Q (\D_0 Q)^\dagger - \D_0 Q
Q^\dagger ] \right) \notag \\
&& \hs{18} - v^2 B_3 + \frac{2}{g^2} \D_i [\Phi B_i] \cos \mu + \sin
\mu \left(
  \frac{2}{g^2}  \D_i [\Phi E_i ] + i  [ Q M (\D_0 Q)^\dagger - \D_0 Q M
  Q^\dagger ] \right)  \bigg]  \notag \\
&\geq & \int dx_3 \, T_v + \mathbf m \cdot \left[ \frac{4\pi}{g^2} \mathbf g
  \cos \mu + \left( \mathbf q_e + \mathbf q_f\right) \sin \mu  \right]\ ,
\eeq
where the charges are defined by
\beq
&\displaystyle\mathbf g \equiv \frac{1}{2\pi} \int d^3 x \, 
\p_i \Tr( B_i \mathbf H )\ , \hs{10} 
\mathbf q_e \equiv \frac{2}{g^2} \int d^3 x \, 
\p_i \Tr( E_i \mathbf H)\ , &\\
&\displaystyle
\mathbf q_f \equiv i \int d^3 x \Tr \left[ Q \mathbf H (\D_0 Q)^\dagger - \D_0 Q
  \mathbf H Q^\dagger \right]\ . & \phantom{\bigg[}
 \label{eq:qf}
\eeq
Since the electric charge $\mathbf q_e$ should be zero for $\theta=0$
(see Fig.\,\ref{fig:SU(2)_dyon}), 
the BPS mass is given by the magnetic charge $\mathbf g$ and 
the $U(1)^n$ global charges $\mathbf q_f$. The relation between
$\mathbf q$ in
Eq.\,(\ref{eq:charge_density}) and $\mathbf q_f$  in Eq.\,(\ref{eq:qf}) is $
\mathbf q\,=\, \mathbf q_e\,+\,\mathbf q_f $.  

The dyonic configurations discussed in this section
have parallel charge vectors: 
$\mathbf q$ and $\mathbf g$ are proportional to the root vectors $\bs \alpha$ 
and the coroot vectors $\tilde{\bs \alpha}$, respectively. 
We can construct dyonic configurations with $\mathbf q \not \propto \mathbf g$
if we introduce complex mass parameters 
\beq
M ~=~ M_1 + i M_2 ~=~ (\mathbf m_1 + i \mathbf m_2) \cdot \mathbf H\ .
\eeq 
By using a method similar to that discussed in section \ref{massdeform}, 
we can easily check that the potential of the effective theory becomes
\beq
V_{\rm eff} ~=~ \frac{4\pi}{g^2} \sum_{a=1}^2 \Tr \left[ X^{-1} \{ \mathbf m_a \cdot
  \mathbf H_n , B^\dagger\} Y^{-1} \{  \mathbf m_a \cdot \mathbf H_n , B \}
\right]\ .  
\eeq
In this case, we can rewrite the energy of the effective theory as
\beq
E &=& \frac{4\pi}{g^2} \int dx_3 \Tr \bigg[ X^{-1} \left( \p_{x_3} B ^\dagger -
 \{ \tilde{\mathbf m}_1 \cdot \mathbf H_n , B^\dagger \} \right) Y^{-1} \left(
 \p_{x_3} B- \{ \tilde{\mathbf m}_1 \cdot \mathbf H_n , B \} \right)  \notag  \\
&& \hs{20} + X^{-1} \left( \p_t B ^\dagger + i \{ \tilde{\mathbf m}_2 \cdot \mathbf H_n , B^\dagger \} \right) Y^{-1} \left( \p_t B - i \{ \tilde{\mathbf m}_2 \cdot \mathbf H_n , B \} \right) \bigg] \notag  \\
&& + \frac{4\pi}{g^2} \tilde{\mathbf m}_1 \cdot \mathbf g + \tilde{\mathbf m}_2 \cdot \left(  \mathbf q + \frac{\theta}{2\pi} \mathbf g \right)\ , 
\label{eq:dyon4}
\eeq
where $\tilde{\mathbf m}_i~(i=1,2)$ are defined by
\beq
\ba{c} \tilde{\mathbf m}_1 \\ \tilde{\mathbf m}_2 \ea = \ba{cc} \cos \mu & - \sin \mu \\ \sin \mu & \cos \mu \ea \ba{c} \mathbf m_1 \\ \mathbf m_2 \ea.
\eeq
Since the quantity in the last line of Eq.\,\eqref{eq:dyon4} is extremized by
\beq
\tan \mu = - \frac{\frac{4\pi}{g^2} \mathbf m_2 \cdot \mathbf g - \mathbf m_1 \cdot \left( \mathbf q + \frac{\theta}{2\pi} \mathbf g \right)}{\frac{4\pi}{g^2} \mathbf m_1 \cdot \mathbf g + \mathbf m_2 \cdot \left( \mathbf q + \frac{\theta}{2\pi} \mathbf g \right)},
\eeq
the lower bound on the energy is given by the central charge 
\beq
E ~\geq~ |Z| ~=~ |(\mathbf m_1 + i \mathbf m_2) \cdot \left( \mathbf g + \tau \mathbf q \right) |.
\eeq
The BPS equations can be easily solved as
\beq
B ~=~ e^{(\tilde{\mathbf m}_1 x_3 + i \tilde{\mathbf m}_2 t) \cdot \mathbf H_n } B_0 \,
e^{(\tilde{\mathbf m}_1 x_3 + i \tilde{\mathbf m}_2 t) \cdot \mathbf H_n }\ .
\eeq
The $n$-by-$n$ constant matrix $B_0$ is related to the conserved charge 
$\mathbf q$ via Eq.\,\eqref{eq:charge_density}.
The right hand side of Eq.\,\eqref{eq:charge_density} is 
a function of the kink positions and phases contained in $B_0$, 
so that the parameters should obey some constraints 
determined by the conserved charge $\mathbf q$. 
This means that some of the kinks form bound states, 
which correspond to the 1/4 BPS dyons \cite{Hashimoto:1998zs} 
attached to vortex strings. 
We can also show that if the angle between 
$\mathbf g$ and $\mathbf q$ takes on a generic value, 
the right hand side of Eq.\,\eqref{eq:charge_density} 
has an upper bound determined by the mass parameters. 
Therefore, there is a finite number of 
elements of the root lattice $\mathbf q$ 
in a generic direction\footnote{
Note that there exist infinite towers of 
the conserved charge vector $\mathbf q$ 
in some specific directions such as $\mathbf q \propto \mathbf g$.}
and the number of BPS bound states changes as we vary the mass parameters. 
This is also analogous to the case of the corresponding $4d$ theory. 
The composite configurations should be 
1/4 BPS \cite{Eto:2005sw} while explicit checks remain a future problem.

\section{Summary}

In this paper we have studied the effects of mass deformation on the vortex
effective worldsheet actions  in $U(1) \times G$ gauge theories with
$G=SO(2n)$, $USp(2n)$, and $SU(N)$. 
The moduli spaces of the vortex solutions, $SO(2n)/U(n)$,
$USp(2n)/U(n)$, and $ SU(N)/U(N-1) \sim \mathbb C P^{N-1}$, respectively,
arising from the exact color-flavor symmetries broken by vortex
configurations, are replaced under mass deformation by a finite number
of minimum-tension vortex solutions (vortex vacua). Kinks develop
along the vortex string, connecting different vortex vacua,  which
turn out to be the ordinary three-dimensional monopoles trapped inside
the vortex core. The structure of these kinks (monopoles) have been
analyzed systematically here (see also \cite{Arai:2011gg}).

In the case of the $U(N)$ theory this $2d$-$4d$ correspondence was
shown to survive quantum mechanically in the theory with no mass
deformation \cite{Hanany:2004ea,Gorsky:2004ad}, thus providing a
highly non-trivial realization of the $2d$-$4d$ duality proposed
earlier by Dorey et.~al.~\cite{Dorey}. In the case explicitly examined
from this point of view, the massless $U(1)\times SU(N)$ theory with
$\NF=N$ flavors, the vortex $\mathbb C P^{N-1}$ fluctuations
dynamically Abelianize at low energies, in perfect agreement
\cite{DKO} with the physics of quantum $r=0$ vacuum of
four-dimensional ${\cal N}=2$ SQCD \cite{CKM}.   
In the case of $SO(2n)$ and $USp(2n)$ gauge theories such a check is
the subject of a future study. 

 We have studied in this paper the properties of  local (i.e., ANO-like) vortices in
$SO(2n) \times U(1)$ and $USp(2n)\times U(1)$ gauge theories
with $\NF=2n$ flavors  in the color-flavor locked vacuum,  Eq.(2.5).       
It has been noted \cite{Eto:2008qw,Eto:2009bg}  that in theories other than $SU(N)$, the system possesses 
more general types of vortex solutions, such as fractional or semi-local vortices (vortex moduli), even with the minimum number of flavors needed for the system to possess the color-flavor locked phase, Eq.~(\ref{eq:vev}). 
Furthermore, the system possesses a large class 
of vacua besides the particular vacuum considered here, Eq.~(\ref{eq:vev}):  the above mentioned general class of vortex solutions is related to the existence of  such nontrivial vacuum moduli. This means that 
 domain walls can also be formed connecting different vacua.   
Accordingly,  equations (\ref{eq:BPS1})--(\ref{eq:BPS2}) themselves admit  more general classes of solutions involving, e.g.,  
 both domain walls and vortices connecting them, such as those studied in Refs.~\cite{Isozumi:2004vg,Eto:2006pg,Eto:2008mf}.    The mass deformation of semi-local vortices induces  domain
walls in the bulk in general, though this is not always the case. A related comment is given in Appendix \ref{appendix:wall}.

\subsection*{Acknowledgments} 

The work of M.~E. is supported in part by Grant-in Aid for Scientific
Research (No. 23740226). 
S.~B.~G gratefully acknowledges a Golda Meir post-doctoral fellowship. 
The work of Y.~J. is supported
by the NSF of China under Grant No.10875129.
The work of M.~N.  is supported in part by 
Grant-in Aid for Scientific Research (No. 23740198) 
and by the ``Topological Quantum Phenomena'' 
Grant-in Aid for Scientific Research 
on Innovative Areas (No. 23103515)  
from the Ministry of Education, Culture, Sports, Science and Technology 
(MEXT) of Japan.  K.~K. thanks IPMU and Keio University for hospitality.  

\appendix

\section{Domain wall}\label{appendix:wall}

As described in the main text, 
the 1/4 BPS equations (\ref{eq:BPS1})--(\ref{eq:BPS2}) 
admit domain walls and 
vortices stretched between domain walls 
\cite{Isozumi:2004vg,Eto:2006pg,Eto:2008mf},
when the VEVs of the scalar fields
are different at $x_3 \rightarrow - \infty$ and $x_3 \rightarrow \infty$. 
In that case, the third term in Eq.~(\ref{eq:bulk_BPS_energy}) 
gives the domain wall charge. 
In the case of $G=SO(2n), USp(2n)$ with $N_f=2n$ and 
non-degenerate mass matrix, $M$,
 there exist many isolated 
$r$-vacua where a subgroup of $G$ remains unbroken, 
in addition to the Higgs vacua in Eq.~(\ref{eq:vev}).  
Therefore we need a special condition 
for prohibiting the creation of domain walls 
interpolating their vacua as a 1/4 BPS solution.
To illustrate this, it is convenient to 
consider the following $G$-invariant quantity which 
parametrizes vacuum moduli for the massless case $M=0$,
\begin{eqnarray}
 R=\frac{2n}{\tr(J^\dagger Q^{\rm T}JQ)} Q^{\rm T}JQ.
\end{eqnarray}
The first two equations in the 1/4 BPS equations can be easily solved 
in terms of the invariant $R$ as 
\begin{eqnarray}
 0=\partial_{\bar z}R,\quad 0=\partial_3 R-M^{\rm T}R-R\,M,\quad
  \Rightarrow \quad R= e^{M^{\rm T}x_3}R_0(z)e^{M x_3},
\end{eqnarray}
where we have used the fact 
that $M$ is an element of $so(2n), usp(2n)$, that is 
$M^{\rm T}J+JM=0$. In the case of $M=0$, $R=R_0(z)$
is the so-called rational map describing lump (semi-local vortex)
solutions. Here, we require that a composite state
becomes  the vacuum given by Eq.~(\ref{eq:vev}) 
 at $|x_3|=|z|\to \infty$, that is, $R\to  J$. 
According to the above solution, 
this requirement can be formulated more strictly as
\begin{equation}
 M^{\rm T}R+RM=0\;.
\end{equation}
Local vortices, which defined by $R=J$, obviously satisfies
this condition, and vortices discussed in this paper are only of local
type. A generic solution for the condition is a set of 
fractional vortices with vanishing size moduli, $R_0(z)_{ab}=r_a(z)
J_{ab}=r_b(z)\,J_{ab}$. A single local vortex is formed by coincident
fractional vortices, $r_a(z)=r_b(z),{a\not=b}$.  One way to guarantee the solution to satisfy the above condition is to 
require invariance 
of the solution under the rotation  $z\to e^{i\theta}z$. This  forces
$R_0(z)$ to be constant:   this turns out to be the condition for  vortices to be local
(see Eq. (35) of \cite{Eto:2008yi}). 
According to the principle of symmetric criticality \cite{Palais}, 
this property guarantees consistency of the 
low-energy effective action on 
the local vortices
 discussed in this paper. 
 If such a condition is not met, on the other hand, 
it is inevitable for non-vanishing size moduli of vortices
to increase indefinitely   along the vortex string and
create a domain wall bending logarithmically.   

\section{Lie algebra}\label{appendix:convention}
In this appendix, we summarize the conventions for
the Lie algebra used in this paper. 
The root vectors of $G=SU(N)$, $SO(2n)$, 
$USp(2n)$ and $SO(2n+1)$ are given in Table \ref{tab:roots}.
\begin{table}[h]
\begin{center}
\begin{tabular}{|c|ll|ll|}
\hline
$G$ & positive roots & & simple roots & \\ \hline
$SU(N)$ & 
$\bs e_i - \bs e_j$ & 
$(1 \leq i < j \leq N)$ & 
$\bs e_i - \bs e_{i+1}$ &
$(i = 1, \cdots, N-1)$ \\ \hline
$SO(2n)$ & 
$\bs e_i \pm \bs e_j$ &
$(1 \leq i < j \leq n)$ & 
$\bs e_i - \bs e_{i+1},$ $\bs e_{n-1} + \bs e_n$ &
$(1 \leq i < j \leq n-1)$ \\ \hline
$USp(2n)$ 
& $\bs e_i \pm \bs e_j, ~2 \bs e_i$
& $(1 \leq i < j \leq n)$ 
& $\bs e_i - \bs e_{i+1},~2 \bs e_{n}$ 
&$(1 \leq i < j \leq n-1)$\\ \hline 
$SO(2n+1)$ 
& $\bs e_i \pm \bs e_j, ~~ \bs e_i$
& $(1 \leq i < j \leq n)$ 
& $\bs e_i - \bs e_{i+1},~~ \bs e_{n}$ 
&$(1 \leq i < j \leq n-1)$ \\ \hline 
\end{tabular}
\caption{The root vectors of $G=SU(N)$, $SO(2n)$ and $USp(2n)$. The set of vectors $\{\bs e_i\}$ is the standard orthonormal basis $\bs e_i \cdot \bs e_j = \delta_{ij}$.}
\label{tab:roots}
\end{center}
\end{table}
The generators of the Lie algebra are decomposed into 
the standard Cartan basis: the generators of the Cartan subalgebra 
$\mathbf H = (H_1,H_2,\cdots,H_r) ,~(r={\rm Rank} \, G)$, 
the raising operators $E_{\bs \alpha}$ and 
the lowering operators $E_{-\bs \alpha} = E_{\bs \alpha}^\dagger$. 
Their commutation relations are given by
\beq
[\mathbf H, E_{\pm \boldsymbol \alpha}] = \pm \boldsymbol \alpha E_{\pm
  \boldsymbol \alpha}\ , \hs{10} 
[E_{\boldsymbol \alpha}, E_{- \boldsymbol \alpha}] =\tilde{\boldsymbol \alpha}
\cdot \mathbf H \ , \hs{10}
[E_{\pm \boldsymbol \alpha}, E_{\pm \boldsymbol \beta}] = N_{\pm \boldsymbol
  \alpha, \pm \boldsymbol \beta} E_{\pm \boldsymbol \alpha \pm \boldsymbol
  \beta}\ ,
\eeq
where $\bs \alpha$ and $\bs \beta$ are positive root vectors
and $N_{\pm \boldsymbol \alpha, \pm \boldsymbol \beta}$ are constants. 
The coroot vectors $\tilde{\bs \alpha}$ are defined by
\beq
\tilde{\bs \alpha} ~\equiv~ 2 \, \frac{\bs \alpha}{\bs \alpha \cdot \bs \alpha}\ .
\eeq
The coroot vectors of $G$ are the root vectors of the dual group $\tilde G$ :
$SU(N)$ and $SO(2n)$ are self-dual while $USp(2n)$ and $SO(2n+1)$ are 
dual to each other.
The generators of the Cartan subalgebra $\mathbf H$ are 
the diagonal matrices whose eigenvalues are the weight vectors 
of the corresponding representation
\beq
{\renewcommand{\arraystretch}{0.7}
{\arraycolsep 0.5mm
\mathbf H = \left( \begin{array}{cccc} \boldsymbol \mu_1 & & & \\ &
  \boldsymbol \mu_2 & & \\ & & \ddots & \\ & & & \boldsymbol
  \mu_N \end{array} \right)}}\ .
\eeq

\subsection{$G=SU(N)$}
For the fundamental representation of $SU(N)$, 
the weight vectors are 
\beq
\boldsymbol \mu_i ~=~ \boldsymbol e_i - \frac{1}{N} \sum_{i=1}^{N} \boldsymbol
e_i\ .
\eeq
Therefore the generators of the Cartan subalgebra 
$\mathbf H = (H_1,\cdots,H_N)$ take the form
\beq
{\renewcommand{\arraystretch}{0.7}
{\arraycolsep 1mm
H_1 = \ba{cccc} 1-\frac{1}{N} & & & \\ & -\frac{1}{N} & & \\ & & \ddots & \\ &
& & -\frac{1}{N} \ea, \hs{5} \cdots, \hs{5} H_N = \ba{cccc} -\frac{1}{N} & & &
\\ & \ddots & & \\ & & -\frac{1}{N} & \\ & & & 1-\frac{1}{N} \ea}}\ .
\eeq
Note that there are $N-1$ independent matrices since $(\bs e_1 + \cdots + \bs e_N) \cdot \mathbf H = 0$.
For the positive root $\alpha = \bs e_i - \bs e_j~(i>j)$, 
the raising operator $E_{\bs \alpha}$ and 
the lowering operator $E_{-\bs \alpha}$ are given by
\beq
E_{\bs e_i - \bs e_j} ~=~ 
{\renewcommand{\arraystretch}{0.7}
{\arraycolsep 1.5mm
\begin{array}{c} \\ i \\ \vb{8} \\ \end{array}
\overset{\displaystyle \begin{array}{cccc} &  & ~~j & \end{array}}
{
\ba{cccc} 
0 &        &        &   \\
  & \ddots & 1      &   \\
  &        & \ddots &   \\
  &        &        & 0 
\ea
}, 
\hs{10} 
E_{-(\bs e_i - \bs e_j)} = 
\begin{array}{c} \\  \\ j \\ \vb{1} \end{array}
\overset{\displaystyle \begin{array}{cccc} & i & ~~~ & \end{array}}
{
\ba{cccc} 
0 &        &        &   \\
  & \ddots &        &   \\
  & 1      & \ddots &   \\
  &        &        & 0 
\ea
}. }}
\eeq
For the highest weight vector $\tilde{\bs \nu}_h$ of 
the $SU(N)$ fundamental representation, 
the matrix $\lambda_h $ 
is given by
\beq
\lambda_h ~=~ \nu_0 \mathbf 1_N + \tilde{\bs \nu}_h \cdot \mathbf H ~=~
\ba{c|c} 1 & \\ \hline & \mathbf 0_{N-1} \ea, \hs{10} \nu_0 = \frac{1}{N},
\hs{5} \tilde{\bs \nu}_h = \bs e_1 - \frac{1}{N} \sum_{i=1}^N \bs e_i\ . 
\eeq
\subsection{$G=SO(2n)$}
The weight vectors of the $2n$-dimensional representation of $SO(2n)$ are
\beq
\boldsymbol \mu_i = \boldsymbol e_i\ , \hs{5} 
\boldsymbol \mu_{i+n} = - \boldsymbol e_i~~~(1<i<n)\ .
\eeq
Therefore the generators of the Cartan subalgebra 
$\mathbf H = (H_1,\cdots,H_n)$ take the form
\beq
{\renewcommand{\arraystretch}{0.7}
H_i = 
\begin{array}{c}
      \\
i     \\
\vb{10}\\
i+n   \\
\vb{6}\\
\end{array}
\overset{\displaystyle 
\begin{array}{cccccc} & & i & & ~~~~~~i+n & \end{array}}{
\ba{ccc|ccc}
\ &   &        \ \ & \        &    & \ \\
\ & 1 &        \ \ & \        &    & \ \\
\ &   & \vb{6} \ \ & \        &    & \ \\ \hline
\ &   &        \ \ & \        &    & \ \\
\ &   &        \ \ & \        & -1 & \ \\
\ &   &        \ \ & \ \vb{6} &    & \ \ea}}.
\eeq
The raising operators $E_{\bs \alpha}$ for the positive roots 
$\bs \alpha = \bs e_i - \bs e_j,~\bs e_i + \bs e_j,~(1 \leq j < i \leq n)$
are 
\beq
{\renewcommand{\arraystretch}{0.6}
E_{\bs e_i - \bs e_j} = 
\begin{array}{c}
    \\
i   \\ \vs{10}
    \\
j+n \\
    \\
\end{array}
\overset{\displaystyle 
\begin{array}{cccccc} & & \ j & & i+n & \end{array}}{
\ba{ccc|ccc}
\ &  &        \ &        & & \\
\ &  & 1      \ &        & & \\
\ &  & \vb{7} \ &        & & \\ \hline
\ &  &        \ & \vb{7} & & \\
\ &  &        \ & \ -1   & & \\
\ &  &        \ &        & & \ea}}, \hs{7}
{\renewcommand{\arraystretch}{0.7}
E_{\bs e_i + \bs e_j} = 
\begin{array}{c} \vb{4}
i       \\
\vs{-2} \\
j       \\ 
        \\
        \\
        \\
\end{array}
\overset{\displaystyle 
\begin{array}{cccccc} & & & & ~~i+n & j+n \end{array}}{
\ba{ccc|ccc}
~~~ &  &        \ &         & & 1 \vb{6} \ \\
~~~ &  &        \ & \vs{-2} & &          \ \\
~~~ &  &        \ & -1      & &          \ \\
~~~ &  &        \ & \vs{-3} & &          \ \\ \hline
~~~ &  &        \ &         & &          \ \\
~~~ &  &        \ &         & &          \ \\
~~~ &  &        \ &         & &          \ \ea}}, 
\eeq
The lowering operators $E_{-\bs \alpha}$ for 
$\bs \alpha = \bs e_i - \bs e_j,~\bs e_i + \bs e_j,~(1 \leq j < i \leq n)$
are
\beq
{\renewcommand{\arraystretch}{0.6}
E_{-(\bs e_i - \bs e_j)} = 
\begin{array}{c} 
    \\
j   \\ \vs{4}
    \\
i+n \\
    \\
\end{array}
\overset{\displaystyle 
\begin{array}{cccccc} \ \ \ i & & & & & j + n \end{array}}{
\ba{ccc|ccc}
\, \vb{7} &  & ~~ & & &    \\
\, 1      &  & \ & & &    \\
\,        &  & \ & & &    \\ \hline
\,        &  & \ & & &    \\
\,        &  & \ & & & -1 \\
\,        &  & \ & & & \vb{7} \ea}}, \hs{2}
{\renewcommand{\arraystretch}{0.7}
E_{-(\bs e_i + \bs e_j)} = 
\begin{array}{c} \vb{4}
       \\
       \\ 
       \\
i+n    \\
\vs{-2}\\
j+n    \\
\end{array}
\overset{\displaystyle 
\begin{array}{cccccc} i & \, ~~ & j& & & ~~\end{array}}{
\ba{ccc|ccc}
~ &       &        \ &        & & \hs{5} \\
~ &       &        \ & & &       \ \\
~ &       &        \ &      &  & \ \\ \hline
~ &       & -1 \vb{6}    ~ &       & &        \ \\
~ &       & \vs{-3} ~ &       & &        \ \\
~\underset{}{1} \vb{6} &  & \ &       & &        \ \ea}}. \notag
\eeq
For the highest weight vector $\tilde{\bs \nu}_h$ of 
the $SO(2n)$ Weyl spinor representation, the matrix $\lambda_h$ is given by
\beq
\lambda_h ~=~ \nu_0 \mathbf 1_{2n} + \tilde{\bs \nu}_h \cdot \mathbf H ~=~ 
\ba{c|c} \mathbf 1_n & \\ \hline & \mathbf 0_n \ea, \hs{10} \nu_0 =
\frac{1}{2}\ , \hs{5} \tilde{\boldsymbol \nu}_h = \frac{1}{2} \sum_{i=1}^n
\boldsymbol e_i\ .
\eeq

\subsection{$G=USp(2n)$}
The weight vectors of the $2n$-dimensional representation of $USp(2n)$
are given by
\beq
\boldsymbol \mu_i = \boldsymbol e_i, \hs{5} 
\boldsymbol \mu_{i+n} = - \boldsymbol e_i~~~(1<i<n)\ .
\eeq
Therefore the generators of the Cartan subalgebra 
$\mathbf H = (H_1,\cdots,H_n)$ take the form
\beq
{\renewcommand{\arraystretch}{0.7}
H_i = 
\begin{array}{c}
      \\
i     \\
\vb{10}\\
i+n   \\
\vb{6}\\
\end{array}
\overset{\displaystyle 
\begin{array}{cccccc} & & i & & ~~~~~~i+n & \end{array}}{
\ba{ccc|ccc}
\ &   &        \ \ & \        &    & \ \\
\ & 1 &        \ \ & \        &    & \ \\
\ &   & \vb{6} \ \ & \        &    & \ \\ \hline
\ &   &        \ \ & \        &    & \ \\
\ &   &        \ \ & \        & -1 & \ \\
\ &   &        \ \ & \ \vb{6} &    & \ \ea}}.
\eeq
The raising operators $E_{\bs \alpha}$ for the positive roots 
$\bs \alpha = \bs e_i - \bs e_j~(1 \leq j < i \leq n)$ and 
$\bs \alpha = \bs e_i + \bs e_j~(1 \leq j \leq i \leq n)$
are 
\beq
{\renewcommand{\arraystretch}{0.6}
E_{\bs e_i - \bs e_j} = 
\begin{array}{c}
    \\
i   \\ \vs{10}
    \\
j+n \\
    \\
\end{array}
\overset{\displaystyle 
\begin{array}{cccccc} & & \ j & & i+n & \end{array}}{
\ba{ccc|ccc}
\ &  &        \ &        & & \\
\ &  & 1      \ &        & & \\
\ &  & \vb{7} \ &        & & \\ \hline
\ &  &        \ & \vb{7} & & \\
\ &  &        \ & \ -1   & & \\
\ &  &        \ &        & & \ea}}, 
\hs{7}
{\renewcommand{\arraystretch}{0.7}
E_{\bs e_i + \bs e_j} = 
\begin{array}{c} \vb{4}
i       \\
\vs{-2} \\
j       \\ 
        \\
        \\
        \\
\end{array}
\overset{\displaystyle 
\begin{array}{cccccc} & & & & ~~i+n & j+n \end{array}}{
\ba{ccc|ccc}
~~~ &  &        \ &         & & 1 \vb{6} \ \\
~~~ &  &        \ & \vs{-2} & &          \ \\
~~~ &  &        \ & ~~ 1     & &          \ \\
~~~ &  &        \ & \vs{-3} & &          \ \\ \hline
~~~ &  &        \ &         & &          \ \\
~~~ &  &        \ &         & &          \ \\
~~~ &  &        \ &         & &          \ \ea}},
\eeq
The lowering operators $E_{-\bs \alpha}$ for 
$\bs \alpha = \bs e_i - \bs e_j~(1 \leq j < i \leq n)$ and 
$\bs \alpha = \bs e_i + \bs e_j~(1 \leq j \leq i \leq n)$
are 
\beq
{\renewcommand{\arraystretch}{0.6}
E_{-(\bs e_i - \bs e_j)} = 
\begin{array}{c} 
    \\
j   \\ \vs{4}
    \\
i+n \\
    \\
\end{array}
\overset{\displaystyle 
\begin{array}{cccccc} \ \ \ i & & & & & j + n \end{array}}{
\ba{ccc|ccc}
\, \vb{7} &  & ~~ & & &    \\
\, 1      &  & \ & & &    \\
\,        &  & \ & & &    \\ \hline
\,        &  & \ & & &    \\
\,        &  & \ & & & -1 \\
\,        &  & \ & & & \vb{7} \ea}, \hs{2}}
{\renewcommand{\arraystretch}{0.7}
E_{-(\bs e_i + \bs e_j)} = 
\begin{array}{c} \vb{4}
       \\
       \\ 
       \\
i+n    \\
\vs{-2}\\
j+n    \\
\end{array}
\overset{\displaystyle 
\begin{array}{cccccc} i &  ~ & j& & & ~~\end{array}}{
\ba{ccc|ccc}
~ &       &        \ &        & & \hs{5} \\
~ &       &        \ & & &       \ \\
~ &       &        \ &      &  & \ \\ \hline
~ &       & 1 \vb{6}    ~~ &       & &        \ \\
~ &       & \vs{-3} ~ &       & &        \ \\
~\underset{}{1} \vb{6} &  & \ &       & &        \ \ea}}.
\eeq

For the highest weight vector $\tilde{\bs \nu}_h$ of 
the $SO(2n+1)$ spinor representation, the matrix $\lambda_h$ is given by
\beq
\lambda_h ~=~ \nu_0 \mathbf 1_{2n} + \tilde{\bs \nu}_h \cdot \mathbf H ~=~ 
\ba{c|c} \mathbf 1_n & \\ \hline & \mathbf 0_n \ea, \hs{10} \nu_0 =
\frac{1}{2}\ , \hs{5} \tilde{\boldsymbol \nu}_h = \frac{1}{2} \sum_{i=1}^n
\boldsymbol e_i\ .
\eeq

\section{The ordering of the lowering operators}\label{appendix:ordering}
In general, the matrix $\Lambda_0$ takes the following form
\beq
\Lambda_0 ~=~ (\Lambda_h w) \exp [ b_{j_1} E_{-\bs \alpha_{j_1}} ] \cdots \exp [
b_{j_q} E_{-\bs \alpha_{j_q}} ] R\ ,
\eeq
where $w$ is an element of the Weyl group and $R$ is a product of the
raising operators 
\beq
w = w_{i_1} \cdots w_{i_p}, \hs{10} R = \exp [ c_{i_1} E_{\bs \alpha_{i_1}} ]
\cdots \exp [ c_{i_p} E_{\bs \alpha_{i_p}} ]\ .
\eeq
In this appendix, we show that the matrix $\Lambda$ can always be 
rewritten as
\beq
\Lambda_0 &\sim& (\Lambda_h w \, w_{j_1}) \exp [ b_{j_2}' E_{-\bs \alpha_{j_2}} ]
\cdots \exp [ b_{j_q}' E_{-\bs \alpha_{j_q}} ] \exp[ {b'}_{j_1}^{-1} E_{\bs
  \alpha_{j_1}} ] R\ .
\label{eq:abcd}
\eeq

By using the decomposition formula \eqref{eq:Bruhat}, 
we can rewrite $\Lambda_0$ as
\beq
\Lambda_0 \sim (\Lambda_h w \, w_{j_1}) \exp [ a_{j_1}^{-1} E_{\bs \alpha_{j_1}}
] \exp [ a_{j_2} E_{-\bs \alpha_{j_2}} ] \cdots \exp [ a_{j_q} E_{-\bs
  \alpha_{j_q}} ] R\ .
\eeq
If $j_1 \not \in \{j_2, \cdots , j_q \}$, we can move 
the operator $\exp [ a_{j_1}^{-1} E_{\bs \alpha_{j_1}} ]$ 
to the right of all the lowering operators.
Then we obtain the matrix $\Lambda_0$ of the form \eqref{eq:abcd}. 

On the other hand, if there exists the same 
lowering operator $(\exists j_r = j_1)$, 
we cannot exchange the positions of the operators
\beq
\exp [ a_{j_1}^{-1} E_{\bs \alpha_{j_1}} ] \exp [ a_{j_r} E_{-\bs
  \alpha_{j_1}} ] \not = \exp [ a_{j_r} E_{-\bs \alpha_{j_1}} ] \exp [
a_{j_1}^{-1} E_{\bs \alpha_{j_1}} ]\ .
\eeq
Instead, we use the following formula
\beq
\exp [ b_{j_1}^{-1} E_{\bs \alpha_{j_1}} ] \exp [ b_{j_r} E_{-\bs
  \alpha_{j_1}} ] \hs{-2} &=& \hs{-2} \exp [ \log (1 + b_{j_1}^{-1} b_{j_r})
\tilde{\bs \alpha}_{j_1} \cdot \mathbf H ] \notag \\
&\times& \exp [ b_{j_r} (1 + b_{j_1}^{-1} b_{j_r}) E_{-\bs \alpha_{j_1}} ]
\exp [ b_{j_1}^{-1} (1 + b_{j_1}^{-1} b_{j_r})^{-1} E_{\bs \alpha_{j_1}} ]\ . 
\label{eq:formula1}
\eeq
Then the matrix $\Lambda_0$ can be rewritten as
\beq
\Lambda_0 &\sim& (\Lambda_h w w_{j_1}) \exp [ b_{j_2}' E_{-\bs \alpha_{j_2}} ] \cdots \exp [ b_{j_{r-1}}' E_{-\bs \alpha_{j_{r-1}}} ] \exp [ a_{j_r} (1 + a_{j_1}^{-1} a_{j_r}) E_{-\bs \alpha_{j_1}} ] \notag \\
&{}& \hs{13} \times \exp[ a_{j_1}^{-1} (1 + a_{j_1}^{-1} a_{j_r})^{-1} E_{\bs
  \alpha_{j_1}} ] \exp [ a_{j_{r+1}} E_{-\bs \alpha_{j_{r+1}}} ] \cdots \exp [
a_{j_q} E_{-\bs \alpha_{j_q}} ] R\ ,
\eeq
where the parameters $b_j'$ are given by
\beq
b_{j_i}' ~=~ b_{j_i} (1 + b_{j_1}^{-1} b_{j_r})^{\tilde{\bs \alpha}_{j_1}
  \cdot \bs \alpha_{j_i}}\ .
\eeq
Repeating this procedure for all $j_i = j_1$, 
we can move the raising operator $\exp[b \, E_{\bs \alpha_{j_1}}]$ to the 
right and consequently we obtain the matrix $\Lambda_0$ of the form 
\eqref{eq:abcd}.

\section{The quadric surface $Q^{2n-2}$ sigma models}\label{appendix:quadric}

As explained in detail in Ref.~\cite{GJK}, the non-Abelian
vortex in $U(1)\times SO(2n)$ has an irreducible orbit in the higher
winding $k=2$ case, which has as an effective low-energy theory on the
worldsheet the sigma model on the Hermitian symmetric space  
$Q^{2n-2} = \frac{SO(2n)}{SO(2)\times SO(2n-2)}$, which has the following
K\"ahler potential
\beq
K ~=~  2\,\beta\log\left(1+2\varphi^\dag\varphi+|\varphi^{\rm
T}\varphi|^2\right) \ ,
\eeq
giving rise to the Lagrangian \cite{Higashijima:1999ki,Higashijima:2000rp,GJK}
\beq 
\mathcal{L} ~=~ 8\,\beta \left\{
\frac{\p_\alpha\varphi^\dag\p_\alpha\varphi
+2\left|\varphi^{\rm T}\p_\alpha\varphi\right|^2}
{1+2\varphi^\dag\varphi + \left|\varphi^{\rm T}\varphi\right|^2}
-\frac{2\left|\varphi^\dag\p_\alpha\varphi +
\left(\varphi^\dag\bar{\varphi}\right)
\left(\varphi^{\rm T}\p_\alpha\varphi\right)\right|^2}
{\left[1+2\varphi^\dag\varphi
+\left|\varphi^{\rm T}\varphi\right|^2\right]^2}\right\} \ ,
\label{eq:LQN}
\eeq
where $\varphi$ is a complex ($2n-2$)-component vector. 
The Lagrangian is symmetric under the following transformation
\beq 
\varphi \to U \varphi \ , 
\eeq
where $U^\dag U = 1$ and $U^{\rm T} U = 1$. Choosing 
\beq 
U ~=~ e^{i M \vartheta} \ , 
\eeq
it is clear that $M$ has to be Hermitian and anti-symmetric and hence
purely imaginary. 
Now keeping only the zero mode upon compactification, we get
\beq
\varphi(t,z,\vartheta) ~=~ e^{i M \vartheta} \varphi_0(t,z) \ . 
\eeq
Inserting this field into the Lagrangian \eqref{eq:LQN} 
we obtain \cite{Arai:2009jd}
\begin{align}
\mathcal{L} ~=~ 8\beta \left\{
\frac{\p_\alpha\varphi^\dag\p_\alpha\varphi
  +2\left|\varphi^{\rm T}\p_\alpha\varphi\right|^2 
  +\varphi^\dag M^2 \varphi}
{1+2\varphi^\dag\varphi + \left|\varphi^{\rm T}\varphi\right|^2}
  -\frac{2\left|\varphi^\dag\p_\alpha\varphi +
  \left(\varphi^\dag\bar{\varphi}\right) 
  \left(\varphi^{\rm T}\p_\alpha\varphi\right)\right|^2
  +\left|\varphi^\dag M \varphi\right|^2}
{\left[1+2\varphi^\dag\varphi
+\left|\varphi^{\rm T}\varphi\right|^2\right]^2}\right\} \ ,
\end{align}
where we have used that $\varphi^{\rm T} M \varphi = 0$ due to the
anti-symmetry of the mass matrix. 
The vacuum equations read
\beq
\varphi^\dag M^2 \varphi ~=~ 0 \ , \qquad
\varphi^\dag M \varphi ~=~ 0 \ , 
\eeq
which for a generic choice of the mass matrix yields the only solution
$\varphi = 0$. Hence, we find the number of vacua to be \cite{Arai:2009jd}
\beq
n_{\rm vacua}^{SO(2n), k=2} ~=~ 2\,n  \,.
\eeq
This result is indeed expected as this irreducible orbit of the
corresponding vortex should transform as an $SO(2n)$  vector.

\end{document}